\documentclass[12pt,epsf]{article}

\usepackage{amsmath}
\usepackage{amssymb}
\usepackage{amsthm}
\usepackage{epsfig}
\usepackage[dvipsnames]{xcolor}

\newcommand{\prt}{\partial}

\def\L{{\mathcal L}}

\def\ni{{\noindent}}

\def\be{\begin{equation}}
\def\ee{\end{equation}}
\numberwithin{equation}{section}

\newcommand{\mja}[1]{{\color{red} \textbf{#1}}}

\theoremstyle{definition}

\newcommand{\comment}[1]{}

\begin{document}
\bibliographystyle{plain}

\title{{\Large\bf On the Whitham system for the radial nonlinear Schr\"odinger equation}}
\author{Mark~J.~Ablowitz, Justin~T.~Cole, Igor~Rumanov\footnote{corresponding author; e-mail: igor.rumanov@colorado.edu} \\
{\small Department of Applied Mathematics, University of Colorado, Boulder, CO 80309 USA} }

\maketitle

\bigskip

\begin{abstract}
Dispersive shock waves (DSWs) of the defocusing radial nonlinear Schr\"odinger (rNLS) equation in two spatial dimensions are studied. This equation arises naturally in Bose-Einstein condensates, water waves and nonlinear optics. A unified nonlinear WKB approach, equally applicable to integrable or nonintegrable partial differential equations, is used to find the rNLS Whitham modulation equation system in both physical and hydrodynamic type variables. The description of DSWs obtained via Whitham theory is compared with direct rNLS numerics; the results demonstrate very good quantitative agreement. On the other hand, as expected, comparison with the corresponding DSW solutions of the one-dimensional NLS equation exhibits significant qualitative and quantitative differences.\footnote{We are pleased to contribute this article in honor of Professor Roger Grimshaw's 80th birthday. Roger Grimshaw is a leading expert in the mathematics of fluid dynamics, nonlinear waves and related approximation methods. He has written numerous important papers on Whitham theory and dispersive shock waves; in fluid dynamics these waves are often termed undular bores. The present article involves these topics and the equation we study is one which arises naturally in fluid dynamics.} 
\end{abstract}

%\newpage

\section{Introduction and main results}

The (1+1)-dimensional defocusing cylindrically symmetric radial nonlinear Schr\"odinger (rNLS) equation,

\be
i\epsilon\prt_t\Psi + \epsilon^2(\prt_{rr}+\frac{1}{r}\prt_r)\Psi - |\Psi|^2\Psi = 0, ~~\Psi(r,t=0) =\Psi_0(r), ~~0 \leq r < \infty  \label{eq:1.1}
\ee

\ni is an exact reduction of the defocusing (2+1)-dimensional nonlinear Schr\"odinger (NLS) equation,

\be
i\epsilon\prt_t\Psi + \epsilon^2(\prt_{xx}+\prt_{yy})\Psi - |\Psi|^2\Psi = 0, ~~\Psi(x,y,t=0) =\Psi_0(x,y), ~~|x|< \infty, |y|< \infty      \label{eq:1.2}   %\eqno(1.2)
\ee
A motivation to study the rNLS equation comes from the Bose-Einstein condensate experiments and the analysis in~\cite{HoeferEtAl06} where it was found numerically  that the rNLS equation provides a good approximation to the dynamics. 

\par In the experiments of~\cite{HoeferEtAl06} an initial hump of the Bose-Einstein condensate (BEC) density in the form of a ring was created by the laser beam effectively pushing the particles away from the center. Then the BEC expanded radially. An additional radial parabolic potential which is absent in eq.~(\ref{eq:1.1}) caused the BEC in experiments of~\cite{HoeferEtAl06} to move away from the center. Relative to this motion, however, the BEC expanded radially towards the {\it inside} which is what we consider analytically and numerically here.  

The dimensionless parameter $\epsilon$ characterizing dispersion, see \cite{HoeferEtAl06}, was very small in these experiments, $\epsilon\approx 0.012$. With such a small value of $\epsilon$ Whitham theory, which is a nonlinear WKB-expansion in $\epsilon$, is expected to be relevant. As pictures of the experiments show, in the expansion large oscillations of the BEC density were created which implies a dispersive shock wave started from the initial density jump. The oscillations formed a number of concentric rings i.e.~the DSW propagation was indeed radial with high degree of accuracy. Similar concentric rings forming a radial DSW were observed in nonlinear optics experiments~\cite{WJF-NatPhys07}.
\par However, in~\cite{HoeferEtAl06} only the one-dimensional NLS equation was treated analytically. Whitham modulation theory for the (1+1)-dimensional NLS (1d NLS) equation (see~\cite{GK87, Pa87}) was applied to this problem. While  qualitative agreement with experiments and direct simulations was found, and the solution/experiments exhibit character of a dispersive shock wave (DSW), the analytical results did not always correspond closely; see e.g.~Fig.~24 of~\cite{HoeferEtAl06}. Numerical results using the rNLS equation indicated that there were quantitative discrepancies of  amplitudes and position shifts from the Whitham 1d NLS DSW theory. 
\par In this paper we develop Whitham theory for the rNLS eq.~(\ref{eq:1.1}) and apply it to the quantitative study of radially symmetric  DSWs. We find significant analytical differences between Whitham theory associated with rNLS equation and that for the 1+1 NLS equation. 
\par  Whitham modulation theory~\cite{Whitham74} for spatially one-dimensional partial differential equations (PDEs) has been extensively studied for many years. Important equations and DSWs analyzed by Whitham theory include the integrable Korteweg-de Vries (KdV)~\cite{Whitham74, GurPit73} and the 1d NLS equation~\cite{ForLee86, Pa87, GK87}. A DSW Riemann-type problem for the 1d NLS equation was first considered in~\cite{GK87}. The theory of NLS DSWs was further developed by classifying the possible types of initial discontinuities~\cite{ElGGK95} and finding more general solutions of NLS Whitham hydrodynamic equations~\cite{ElKr95}. Interesting studies of mathematical and physical conditions associated with various (1+1) NLS DSW structures were also undertaken in~\cite{Kodama99, GBYK06}. % in the context of nonlinear optics applications.
The recent review paper~\cite{ElHoefRev16} contains more information on these and other important developments and a valuable and comprehensive list of references.
 
\par Whitham theory and associated DSWs in higher spatial dimensions has developed more slowly until recent years. The approach taken here was initiated when Whitham equations were obtained for the cylindrical Korteweg-de Vries (cKdV) as a reduction of the (2+1)-dimensional Kadomtsev-Petviashvili (KP) equation~\cite{ADM}. This spawned further development of Whitham theory for the KP equation, the (2+1)-dimensional Benjamin-Ono equation and more generally for equations of KP type~\cite{ABW-KP, ABW-BO, ABR}. See also earlier work~\cite{KamPit08, HoefIl09} about plane dark solitons and oblique DSWs of (2+1)-dimensional NLS equation in supersonic fluid (BEC) flow around obstacles. The present study provides another example of Whitham theory applied to (2+1)-dimensional PDEs and their reductions; as indicated below, our rNLS results are quite different from the corresponding 1d NLS system. More generally  Whitham modulation systems related to (2+1)-dimensional equations obtained to date demonstrate a number of important differences from their (1+1)-dimensional counterparts.
\par Transforming Whitham equations from physical variables into hydrodynamic Riemann variables is a common feature to the Whitham procedure and is carried out for the rNLS equation discussed here. In this paper we derive a rNLS Whitham system in terms of such hydrodynamic Riemann-type variables; indeed these are the analogs of the Riemann invariants found for the 1d NLS Whitham system~\cite{ForLee86, Pa87, GK87}. The rNLS Whitham system in Riemann variables below is one of the main results of this paper:
\be
\prt_tr_j + v_j\prt_rr_j + \frac{g_j(r_1,r_2,r_3,r_4)}{r} = 0,   \quad j=1,2,3,4  \label{eq:Wr}    %\eqno(Wr)
\ee

\ni where

{\small$$
v_1 = \frac{r_1+r_2+r_3+r_4}{4} - \frac{(r_2-r_1)(r_4-r_1)}{2(r_4-r_1 - (r_4-r_2)E(m)/K(m))}, 
$$

$$
v_2 = \frac{r_1+r_2+r_3+r_4}{4} + \frac{(r_2-r_1)(r_3-r_2)}{2(r_3-r_2 - (r_3-r_1)E(m)/K(m))},
$$

$$
v_3 = \frac{r_1+r_2+r_3+r_4}{4} - \frac{(r_4-r_3)(r_3-r_2)}{2(r_3-r_2 - (r_4-r_2)E(m)/K(m))},
$$

$$
 v_4 = \frac{r_1+r_2+r_3+r_4}{4} + \frac{(r_4-r_3)(r_4-r_1)}{2(r_4-r_1 - (r_3-r_1)E(m)/K(m))},
$$

$$
m = \frac{(r_1-r_2)(r_3-r_4)}{(r_1-r_3)(r_2-r_4)},    %\label{eq:4.5}   %\eqno(4.5) 
$$

$$
g_1 = -\frac{(r_2+r_3+r_4)v_1}{3} + \frac{(r_1+r_2+r_3+r_4)^2}{8} - \frac{(r_2+r_3)^2+(r_3+r_4)^2+(r_4+r_2)^2}{12},   
$$

$$
g_2 = -\frac{(r_1+r_3+r_4)v_2}{3} + \frac{(r_1+r_2+r_3+r_4)^2}{8} - \frac{(r_1+r_3)^2+(r_3+r_4)^2+(r_4+r_1)^2}{12},   
$$

$$
g_3 = -\frac{(r_1+r_2+r_4)v_3}{3} + \frac{(r_1+r_2+r_3+r_4)^2}{8} - \frac{(r_1+r_2)^2+(r_2+r_4)^2+(r_4+r_1)^2}{12},   
$$

$$
g_4 = -\frac{(r_1+r_2+r_3)v_4}{3} + \frac{(r_1+r_2+r_3+r_4)^2}{8} - \frac{(r_1+r_2)^2+(r_2+r_3)^2+(r_3+r_1)^2}{12},   
$$}
and $K(m)$ and $E(m)$ are the complete elliptic integrals of 1st and 2nd kind, respectively. Knowledge of the slow $r_j$-variables allows us to reconstruct the leading order modulated periodic solution of the rNLS equation which describes a radial DSW converging toward the center.
\par The only difference of the $r_j$ system above from the 1d NLS $r_j$ system is due to additional terms $g_j/r$ which explicitly depend on the radial coordinate $r$. The modulated periodic solution of the rNLS equation has the same form as that of the 1d NLS equation. The velocities $v_j$ here are also the same as for 1d NLS Whitham system~\cite{GK87, Pa87, ElKr95, HoeferEtAl06}. 
\par We study the initial value problem associated with suitable jump initial conditions (ICs) -- see section 5. The chosen ICs lead to the formation of a (arguably simplest) single dispersive shock wave (DSW) moving radially inward, analogous to the well-known simple DSW of the 1d NLS equation~\cite{GK87, HoeferEtAl06}. However, in other respects this radial DSW is very different from its one-dimensional counterpart. 
\par The Whitham system for the rNLS eq.~(\ref{eq:1.1}) is characterized by radial dependence in the boundary conditions (BCs); this is in contrast with the corresponding 1d NLS or KdV problems where the BCs are constant. In the DSW problem for the cylindrical KdV (cKdV) equation the BCs are dependent on time~\cite{ADM}. Also, the trailing and leading edge velocities of the cKdV DSW are time-dependent, see appendix \ref{a:cKdV} of the present paper. For the rNLS DSW, however, we observe that both the velocities are approximately constant to leading order. The leading `linear' edge of the rNLS DSW has the same velocity as that of the 1d NLS case. 
\par Interesting phenomena are observed numerically at the trailing edge which is the analog of the `solitonic' edge of the 1d NLS DSW. In the rNLS problem the trailing edge structure is observed to move approximately with the 1d NLS dark soliton velocity; this is the same velocity as that of the DSW trailing edge of the 1d NLS equation even though other quantities, e.g.~the amplitudes of the maxima and minima, are changing with time. Here we also observe noticeable higher curvatures of the DSW envelope profiles, in contrast to the well-known 1d NLS and KdV DSWs.  %where the envelopes are straight lines.
%One can recall in this connection that {\it plane} dark solitons of (2+1)-dimensional NLS equation can effectively stabilize supersonic fluid (BEC) flow making their transverse instability convective~\cite{KamPit08, HoefIl09}. %In our case, however, amplitude growing with time implies something substantially different from a dark soliton.
\par While the leading (linear, harmonic) edge of the rNLS DSW is identical to that of 1d NLS equation the trailing edge (the analog of solitonic edge for 1d NLS) has very different structure. The surprising result is the numerical observation that the trailing edge velocity of the radial DSW is changing so slowly that it remains with good accuracy close to the corresponding constant value for 1d NLS, which is the 1d NLS dark soliton velocity. Besides, a slowly varying (with $r$) shelf type structure is observed to form behind the trailing edge. This phenomenon appears similar in spirit to the shelfs studied in perturbed soliton equations~\cite{Grimshaw79KdV, Grimshaw79NLS, KodAb81, GriMi93, ElGr02, Abls08, AbNiHorFr11} although there the shelf formation was often caused by variation in media parameters or by dissipation.  
\par As mentioned earlier, the Whitham theory for cylindrical KdV considered as an exact reduction of the KP equation~\cite{ADM} set the stage for the subsequent development of the full (2+1)-dimensional theory resulting in the `hydrodynamic' form of the Whitham system (i.e analog of the $r_j$-equations) for the KP equation itself~\cite{ABW-KP} and more generally for (2+1)-dimensional PDEs of KP type~\cite{ABR}. Our present rNLS Whitham theory can be considered as a similar predecessor to the derivation and study of the Whitham system for the NLS equation in two spatial dimensions and other (2+1)-dimensional PDEs related to it. It is important to emphasize that we use a unified approach to derive Whitham systems that is equally applicable to both integrable and non-integrable PDEs. It is based on a WKB expansion using multiple scales and singular perturbation theory, see e.g.~\cite{Ab2011, ElHoefRev16} and references therein.
\par While our derivation of the Whitham-rNLS system eq.~(\ref{eq:Wr}) in many respects parallels that of the well-known 1d NLS Whitham system, we give it in full detail in this paper and present in two different ways. We consider it valuable since such details are hard to find in the huge literature associated with the 1d NLS equation with small dispersion and there are some conceptual issues that we believe are important. In particular, the relevant asymptotic (the weak limit of small dispersion) solution in general involves two fast phases for genus one, while only one fast phase would correspond to genus zero solution. This is different from the KdV case and related (2+1)-dimensional equations of KP type and their reductions where the basic single DSW genus one solutions have only one fast phase. We carefully restore these phases as well as the total phase of the BEC wavefunction to leading order. The modulated leading order solution for $\Psi=\rho^{1/2}e^{i\Theta}$, $\rho=|\Psi|^2$, has the form

$$
\Psi_0 = \rho_0^{1/2}(\theta, r, t; \epsilon)e^{i\Theta_0(\theta, r,t; \epsilon)},
$$

\be
\rho_0 = \frac{(r_2+r_4-r_1-r_3)^2}{32} - \frac{(r_1-r_2)(r_3-r_4)}{8}\;\text{cn}^2\left(2K(m)(\theta - \theta_*); m\right),   \label{eq:rho}   %\eqno(2.11)
\ee

\ni where the fast phase $\theta$ is determined by the formula

\be
\theta(r,t) = \int_{r_b}^r\frac{k(z,t)dz}{\epsilon} - \int_0^t\frac{k(r_b,\tau)V(r_b,\tau)d\tau}{\epsilon}   \label{eq:theta}
\ee
where $r_b$ is the location of the initial jump of the BEC density $\rho$, and

\be
k^2 = \frac{(r_1-r_3)(r_2-r_4)}{64K^2(m)},   \qquad  V = \frac{r_1+r_2+r_3+r_4}{4};   %\label{eq:2.14}   %\eqno(2.14)
\ee
$\theta_*=\theta_*(r,t)$ is an order one phase shift which we do not find here. The associated analog of the hydrodynamic velocity $u = \epsilon\prt_r\Theta$ is to leading order
\be
u_0 = \frac{V}{2} + \frac{(r_1+r_4-r_2-r_3)(r_1+r_3-r_2-r_4)(r_1+r_2-r_3-r_4)}{256\rho_0},   \label{eq:u0r}   %\qquad \sigma \equiv -\text{sign}(kC_0),  \label{eq:3.6}   %\eqno(3.6)  
\ee
in terms of the introduced variables $r_j$~\footnote{This formula is true for the case of a not too large density jump in the initial condition. Otherwise a vacuum point is formed and the sign before the second term in eq.~(\ref{eq:u0r}) is not constant; see section 5 for more details.}. Then the total leading order phase of the wave function $\Theta_0$, valid for times before the DSW reaches the origin, is determined by the formula below:

\be
\Theta_0 = \frac{1}{\epsilon}\int_0^ru_0(z, t)dz - \frac{t}{\epsilon}.   \label{eq:Phase}
\ee 

\par The outline of the paper is the following. In section 2 we apply the direct perturbation version of Whitham approach with fast and slow space and time scales and describe the leading order solution. In the next section we use the Euler hydrodynamic representation of the rNLS equation to derive the Whitham modulation system in physical varaibles. Then, in section 4, together with technical details in appendix B, we express the system in terms of 1d NLS Riemann variables and obtain the main equations eq.~(\ref{eq:Wr}). We describe in detail how we choose the appropriate ICs and BCs for the rNLS Whitham system in section 5. In section 6 we present numerical results comparing direct simulation of the rNLS DSW evolution with that from rNLS Whitham theory and also with the corresponding 1d NLS DSW simulation. Section 7 contains analytical results partially explaining the numerical observations of rNLS trailing edge and the shelf behind it. Section 8 presents the conclusions. Appendix A contains some auxiliary formulas from the theory of elliptic functions, see e.g.~\cite{BF71}. In appendix C we give an alternative, explicitly two-phase approach to derive the rNLS Whitham system and its relations with the approach of the main text. In appendix D we retrieve the leading and trailing edge dynamics for the cKdV DSW of~\cite{ADM} which were not determined in that paper.
\par This rNLS Whitham theory can be considered as a first step in development of nonlinear modulation theory for NLS-type systems in more than one spatial dimension.

\section{Whitham approach: setup, leading order solution}

As is very common, see e.g.~the recent review~\cite{ElHoefRev16} and references therein, it is convenient to represent the rNLS eq.~(\ref{eq:1.1}) in the form of Euler type hydrodynamics equations for the ``density" and ``velocity". For this purpose, we express $\Psi = \sqrt\rho e^{i\Theta}$, with $\rho$ and $\Theta$ real. We will employ the version of Whitham theory~\cite{Whitham74} involving singular perturbations and multiple scales, see e.g.~\cite{Ab2011} and references therein. This means we will consider an $\epsilon$-expansion for $\epsilon \ll 1$ of the solution to rNLS equation of the form

\be
\Psi = \rho^{1/2}(\theta, r, t; \epsilon)e^{i\Theta(\theta, r,t; \epsilon)},   \label{eq:2.1}   %\eqno(2.1)
\ee

\ni where we impose the following condition for the fast phase $\theta$:   %and $\phi$:

\be
\prt_r\theta = \frac{k}{\epsilon},  \qquad \prt_t\theta = -\frac{\omega}{\epsilon},  \label{eq:2.2}   %\qquad   \prt_r\phi = \frac{\alpha}{\epsilon},  \qquad \prt_t\phi = -\frac{\beta}{\epsilon},   \label{eq:2.2-3}   %\eqno(2.2-2.3)
\ee

\ni and $k$, $\omega$, $\alpha$ are slowly varying quantities and we denote

$$\omega=kV         $$  %\eqno(2.2A)$$ 
where $V$ is called the phase velocity. The hydrodynamic velocity $u$ is introduced as

\be
u = \epsilon\prt_r\Theta.    \label{eq:3.3}    %= \alpha + \epsilon\prt_r\Phi. \label{eq:3.3}    %\eqno(3.3)
\ee

\ni Then the imaginary part of eq.~(\ref{eq:1.1}) is

\be
\epsilon\prt_t\rho + \left(\epsilon\prt_r + \frac{\epsilon}{r}\right)(2\rho u) = 0,  \label{eq:3.4}   %\eqno(3.4)
\ee

\ni while its real part can be written as

\be
\epsilon\prt_t\Theta + u^2 + \rho = \frac{\epsilon^2\prt_{rr}\rho}{2\rho} - \frac{\epsilon^2(\prt_r\rho)^2}{4\rho^2} + \frac{\epsilon^2\prt_r\rho}{2\rho r}.  \label{eq:3.5}  %\eqno(3.5)
\ee

\ni Unlike the imaginary part, it contains the total phase $\Theta$, besides the hydrodynamic density $\rho$ and velocity $u$. Using multiple scales, we calculate the derivatives as the sum of fast and slow derivatives,
$$\epsilon\prt_rf = (k\prt_{\theta} + \epsilon\tilde\prt_r)f,    \qquad   \epsilon\prt_tf = (-kV\prt_{\theta} + \epsilon\tilde\prt_t)f,$$
where $\prt_{\theta}/\epsilon$ is the fast derivative and $\tilde\prt_r$ and $\tilde\prt_t$ are the slow space and time derivatives at fixed $\theta$, for $f=\rho$, $u$ or $\Theta$. We also denote $f' \equiv \prt_\theta f$ here and further on. Then we expand in $\epsilon$:

\[ \rho = \rho_0 + \epsilon\rho_1 + \dots, ~~u = u_0 + \epsilon u_1 + \dots,  ~~\Theta = \Theta_0 + \epsilon\Theta_1 + \dots. \]
 
\ni However, since $\Theta$ itself is a fast variable, i.e.~is of order $1/\epsilon$, the leading order of the expansion of its derivatives will have forms
$$\epsilon\prt_r\Theta_0 = k\Theta_0' + \alpha,   \qquad  \epsilon\prt_t\Theta = -kV\Theta_0' - \beta, $$
where $\alpha$ and $\beta$ are additional slow variables, the leading orders of ``slow derivatives" $\epsilon\tilde\prt_r\Theta$ and $\epsilon\tilde\prt_t\Theta$, respectively. This prescription is in fact equivalent to having two fast phases, see appendix \ref{a:dpha}. Note that, while $\Theta_0 \sim 1/\epsilon$, the derivative $\prt_\theta\Theta_0 \equiv \Theta_0' \sim 1$. 
\par At the leading order eq.~(\ref{eq:3.4}) yields

$$
-kV\rho_0' + 2k(\rho_0u_0)' = 0,
$$

\ni and its first integral is

\be
u_0 = \frac{V}{2} + \frac{kC_0}{\rho_0},  \label{eq:3.6A}   %\eqno(3.6)  
\ee

\ni where an integration `constant' $C_0$~\footnote{$C_0$ and other `constants' of integration in $\theta$ appearing later are slow variables depending on $r$ and $t$.} is introduced.
\par We now turn to the real part of rNLS, eq.~(\ref{eq:3.5}). Its leading order yields

\be
- kV\Theta_0' - \beta + u_0^2 + \rho_0 = k^2\left(\frac{\rho_0''}{2\rho_0} - \frac{(\rho_0')^2}{4\rho_0^2}\right).  \label{eq:3.10}   %\eqno(3.10)  
\ee

\par From the definition (\ref{eq:2.2}), it follows immediately that the slow variables satisfy conservation of waves

\be
\prt_tk + \prt_r\omega = \prt_tk + \prt_r(kV) = 0.   \label{eq:k}   %\eqno(2.15)  
\ee
This is an important first and it is the simplest Whitham equation -- in physical variables. To get the other Whitham equations which we will find as secularity conditions, the derivative of eq.~(\ref{eq:3.5}) with respect to (w.r.t.)~$r$ is helpful. The $r$-derivative gives another hydrodynamic equation (i.e.~containing $\rho$ and $u$ only as dependent variables),

\be
\prt_tu + \prt_r(u^2 + \rho) = \prt_r\left(\frac{\epsilon^2\prt_{rr}\rho}{2\rho} - \frac{\epsilon^2(\prt_r\rho)^2}{4\rho^2} + \frac{\epsilon^2\prt_r\rho}{2\rho r}\right).  \label{eq:3.13}  %\eqno(3.13)  
\ee
The leading order of eq.~(\ref{eq:3.13}) is a total derivative in $\theta$ which integrates to
\be
\rho_0 + u_0^2 - Vu_0 + H_0 = k^2\left(\frac{\rho_0''}{2\rho_0} - \frac{(\rho_0')^2}{4\rho_0^2}\right),   \label{eq:3.130}
\ee
where $H_0$ is a second integration `constant'. Using eqs.~(\ref{eq:3.6A}), (\ref{eq:3.10}) and (\ref{eq:3.130}) we find

$$
\Theta_0' = \frac{V}{2k} - \frac{H_0+\beta}{kV} + \frac{C_0}{\rho_0}.
$$
On the other hand, taking the leading order of the definition eq.~(\ref{eq:3.3}) one gets
$$
u_0 = k\Theta_0' + \alpha.
$$
Comparing with the previous equation and eq.~(\ref{eq:3.6A}) one finds the relation
$$
H_0 = V\alpha-\beta= V^2/4 - \Omega.
$$
where we define  $\Omega = V^2/4 - H_0$. Substituting eq.~(\ref{eq:3.6A}) into eq.~(\ref{eq:3.130}) one obtains a second order ODE for $\rho_0$,

$$
k^2(2\rho_0\rho_0'' - (\rho_0')^2) = 4\rho_0^3 - 4\Omega\rho_0^2 + 4k^2C_0^2.   
$$
One can recognize in it an ODE for an elliptic function $\rho_0$ (e.g. assume $(\rho_0')^2=F(\rho_0)$) and verify that as a consequence $\rho_0$ satisfies

\be
2k^2(\rho_0')^2 = 4(\rho_0^3-2\Omega\rho_0^2+2C_1\rho_0-2k^2C_0^2) = 4(\rho_0-\lambda_1)(\rho_0-\lambda_2)(\rho_0-\lambda_3),  \label{eq:2.10}  %\eqno(2.10)
\ee
with another integration constant (slow variable) $C_1$. Its general solution for $\rho_0$ can be written as

\be
\rho_0 = a + b\;\text{cn}^2\left(2K(m)(\theta - \theta_*); m\right),   \label{eq:2.11}   %\eqno(2.11)
\ee

\ni where $\theta_*$ is another integration constant (slow variable in general) and

\be
a = \lambda_2,  \qquad b = -(\lambda_2-\lambda_1),  \qquad  m = \frac{\lambda_2-\lambda_1}{\lambda_3-\lambda_1}.   \label{eq:2.12}   %\eqno(2.12)
\ee

\ni We take $\lambda_1\le\lambda_2\le\lambda_3$, then we have the following relations among the parameters (slow variables):

\be
e_1 \equiv \lambda_1 + \lambda_2 + \lambda_3 = 2\Omega,  \quad  e_2 \equiv \lambda_1\lambda_2+\lambda_2\lambda_3+\lambda_3\lambda_1 = 2C_1,  \quad e_3\equiv \lambda_1\lambda_2\lambda_3 = 2k^2C_0^2;  \label{eq:2.13} %\eqno(2.13)
\ee

\ni and the normalization of the elliptic function with fixed period $1$ implies that

\be
k^2 = \frac{|b|}{8mK^2(m)} = \frac{\lambda_3-\lambda_1}{8K^2(m)},   \label{eq:2.14}   %\eqno(2.14)
\ee

\ni where $K(m)$ is the first complete elliptic integral. Also it follows that the leading order hydrodynamic velocity is 
\be
u_0 = \frac{V}{2} + \frac{kC_0}{\rho_0} = \frac{V}{2} - \sigma\frac{\sqrt{2\lambda_1\lambda_2\lambda_3}}{2\rho_0},   \qquad \sigma \equiv -\text{sign}(kC_0),  \label{eq:3.6}   %\eqno(3.6)  
\ee
in terms of the introduced roots of the cubic $\lambda_i$, $i=1,2,3$. The above is the same as the well-known leading order solution for 1d NLS as expected. At this stage we expect that the variables $\lambda_1,\lambda_2,\lambda_3,V$ are going to be the key dependent variables. Knowledge of these variables allows us to reconstruct $\rho_0, u_0$ and $\theta$ and hence $\Theta_0$ and the leading order approximation for $\Psi$. The fast phase $\theta$ is determined by the formula (\ref{eq:theta}). This gives the complete leading order fast phase unlike the expressions proportional to $x-Vt$ which can often be seen in older literature on the subject~\footnote{The leading order difference between true $\theta$ and the unmodulated traveling wave fast phase $k(x-Vt)$ is often restored from the solutions of Whitham equations, see e.g.~\cite{ElHoefRev16, Grava17} and references therein, but formula (\ref{eq:theta}) always incorporates that.}. The additional next order (order one) phase shift $\theta_*$ is also important being a part of leading order solution but we do not determine it here.
\par As for the total phase $\Theta$ of the condensate wavefunction $\Psi$, its leading order $\Theta_0$ can be found from eq.~(\ref{eq:Phase}). This formula can be obtained if one takes into account that $i\epsilon\prt_t\Psi = |\Psi|^2\Psi$ for $r=0$ initially with $\rho=1$ and $\Theta=0$ there (see section 5). The above conditions hold only before the DSW front reaches the origin, afterwards the whole character of the solution must change.

\section{Whitham system for the rNLS equation}   

The next order ($\sim\epsilon$) of eq.~(\ref{eq:3.4}) reads~\footnote{\mja{Here} and further on we write $\prt_r$ for $\tilde\prt_r$ and $\prt_t$ for $\tilde\prt_t$ but they denote the derivatives at fixed $\theta$ in all equations at order $\epsilon$ while they have usual meaning in exact equations.}

\be
-kV\rho_1' + 2k(\rho_0u_1 + u_0\rho_1)' + \prt_t\rho_0 + \left(\prt_r + \frac{1}{r}\right)(2\rho_0u_0) = 0.  \label{eq:3.8}   %\eqno(3.8)  
\ee

\ni Requiring that the first-order corrections $\rho_1$ and $u_1$ do not grow but are periodic in $\theta$, integrating eq.~(\ref{eq:3.8}) over the period and using eq.~(\ref{eq:3.6}), we find a secularity condition

\be
\prt_tQ + \left(\prt_r + \frac{1}{r}\right)\left(VQ + 2kC_0\right) = 0,   \label{eq:3.9}   %\eqno(3.9)  
\ee

%\ni which is exactly the previous eq.~(2.23); we recall that $Q=\int_0^1\rho_0 d\theta$. 
\ni the analog of mass conservation law in the 1d case. Here (using eq.~(\ref{eq:2.11}))

\be
Q=\int_0^1\rho_0(\theta) d\theta = a + b\left(\frac{E(m)}{mK(m)} - \frac{1-m}{m}\right) = \lambda_3 - (\lambda_3-\lambda_1)\frac{E}{K},   \label{eq:Ql}  
\ee

\ni and $K=K(m)$ and $E=E(m)$ are the first and second complete elliptic integrals, respectively. 
\par Eq.~(\ref{eq:3.13}) itself turns out to lead to eq.~(\ref{eq:a}), obtained here as a secularity condition, see appendix \ref{a:dpha}. Instead we use the combination of eq.~(\ref{eq:3.13}) and eq.~(\ref{eq:3.4}), $2\rho\cdot($\ref{eq:3.13}$) + 2u\cdot($\ref{eq:3.4}$)$, to get what would be the (hydrodynamic) momentum conservation law in the 1d NLS case (i.e.~without $1/r$ terms), 

\be
\epsilon\prt_t(2\rho u) + \epsilon\left(\prt_r + \frac{1}{r}\right)(4\rho u^2) + \epsilon\prt_r(\rho^2) =  \epsilon\left(\prt_r + \frac{1}{r}\right)\left(\epsilon^2\prt_{rr}\rho - \frac{\epsilon^2(\prt_r\rho)^2}{\rho}\right) - \frac{\epsilon^3\prt_r\rho}{r^2}.   \label{eq:3.15}  %\eqno(3.15)
\ee

\ni Its leading order integrates to

\be
k^2\left(\rho_0'' - \frac{(\rho_0')^2}{\rho_0}\right) = \rho_0^2 + 2\rho_0u_0(2u_0-V) + H_1,  \label{eq:3.16}   %\eqno(3.16)
\ee

\ni where $H_1$ is the constant of integration. Upon using eqs.~(\ref{eq:3.6}), (\ref{eq:A1}) and (\ref{eq:A2}), one finds $H_1 = -2C_1-2VkC_0$. Then at order $\epsilon$ eq.~(\ref{eq:3.15}) reads:

$$
-2kV(\rho_0u_1+u_0\rho_1)' + k(8\rho_0u_0u_1+ (4u_0^2+2\rho_0)\rho_1)' - k\left(k^2\rho_1'' - \frac{2k^2\rho_0'\rho_1'}{\rho_0} + \frac{k^2(\rho_0')^2\rho_1}{\rho_0^2}\right)' +
$$

\be
+ \prt_t(2\rho_0u_0) + \left(\prt_r + \frac{1}{r}\right)(4\rho_0u_0^2) + \prt_r(\rho_0^2) - \left(\prt_r + \frac{1}{r}\right)\left(k^2\left(\rho_0'' - \frac{(\rho_0')^2}{\rho_0}\right)\right) = 0.  \label{eq:3.17}   %\eqno(3.17)
\ee

\ni (Note that the last term $\frac{\epsilon^3\prt_r\rho}{r^2}$ in eq.~(\ref{eq:3.15}) contributes only at order $\epsilon^2$ and higher). Integrating eq.~(\ref{eq:3.17}) over the period in $\theta$ and using eqs.~(\ref{eq:3.16}) and (\ref{eq:A3}), one finds a secularity condition,

\be
\prt_t(VQ + 2kC_0) + \left(\prt_r + \frac{1}{r}\right)(V^2Q + 4VkC_0 + 2C_1) - \frac{4\Omega Q - 2C_1}{3r} = 0.  \label{eq:3.18}   %\eqno(3.18)
\ee
This is the analog of momentum conservation law for 1d case. One needs one more secularity condition to close the system. We are led to it guided by the energy conservation law for the 1d NLS equation. To obtain its analog in our case, we need the relation

\be
\epsilon^3\prt_t\left(\frac{(\prt_r\rho)^2}{\rho}\right) = \epsilon^3\left(\frac{\prt_r\rho}{\rho}\right)^2\left(\prt_r + \frac{1}{r}\right)(2\rho u) - 2\epsilon^3\frac{\prt_r\rho}{\rho}\prt_r\left(\prt_r + \frac{1}{r}\right)(2\rho u),  \label{eq:3.20}  %\eqno(3.20)
\ee

\ni a consequence of eq.~(\ref{eq:3.4}) and its $r$-derivative. Taking now the combination of the equations $4(\rho+u^2)\cdot($\ref{eq:3.4}$) + 8\rho u\cdot($\ref{eq:3.13}$) + ($\ref{eq:3.20}$)$, we derive the needed energy equation:

$$
\epsilon\prt_t\left(2\rho^2 + 4\rho u^2 + \frac{\epsilon^2(\prt_r\rho)^2}{\rho}\right) + 
$$

\be
+\epsilon\left(\prt_r + \frac{1}{r}\right)\left(8\rho u^3 + 8\rho^2u + 6u\frac{\epsilon^2(\prt_r\rho)^2}{\rho} + 4\epsilon^2\prt_r\rho\prt_ru - 4u\epsilon^2\prt_{rr}\rho\right) = 0.   \label{eq:3.21}   %\eqno(3.21)
\ee

\ni Its leading order is again a total derivative in $\theta$, and, integrating, we obtain

\be
-V\left(2\rho_0^2+4\rho_0u_0^2+k^2\frac{(\rho_0')^2}{\rho_0}\right) + 8\rho_0 u_0^3 + 8\rho_0^2u_0 + 6u_0k^2\frac{(\rho_0')^2}{\rho_0} + 4k^2\rho_0'u_0' - 4u_0k^2\rho_0'' + H_2 = 0,   \label{eq:3.22}   %\eqno(3.22)
\ee

\ni with another integration constant $H_2$. Using again eq.~(\ref{eq:3.6}) and also eq.~(\ref{eq:3.16}) and leading order of eq.~(\ref{eq:3.13}), we find 

$$H_2 = 2kC_0(V^2-4\Omega) + 2VH_1.$$

 The next order (order $\epsilon$) of the energy equation (\ref{eq:3.21}) can be presented as

$$
0 = -kV\left[2\rho^2 + 4\rho u^2 + \frac{\epsilon^2(\prt_r\rho)^2}{\rho}\right]_1' + k\left[8\rho u^3 + 8\rho^2u + 6u\frac{\epsilon^2(\prt_r\rho)^2}{\rho} + 4\epsilon^2\prt_r\rho\prt_ru - 4u\epsilon^2\prt_{rr}\rho\right]_1' +
$$

\be
\prt_t\left(2\rho_0^2+4\rho_0u_0^2+k^2\frac{(\rho_0')^2}{\rho_0}\right) + \left(\prt_r + \frac{1}{r}\right)\left(8\rho_0 u_0^3 + 8\rho_0^2u_0 + 6u_0k^2\frac{(\rho_0')^2}{\rho_0} + 4k^2\rho_0'u_0' - 4u_0k^2\rho_0''\right),   \label{eq:3.23}  %\eqno(3.23)
\ee

\ni where notation $[\dots]_1$ means that the part of expression in brackets at order $\epsilon$ is taken. Then, upon integration over the period, the first line of eq.~(\ref{eq:3.23}) being a $\theta$-derivative integrates to zero and we obtain a secularity condition from the integral of the second line. Applying eqs.~(\ref{eq:3.6}), (\ref{eq:A3}), (\ref{eq:A4}) and (\ref{eq:A5}) (see Appendix A), we finally obtain

$$
\prt_t\left((V^2+4\Omega/3)Q + 4VkC_0 + 4C_1/3\right) + 
$$

\be
+ \left(\prt_r + \frac{1}{r}\right)\left(V(V^2+4\Omega/3)Q + 2(3V^2+4\Omega)kC_0 + 16VC_1/3\right) = 0.   \label{eq:3.24}   %\eqno(3.24)  
\ee

\ni We observe that eqs.~(\ref{eq:k}), (\ref{eq:3.9}), (\ref{eq:3.18}) and (\ref{eq:3.24}) comprise a complete system of Whitham equations for rNLS. The last three of the four Whitham equations (``mass", ``momentum" and ``energy") have been obtained as secularity conditions here. 
\par From now on we will use the elementary symmetric functions $e_1$, $e_2$ and $e_3$ of the roots of the cubic in eq.~(\ref{eq:2.10}) which were introduced in eq.~(\ref{eq:2.13}), instead of the variables $\Omega$, $C_0$ and $C_1$. Using also eq.~(\ref{eq:3.6}), the obtained system of four Whitham equations is

\be
\prt_tk + \prt_r(Vk) = 0,   \label{eq:(k)}  %\eqno(k)   
\ee

\be
\prt_tQ + \left(\prt_r + \frac{1}{r}\right)\left(VQ - \sigma\sqrt{2e_3}\right) = 0,   \label{eq:(Q)}   %\eqno(Q)   
\ee

\be
\prt_t(VQ - \sigma\sqrt{2e_3}) + \left(\prt_r + \frac{1}{r}\right)(V^2Q - 2V\sigma\sqrt{2e_3} + e_2) + \frac{e_2 - 2e_1Q}{3r} = 0,  \label{eq:P}   %\eqno(P)   
\ee

$$
\prt_t\left((V^2+\frac{2e_1}{3})Q - 2V\sigma\sqrt{2e_3} + \frac{2e_2}{3}\right) + 
$$

\be
+ \left(\prt_r + \frac{1}{r}\right)\left(V(V^2+\frac{2e_1}{3})Q - (3V^2+2e_1)\sigma\sqrt{2e_3} + \frac{8Ve_2}{3}\right) = 0.   \label{eq:E}   %\eqno(E)   
\ee
We remark that this form of the Whitham system is different from the one used e.g.~in~\cite{HoeferEtAl06} for 1d NLS which involved first four hydrodynamic conservation laws but not the simplest and most fundamental conservation of waves eq.~(\ref{eq:(k)}).
\par It is convenient here to introduce and use the operator 

\be
D = \prt_t + V\prt_r        \label{eq:D}   %\eqno(D)
\ee

\ni rather than partial time derivative $\prt_t$. After transformations described in Appendix~\ref{a:transf} the system of Whitham PDEs becomes 

\be
\frac{Dk}{k} + \prt_rV = 0,   \label{eq:k1}   %\eqno(k)   
\ee

\be
DV - \frac{1}{3\sigma\sqrt{2e_3}}\left(QDe_1 - 2e_1(DQ - Q\frac{Dk}{k}) + De_2 - 4e_2\frac{Dk}{k}\right) + \prt_re_1 = 0,   \label{eq:E1}  %\eqno(E1)
\ee

\be
DQ - Q\frac{Dk}{k} - \prt_r(\sigma\sqrt{2e_3}) + \frac{VQ - \sigma\sqrt{2e_3}}{r} = 0,   \label{eq:Q1}  %\eqno(Q1)   
\ee

\be
QDV - D(\sigma\sqrt{2e_3}) + 2\sigma\sqrt{2e_3}\frac{Dk}{k} + \prt_re_2 + \frac{-3V\sigma\sqrt{2e_3} + 4e_2 - 2e_1Q}{3r} = 0.   \label{eq:P1}   %\eqno(P1)
\ee

\ni The system of four PDEs (\ref{eq:k1}), (\ref{eq:E1}), (\ref{eq:Q1}) and (\ref{eq:P1}) can be expressed in terms of the four key dependent variables -- phase velocity $V$ and the roots of the cubic in eq.~(\ref{eq:2.10}), $\lambda_i$, $i=1,2,3$, (see eqs.~(\ref{eq:2.10})--(\ref{eq:2.14}) and (\ref{eq:Ql})), which in turn depend on the variables $\Omega,C_0,C_1$. Variables $k$ and $Q$ are the functions of the roots given by eq.~(\ref{eq:2.14}) and (\ref{eq:Ql}), and $e_1$, $e_2$ and $e_3$ are the elementary symmetric polynomials of the roots. 

\section{Whitham equations in Riemann-type variables}

The one-dimensional (1d) NLS equation (eq.~(\ref{eq:1.1}) without $\frac{1}{r}\prt_r$ term and derivatives $\prt_x$ instead of $\prt_r$) is known~\cite{GK87, Pa87, ElKr95, HoeferEtAl06} to have Whitham equations diagonal in the spatial and temporal derivatives of the Riemann variables $\{r_j\}$, $j=1,\dots, 4$,

\be
\prt_tr_j + v_j\prt_xr_j = 0,   \qquad   r_1 \le r_2 \le r_3 \le r_4,   \label{eq:4.1}   %\eqno(4.1)
\ee

\ni or, equivalently, see e.g.~\cite{ElKr95},

\be
\frac{4\prt_jk}{k}Dr_j + \prt_xr_j = 0,   \quad \prt_j \equiv \frac{\prt}{\prt r_j},   \qquad  v_j = V + \frac{k}{4\prt_jk},   \label{eq:4.2}   %\eqno(4.2)  
\ee

\ni where the relation with variables $V$, $\lambda_i$, $i=1,2,3$, is, in our normalization,

$$
V = \frac{r_1+r_2+r_3+r_4}{4},    \qquad   \lambda_1 = \frac{(r_1+r_4-r_2-r_3)^2}{32},  
$$

\be
\lambda_2 = \frac{(r_2+r_4-r_1-r_3)^2}{32},   \qquad  \lambda_3 = \frac{(r_3+r_4-r_1-r_2)^2}{32},   \label{eq:4.3}   %\eqno(4.3)   
\ee

\ni Thus, the velocities $v_j$ are determined by log-derivatives of $k$ w.r.t.~the Riemann variables. The log-derivatives of $k$ are found from formula (\ref{eq:A12}) and definitions (\ref{eq:4.3}), which imply the facts

$$
\lambda_2-\lambda_1 = \frac{(r_2-r_1)(r_4-r_3)}{8},   \qquad  \lambda_3-\lambda_1 = \frac{(r_3-r_1)(r_4-r_2)}{8},   
$$

\be
m = \frac{\lambda_2-\lambda_1}{\lambda_3-\lambda_1} = \frac{(r_2-r_1)(r_4-r_3)}{(r_3-r_1)(r_4-r_2)}   \label{eq:4.5}   %\eqno(4.5) 
\ee

\ni and 

$$
\frac{\prt_1k}{k} = -\frac{1}{2(r_2-r_1)}\left(1 - \frac{r_4-r_2}{r_4-r_1}\frac{E}{K}\right),   \qquad  \frac{\prt_2k}{k} = \frac{1}{2(r_2-r_1)}\left(1 - \frac{r_3-r_1}{r_3-r_2}\frac{E}{K}\right),
$$

\be
\frac{\prt_3k}{k} = -\frac{1}{2(r_4-r_3)}\left(1 - \frac{r_4-r_2}{r_3-r_2}\frac{E}{K}\right),   \qquad  \frac{\prt_4k}{k} = \frac{1}{2(r_4-r_3)}\left(1 - \frac{r_3-r_1}{r_4-r_1}\frac{E}{K}\right).   \label{eq:4.6}   %\eqno(4.6)   
\ee

\ni In the above and further on we use $K \equiv K(m)$ and $E\equiv E(m)$.
 
\par Later we also obtain the rNLS Whitham equations in terms of the Riemann-type variables $r_j$ defined by eqs.~(\ref{eq:4.3}) and the inequalities 

\be 
r_1 \ge r_2 \ge r_3 \ge r_4.     \label{eq:r1r4}
\ee
The relevance of the same variables here is due to the fact that the leading order solution in $\epsilon$ is the same for rNLS and for 1d NLS. However, the DSW used in \cite{HoeferEtAl06} and described by eq.~(\ref{eq:4.1}) is moving in the positive $x$-direction while here the DSW moving toward smaller radius $r$ is considered, therefore all velocities have the opposite signs leading to the opposite inequalities for the corresponding Riemann variables; see the discussion of initial and boundary conditions below.
\par We do the transformation from $V$ and the roots $\lambda_i$ to the Riemann variables in two steps. We introduce the variables $R_i$ by 

\be
R_i = \frac{r_i + r_4 - r_l - r_m}{4}, \quad i=1,2,3,  \quad  i\neq l \neq m \neq i,  \qquad V = \frac{\sum_{j=1}^4r_j}{4}.   \label{eq:Rr}   %\eqno(Rr)   
\ee

\ni Then 

\be
r_i = V + R_i - R_l - R_m,  \quad i=1,2,3,  \quad  i\neq l \neq m \neq i,  \qquad r_4 = V + R_1 + R_2 + R_3,   \label{eq:rR}   %\eqno(rR)
\ee

\ni and we see that the inequalities $R_1\ge R_2 \ge R_3$ always hold (this will be important later in the discussion of initial and boundary conditions). First, we express everything in terms of variables $R_i$ and $V$, using
%$R_1\ge R_2 \ge R_3 \ge V$ always hold. 

\be
\lambda_i = \frac{R_i^2}{2}, \quad  i=1,2,3,   \label{eq:4.7}   %\eqno(4.7)
\ee

\ni and 

\be
e_1 = \frac{R_1^2+R_2^2+R_3^2}{2},   \qquad  e_2 = \frac{R_1^2R_2^2+R_2^2R_3^2+R_3^2R_1^2}{4},  \qquad  -\sigma\sqrt{2e_3} = \frac{R_1R_2R_3}{2}.  \label{eq:4.8}   %\eqno(4.8)  
\ee

\ni (As for the last formula in eq.~(\ref{eq:4.8}), see the discussion of initial and boundary conditions below.) As the second step of the transformation, we express everything in terms of 1d NLS Riemann variables $r_j$, $j=1,2,3,4$. The details are given in Appendix B. Using Appendix B, the final form of the Whitham equations is given by  

$$
~~~~~~~~\prt_tr_j + v_j\prt_rr_j + \frac{g_j(r_1,r_2,r_3,r_4)}{r} = 0,   \quad j=1,2,3,4  ~~~~~~~~~~~~~~~~~~~~~~~~~~~~~~~~~~~~~~~\text{(1.3)}%\eqno(Wr) 
$$

\ni This is eq.~(\ref{eq:Wr}); here the velocities $v_j$ are the 1d NLS velocities given by eqs.~(\ref{eq:4.2}) and (\ref{eq:4.6}) and the functions $g_j$ in terms of Riemann variables $r_j$ read:

{\small$$
g_1 = \frac{1}{6}\left( \frac{(r_4-r_1)(r_2-r_1)(r_4+r_3+r_2)}{r_4-r_1 - (r_4-r_2)E/K} - \right.
$$

$$
\left. - \frac{3}{4}(r_4^2+r_3^2+r_2^2-r_1^2) - \frac{r_4r_3+r_4r_2+r_3r_2}{2} + r_1(r_4+r_3+r_2)\right),   %\eqno(g1r)
$$

$$
g_2 = -\frac{1}{6}\left( \frac{(r_3-r_2)(r_2-r_1)(r_4+r_3+r_1)}{r_3-r_2 - (r_3-r_1)E/K} + \right.
$$

$$
\left. + \frac{3}{4}(r_4^2+r_3^2+r_1^2-r_2^2) + \frac{r_4r_3+r_4r_1+r_3r_1}{2} - r_2(r_4+r_3+r_1)\right),   %\eqno(g2r)
$$

$$
g_3 = \frac{1}{6}\left( \frac{(r_3-r_2)(r_4-r_3)(r_4+r_2+r_1)}{r_3-r_2 - (r_4-r_2)E/K} - \right.
$$

$$
\left. - \frac{3}{4}(r_4^2+r_2^2+r_1^2-r_3^2) - \frac{r_4r_2+r_4r_1+r_2r_1}{2} + r_3(r_4+r_2+r_1)\right),   %\eqno(g3r)
$$

$$
g_4 = -\frac{1}{6}\left( \frac{(r_4-r_1)(r_4-r_3)(r_3+r_2+r_1)}{r_4-r_1 - (r_3-r_1)E/K} + \right.
$$

$$
\left. + \frac{3}{4}(r_3^2+r_2^2+r_1^2-r_4^2) + \frac{r_3r_2+r_3r_1+r_2r_1}{2} - r_4(r_3+r_2+r_1)\right).  %\eqno(g4r)
$$}
In this form all coefficient functions $v_j$ and $g_j$ are nonsingular at the edges of the DSW where $m=0$ and $m=1$; i.e.~they have finite limits there. It is easy to verify that in the limit $m\to0$, characterized by $r_3\to r_4$, $E(m)/K(m)\to 1$ and $r_2\to -r_1$, all functions $g_j$ tend to zero. This is as it should be since they must equal zero at the origin $r=0$, and they are zero everywhere from $r=0$ to the front edge of DSW where $m$ starts increasing from $0$. As for the trailing edge of DSW, $m\to 1$, characterized by $r_3\to r_2$ and $E(m)/K(m)\to 0$, there $g_j$ have the following finite limits:

$$
m\to 1: \qquad g_1 \to -\frac{r_4^2-r_1^2}{8},  \qquad  g_4 \to \frac{r_4^2-r_1^2}{8},  
$$

%\be
%g_2, g_3 \to \frac{1}{6}\left(r_2^2 + \frac{(r_4+r_1)r_2}{2} - \frac{3(r_4+r_1)^2}{4} + r_4r_1\right).    \label{eq:gm1}   %\eqno(gm1)
%\ee

\be
g_2, g_3 \to \frac{1}{6}\left(r_2^2 + \frac{(r_1+r_4)r_2}{2} - \frac{(r_1+r_4)^2}{2} - \frac{(r_1-r_4)^2}{4}\right).    \label{eq:gm1}   %\eqno(gm1)
\ee

\section{Initial and boundary conditions for the BEC expansion problem}

\subsection{IC and BC from dispersionless rNLS}

The ICs and BCs for the classical 1d NLS DSW problem were  obtained from the Riemann simple wave problem associated with the {\it dispersionless} 1d NLS eq., see~\cite{Kodama99, HoeferEtAl06} or the discussion in the next subsection. Motivated by this, for the study of the rNLS equation we derive the ICs, BCs from the {\it dispersionless} rNLS equation. To properly define it, let us look at imaginary and real parts of rNLS,

\be
\prt_t\rho + \left(\prt_r + \frac{1}{r}\right)(2\rho u) = 0,  \label{eq:Im}   %\eqno(Im)
\ee

\be
\epsilon\prt_t\Theta + u^2 + \rho = \epsilon^2\left(\frac{\prt_{rr}\rho}{2\rho} - \frac{(\prt_r\rho)^2}{4\rho^2} + \frac{\prt_r\rho}{2\rho r}\right),  \label{eq:Re}   %\eqno(Re)
\ee

\ni respectively, i.e.~eqs.~(\ref{eq:3.4}) and (\ref{eq:3.5}). Recall that $u=\epsilon\prt_r\Theta$ here. The dispersionless rNLS system is the one given by eq.~(\ref{eq:Im}) and eq.~(\ref{eq:Re}) without its right-hand side. The $r$-derivative of equation (\ref{eq:Re}) without the right-hand side is

\be
\prt_tu + \prt_r(u^2 + \rho) = 0,   \label{eq:u}   %\eqno(u)
\ee

\ni which is the same as for dispersionless 1d NLS. Thus, the only difference between the dispersionless NLS and rNLS equations is the last term $2\rho u/r$ in eq.~(\ref{eq:Im}). The appropriate IC and BC for the DSW problem come from the stationary solution of eqs.~(\ref{eq:Im}) and (\ref{eq:u}), as was the case for 1d NLS (with eq.~(\ref{eq:Im}) without the last term). Now the general stationary solution is given by equations

$$
\rho u = C_I\frac{r_b}{r},  \qquad u^2 + \rho = C_R,  %\eqno(ru-st)
$$

\ni involving two arbitrary constants $C_R \geq 0$ and $C_I$. Here we introduced the radius $r_b$ which is the location of the initial discontinuity. Taking $C_I=0$ corresponds to $u=0$ for the DSW problem, which also implies $\rho = \text{const.}$ and these are the proper boundary conditions ahead of the DSW leading (or front) edge. Here we consider a DSW propagating towards $r=0$; in this case the leading edge is closer to the origin than the trailing edge. They are the same as for a left propagating DSW in 1d NLS. However, for the rNLS eq. behind the DSW trailing edge both $\rho$ and $u$ will not be constant. Since we consider a DSW moving toward the center $r=0$, $u<0$ and $C_I<0$ behind the trailing edge for $r>r_b$. We set 

$$
\rho \to \rho_b,     \qquad  u \to u_b<0  \qquad \text{as } r \to r_b+0,   %\eqno(urb)
$$

\ni then $C_R=\rho_b+u_b^2$, $C_I=\rho_bu_b$ and the solution can be written as

\be
u=u_{st}(r) = -(\rho_b+u_b^2)^{1/2}s(r/r_b),  \qquad \rho=\rho_{st}(r) = (\rho_b+u_b^2)(1-s^2(r/r_b)),   \qquad  r>r_b, \label{eq:rst}   %\eqno(rst)
\ee

\ni where $s(x)$ is a nonnegative root of the cubic equation

\be
s^3 - s + g(x) = 0,   \qquad  g(x) = -\frac{\rho_bu_b}{(\rho_b + u_b^2)^{3/2}x} \equiv \frac{b}{x} > 0 \text{ for } x>0.   \label{eq:s}  %\eqno(s)
\ee

\ni If we consider the situation where $\rho_{st}(r)$ is nondecreasing (the analog of simple wave here), then we should take the root $s(x)$ approaching $0$ as $x\to\infty$. Explicitly, as function of $x$, it is 

\be
s(x) = \frac{2}{\sqrt3}\cos\left[\frac{\pi}{3} + \frac{1}{3}\arccos\left(\frac{3\sqrt3b}{2x}\right)\right].  \label{eq:s-sol}   %\eqno(s-sol)
\ee

\ni As $x \rightarrow \infty$, $\arccos(\frac{3\sqrt3b}{2x}) \rightarrow \pi/2$, hence $s(x) \rightarrow 0$ and, for large $x$,

\be
s(x) = g + g^3 + 3g^5 + \dots, \qquad s^2(x) = g^2(1 + 2g^2 + 7g^4 + \dots), \qquad g\equiv g(x) = \frac{b}{x}.   \label{eq:s-exp}   %\eqno(s-exp)
\ee

\ni The nonnegative solution $s(x)$ exists as long as $g(x) \le 2/3^{3/2}$. One can check that this condition is not really restrictive here: the inequalities $g(r/r_b)\le b \le 2/3^{3/2}$ always hold for $r\ge r_b$ and constant $b$ defined in eq.~(\ref{eq:s}) since $\rho_b>0$. Thus, we assume the following initial condition (IC) for the density $\rho=|\Psi|^2$ and hydrodynamic velocity $u=\epsilon\prt_r\text{arg}\Psi$,

\be
\rho(r,0) = \left\{ \begin{array}{cc} 1, & 0\le r<r_b \\ \rho_{st}(r), & r>r_b \end{array} \right.   \label{eq:ICd}   %\eqno(ICd)
\ee

\be
u(r,0) = \left\{ \begin{array}{cc} 0, & 0\le r<r_b \\ u_{st}(r)<0, & r>r_b \end{array} \right.   \label{eq:ICu}   %\eqno(ICu)
\ee

\begin{figure} [ht]
\centering
\includegraphics[scale=.6]{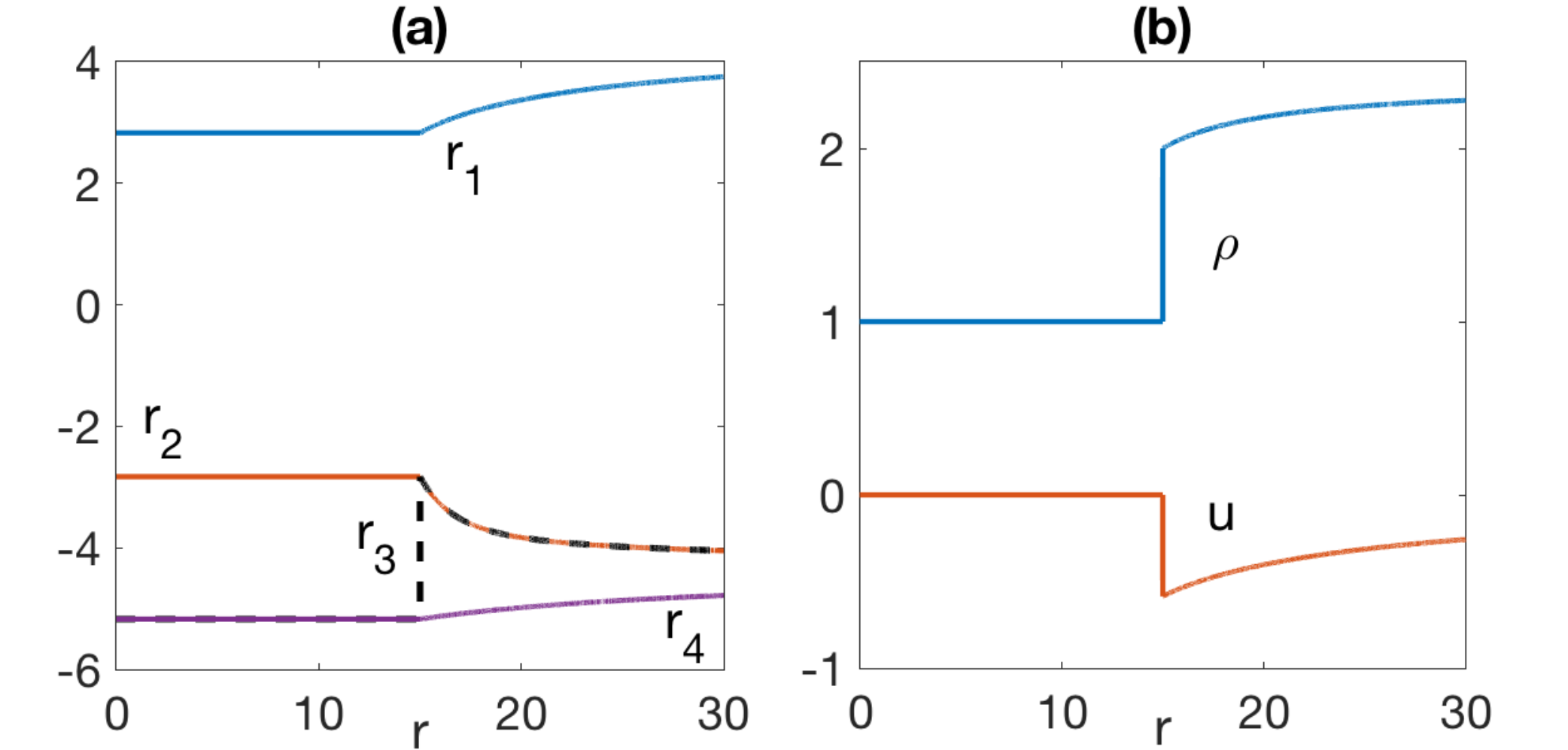}
\caption{Initial profiles of the (a) Riemann variables (\ref{eq:ICr}), and (b) density (\ref{eq:ICd}) and velocity (\ref{eq:ICu}) when $\rho_b = 2$ and $r_b$ near $15$.}
\label{r_initial_profiles}
\end{figure}

\ni where functions $\rho_{st}(r)$ and $u_{st}(r)$ with $\rho_{st}(r=r_b+)=\rho_b$ and  $u_{st}(r=r_b+)=u_b$ are given by eq.~(\ref{eq:rst}). For the DSW solutions we consider here, we need $\rho_b >1$ (in the opposite case $\rho_b<1$ the solution will be a rarefaction wave)~\cite{HoeferEtAl06}, see Fig.~\ref{r_initial_profiles}(b). As indicated above, the solution we study here is an inward propagating DSW. 
\par The corresponding IC for the total phase $\Theta$ of the wave function $\Psi$ is then

\be
\psi = \rho^{1/2}e^{i\Theta},  \qquad  \Theta(r,0) = \left\{ \begin{array}{cc} C, & 0\le r<r_b \\ \frac{\int_{r_b}^ru_{st}(z)dz}{\epsilon} + C, & r>r_b \end{array} \right.  \label{eq:ICwf}   %\eqno(ICwf)
\ee

\ni where $C$ is an arbitrary constant which we take to be zero without loss of generality; note that the total phase $\Theta$ is continuous at $r=r_b$. While the solution behind the trailing edge for rNLS is nonstationary in general (see below), the trailing edge here is also moving toward the origin, and therefore it is reasonable to assume that the solution will be approximately stationary for $r$ much larger than the initial jump location $r_b$.

%%%%%%%%%%%%%%%%%%%%%%%%%%%%%%%%%%%%%%%%%%%%%%%%%
%\comment{
%}
%%%%%%%%%%%%%%%%%%%%%%%%%%%%%%%%%%%%%%%%%%%%%%%%%

\par Let us see how the Whitham equations (\ref{eq:Wr}) fit the above consideration. First, they are consistent with Riemann variables $r_j$ being constant ahead of the front edge and equal to their corresponding values for the 1d NLS because as $m \rightarrow 0$ the functions $g_j$ vanish. 
\par Next, let us consider the conditions behind the trailing edge. There $r_3=r_2$, so $m=1$ and $E(m)/K(m)=0$. According to eqs.~(\ref{eq:4.2}), (\ref{eq:4.6}) and (\ref{eq:gm1}), the Whitham equations (\ref{eq:Wr}) reduce to the following three PDEs (the equation for $r_3$ becomes identical to the one for $r_2$):

\be
\prt_tr_1 + \frac{(r_4+3r_1)}{4}\prt_rr_1 - \frac{r_4^2-r_1^2}{8r} = 0,   \label{eq:r1d}   %\eqno(r1d)
\ee

\be
\prt_tr_4 + \frac{(3r_4+r_1)}{4}\prt_rr_4 + \frac{r_4^2-r_1^2}{8r} = 0,   \label{eq:r4d}   %\eqno(r4d)
\ee

\be
\prt_tr_2 + \frac{(2r_2+r_4+r_1)}{4}\prt_rr_2 + \frac{1}{6r}\left(r_2^2 + \frac{(r_4+r_1)r_2}{2} - \frac{(r_1+r_4)^2}{2} - \frac{(r_1-r_4)^2}{4}\right) = 0.   \label{eq:r2d}   %\eqno(r2d)
\ee

\ni One sees that the first two PDEs here make a closed subsystem, which determines the functions $r_1(r,t)$ and $r_4(r,t)$ while $r_2(r,t)$ can be found from the last equation afterwards in terms of them. One can verify that the closed subsystem of eqs.~(\ref{eq:r1d}) and (\ref{eq:r4d}) is exactly equivalent to the dispersionless rNLS system of eqs.~(\ref{eq:Im}) and (\ref{eq:u}) considered above, upon the identification

\be
r_4+r_1 = 4u,  \qquad  (r_1-r_4)^2 = 32\rho.  \label{eq:RP}   %\eqno(RP)
\ee

\ni This demonstrates the consistency of the considerations here. Thus, this system is indeed the analog of Riemann problem for the rNLS equation.
\par In terms of stationary solution described by function $s(x)$ in eqs.~(\ref{eq:s}) and (\ref{eq:s-sol}), the {\it stationary} eq.~(\ref{eq:r2d}) can be rewritten as

\be
(s_2+s)\frac{ds_2}{dx} + \frac{s_2^2+ss_2-2}{3x} = 0,  \qquad   r_2(r) = -2\sqrt{\rho_b+u_b^2}\cdot s_2(r/r_b),   \label{eq:s2st}   %\eqno(s2st)
\ee

\ni where $s=s(x)$ and $x=r/r_b$. Its solution $r_2=r_{2st}(r)$ may be hard to find exactly, but for large $x$ one obtains the expansion

\be
r_{2st}(r) = -2(\rho_b+u_b^2)^{1/2}(\sqrt2 - C_{r2}/x^{2/3} + g + \dots),   \qquad g=\frac{b}{x} = -\frac{\rho_bu_b}{(\rho_b + u_b^2)^{3/2}}\frac{r_b}{r},   \label{eq:r2st}   %\eqno(r2st)
\ee
where $C_{r2}>0$ is the unique constant such that the function $r_{2st}(r)$ takes the value $-2\sqrt2$ at $r=r_b$ making $r_2(r,0)$ continuous there.

\subsection{Derivation of IC and BC}

We impose the same IC and BC for the corresponding leading-order quantities $\rho_0$ and $u_0$ as they are for the full functions $\rho$ and $u$. To derive the IC and BC for the variables $r_j$, we recall the leading-order solution eq.~(\ref{eq:2.11}) for the density and note how $\rho_0$ changes throughout the DSW structure. At the lower density (front) edge of DSW, $m=0$ and so $\rho_0=\lambda_2=\lambda_1=1$ there. At the higher density (trailing) edge, $m=1$ and so $\rho_0=\lambda_2=\lambda_3$ and $\rho_0=\rho_b>1$ initially.
\par Consider now the implication of eq.~(\ref{eq:ICu}), the IC for the velocity $u_0$ determined by eq.~(\ref{eq:3.6}). Ahead of the front edge we have

$$
m=0:  \qquad  u_0 = \frac{V}{2} - \sigma\frac{\sqrt{2\lambda_1^2\lambda_3}}{2\lambda_1} = \frac{V}{2} - \sigma\frac{\sqrt{2\lambda_3}}{2} = 0,
$$

\ni which, since $V<0$, implies

\be
m=0: \qquad \sigma=-1,  \qquad V=V_+ = -\sqrt{2\lambda_3}   \label{eq:fe}
\ee

\ni at the front edge. Behind the trailing edge, we have

\be
m=1:  \qquad  u_0 = \frac{V}{2} - \sigma\frac{\sqrt{2\lambda_1\lambda_3^2}}{2\lambda_3} = \frac{V}{2} - \sigma\frac{\sqrt{2\lambda_1}}{2}.  \label{eq:te}  
\ee

\ni The condition $r_1\ge r_2 \ge r_3 \ge r_4$,  which needs to be satisfied everywhere and the definitions eq.~(\ref{eq:Rr}) imply also that $R_1\ge R_2 \ge R_3$ always hold.
%$R_1\ge R_2 \ge R_3 \ge V$ always hold ?
Recall that the sign $\sigma$ was defined by eq.~(\ref{eq:4.8}) so that $-2\sigma\sqrt{2\lambda_1\lambda_2\lambda_3} = R_1R_2R_3$. Since $\lambda_1\leq\lambda_2\leq\lambda_3$ we have $R_1^2\le R_2^2\le R_3^2$; then the possible signs are 

\be
\sigma=-1:  \quad R_1\ge 0 \ge R_2 \ge R_3;   \quad  \sigma=1:  \quad 0\ge R_1\ge R_2 \ge R_3,    \label{eq:sR}   %\eqno(sR)
\ee

\ni so that $R_1=-\sigma\sqrt{2\lambda_1}$. Thus, from eqs.~(\ref{eq:fe}), (\ref{eq:te}) and (\ref{eq:sR}), we have at the edges: 

%$$
%m=0:  \quad  R_1=\sqrt2  \mja{~($\lambda_1=1$)}, \quad R_2=-\sqrt2  \mja{~($\lambda_2=\lambda_1=1; EXPLAIN THE MINUS SIGN$)}, \ R_3=V=V_+  \mja{(from $u_0=0$)};   
%$$

$$
m=0:  \quad  R_1=\sqrt2, \quad R_2=-\sqrt2, \quad R_3=V=V_+;   
$$

\be
m=1:  \quad R_2=R_3=-\sqrt{2\rho_0}, \quad V+R_1 = V_-+R_1= 2u_0.  \label{eq:edgR}   %\eqno(edgR)
\ee

\ni This in turn translates into the following DSW edge conditions for the Riemann variables $r_j$, $j=1,2,3,4$,

$$
m=0:  \quad  r_1=2\sqrt2, \quad r_2=-2\sqrt2, \quad r_3=r_4= 2V_+;   
$$

\be
m=1:  \quad r_1 = 2(u_0+\sqrt{2\rho_0}), \qquad r_4 = 2(u_0-\sqrt{2\rho_0}), \qquad  r_2=r_3=2(V_- - u_0).  \label{eq:edgr}   %\eqno(edgr)
\ee

\ni From the last equations we also see that the identification eq.~(\ref{eq:RP}) fits exactly what we get here for the leading order solution $\rho_0$, $u_0$ at the trailing edge $m=1$. We impose the initial conditions for $r>r_b$ corresponding to the stationary solution of the dispersionless rNLS system described in the previous subsection. This means that {\it initially} we should have 

$$
r>r_b:  \quad \rho(r,0) = \rho_{st}(r),   \qquad  u(r,0) = u_{st}(r).
$$
 
\par Now we take into account that the DSW is the analog (dispersive regularization) of the simple wave solution for the dispersionless rNLS system in the case of small dispersion. Therefore we impose the continuity of $r_1$ and $r_4$ initially at $r=r_b$. First of these initial conditions, $r_1(r_b-0)=r_1(r_b+0)$, together with eq.~(\ref{eq:edgr}) implies

\be
u_b = -\sqrt{2}(\sqrt{\rho_b}-1)      \label{eq:ub} %\eqno(ub)
\ee
where $\rho_b \equiv \rho_{st}(r_b), \ u_b \equiv u_{st}(r_b)$. The second IC, $r_4(r_b-0)=r_4(r_b+0)$, implies that $r_4 = -2\sqrt2(2\sqrt{\rho_b} - 1)$ for $r\le r_b$ initially, which also fixes the value of $V$, $V=V_+ = -\sqrt2(2\sqrt{\rho_b} - 1)$ at the front edge.
\par Recall that $r_3=r_4$ at the front edge while $r_3=r_2$ at the trailing edge. The variable $r_3$ is the only one which has a jump initially and changes in the corresponding 1d NLS solution afterwards~\cite{HoeferEtAl06}. Therefore we should make also $r_2$ continuous initially which fixes the initial values $r_2(r_b+0)=r_2(r_b-0)=-2\sqrt2$ and $V_-(r_b+0) = u_b-\sqrt2 = -\sqrt{2\rho_b}$ on the outer side. Thus, from the above discussion starting with eq.~(\ref{eq:edgr}) and recalling also eq.~(\ref{eq:RP}) together with the ordering $r_1\ge r_2 \ge r_3 \ge r_4$, we obtain the ICs

{\small$$
r_1(r,0) = \left\{ \begin{array}{cc} 2\sqrt2, & 0\le r<r_b \\ 2(u_{st}(r)+\sqrt{2\rho_{st}(r)}), & r>r_b \end{array} \right.  \qquad r_2(r,0) = \left\{ \begin{array}{cc} -2\sqrt2, & 0\le r<r_b \\ r_{2st}(r), & r>r_b \end{array} \right.
$$

\be
r_3(r,0) = \left\{ \begin{array}{cc} -2\sqrt2(2\sqrt{\rho_b} - 1), & 0\le r<r_b \\ r_{2st}(r), & r>r_b \end{array} \right.  \quad  r_4(r,0) = \left\{ \begin{array}{ll} -2\sqrt2(2\sqrt{\rho_b} - 1), & 0\le r<r_b \\ 2(u_{st}(r)-\sqrt{2\rho_{st}(r)}), & r>r_b \end{array} \right. ,   \label{eq:ICr}   %\eqno(ICr)
\ee}

\ni where $u_{st}(r)$, $\rho_{st}(r)$ and $r_{2st}(r)$ are found from eqs.~(\ref{eq:rst}), (\ref{eq:s-sol}) and the solution of eq.~(\ref{eq:s2st}), respectively. The ICs eq.~(\ref{eq:ICr}) and those for $\rho$ and $u$ in eqs.~(\ref{eq:ICd}) and (\ref{eq:ICu}) are shown in Fig.~\ref{r_initial_profiles}. They are to be compared with the ICs for the corresponding 1d NLS initial step problem in Fig.~\ref{ic_1dnls} in the next section. 

\par {\it Remark.} An alternative to the above explanation of the ICs for Riemann variables relevant here comes from the concept of two Riemann variables, $2(u \pm \sqrt{2\rho})$, for dispersionless problem on each side of the DSW region. In our case these are $r_1$ and $r_2$ to the left of the front edge and $r_1$ and $r_4$ to the right of the trailing edge. The dispersionless analog of $r_1$ remains single-valued through the DSW region while the second dispersionless variable is triple-valued there, making up $r_2$, $r_3$ and $r_4$. The $r_3$ is the middle value merging with $r_4$ on the left and with $r_2$ on the right. The merged variables $r_4=r_3$ on the left and $r_2=r_3$ on the right satisfy the stationary equations there. The stationary solution for $r_4$ on the left happens to be a constant in our case while that for $r_2$ on the right is $r_{2st}(r)$. This gives the formulas for $r_j(r,0)$ in terms of $\rho(r,0)$ and $u(r,0)$ valid for more general step initial conditions and reducing to eq.~(\ref{eq:ICr}) for the case of ICs in eqs.~(\ref{eq:ICd}) and (\ref{eq:ICu}):
$$
r_1(r,0) = 2(u(r,0) + \sqrt{2\rho(r,0)}),   \qquad  r_2(r,0) = \left\{ \begin{array}{cc} 2(u - \sqrt{2\rho}), & r<r_b \\ r_{2st}(r), & r>r_b \end{array} \right.
$$

\be
r_3(r,0) = \left\{ \begin{array}{cc} r_4(r,0), & r<r_b \\ r_2(r,0), & r>r_b \end{array} \right.   \qquad r_4(r,0) = \left\{ \begin{array}{cc} r_{4st}(r)=2(u_b - \sqrt{2\rho_b}), & r<r_b \\ 2(u(r,0) - \sqrt{2\rho(r,0)}), & r>r_b \end{array} \right.  \label{eq:rho-u-to-r}
\ee

\par Then at later times we will have the same conditions ahead of the front (inner) edge of DSW $r<r_f(t)$ which determine the BC at the origin $r=0$. These conditions will no longer hold once the DSW reaches $r=0$. Also it is reasonable to assume that for large $r\gg r_b$ we still approximately have the stationary solution $\rho_{st}(r)$ and $u_{st}(r)$ at later times. This yields the boundary conditions for $r_1$, $r_4$ and $r_2$ ($r_3=r_2$ there) 

{\small$$
 \left\{ \begin{array}{l} r_1(r=0,t) = 2\sqrt2,  \\  r_1(r=\infty,t) = 2\sqrt{2(\rho_b+u_b^2)},  \end{array} \right.  \qquad  \left\{ \begin{array}{l} r_2(r=0,t) = -2\sqrt2,  \\ r_2(r=\infty,t) = -2\sqrt{2(\rho_b+u_b^2)},  \end{array} \right.
$$

\be
 \left\{ \begin{array}{l} r_3(r=0,t) = -2\sqrt2(2\sqrt{\rho_b} - 1),  \\  r_3(r=\infty,t) = -2\sqrt{2(\rho_b+u_b^2)},  \end{array} \right.  \quad  \left\{ \begin{array}{l} r_4(r=0,t) = -2\sqrt2(2\sqrt{\rho_b} - 1),  \\  r_4(r=\infty,t) = -2\sqrt{2(\rho_b+u_b^2)}, \end{array} \right.    \label{eq:BCr}   %\eqno(BCr)
\ee}

\ni The time $T_{max}$ needed for the front edge to reach the origin, after which the character of solution must change, is determined by

\be
T_{max} = \frac{r_b}{|v_f|},  \qquad  v_f = \left.v_3\right|_{m=0} = \left.\left(r_4 - \frac{r_1^2}{2r_4}\right)\right|_{m=0} = -\frac{\sqrt2(8\rho_b-8\sqrt{\rho_b}+1)}{2\sqrt{\rho_b}-1},   \label{eq:tmax}
\ee

\ni where $|v_f|$ is the front edge speed (the DSW is moving inside toward the origin $r=0$ so $v_f<0$).
\par The above consideration determines also the sign $\sigma$ for $r>r_b$ initially and at the trailing edge. Initially at $r=r_b+0$, we get $R_1 = 2u_b-V_-= -\sqrt2(\sqrt{\rho_b}-2)$. This implies

\[V_-=2u_b+\sqrt2(\sqrt{\rho_b}-2) = - \sqrt{2\rho_b}. \]

Thus, there are two cases: if $\rho_b<4$, then $R_1>0$ and $\sigma=-1$; if $\rho_b>4$, then $R_1<0$ and $\sigma=1$. In the first case, $\sigma(r,0) \equiv -1$ and it remains constant later. In the second case, for large initial density jump such that $\rho_b>4$, we have $\sigma(r,0) = -1$ for $r<r_b$ and $\sigma(r,0) = 1$ for $r>r_b$ close to $r_b$. This implies the corresponding different boundary conditions for $\sigma$ at the front and the trailing edge. Therefore there is always a radius $r=r_{vac}$ inside the DSW structure, where $\sigma$ jumps from one value to the other. The jump occurs when $R_1=0$, and then the minimum value of $\rho_0$ as a function of $\theta$ is $(\rho_0)_{min}=(\lambda_1)_{min}=0$, according to eq.~(\ref{eq:2.11}). This corresponds to the {\it vacuum point}~\cite{GK87, HoeferEtAl06} (or rather vacuum ring in our two-dimensional case) where density falls to zero and hydrodynamic velocity (phase gradient) has infinite jump, see eq.~(\ref{eq:3.6}). The first of eqs.~(\ref{eq:Rr}) gives 

$$
r_3 = r_1 + r_4 - r_2   %\eqno(rv)
$$

\ni at the vacuum point. For 1d NLS, this implies $r_3= -2\sqrt2(2\sqrt{\rho_b} - 3)$ at $r=r_{vac}$, since only $r_3$ changes in DSW and $r_1$, $r_2$ and $r_4$ remain constant at all times. This is, however, not true for the DSW solution of rNLS we study. Here all $r_j$ change inside the DSW and even behind its trailing (solitonic) edge due to the additional $\sim1/r$ terms in the equations.

\section{Numerical results}   

In this section we compare the numerical evaluation of the Whitham theory asymptotic results with direct numerical simulations. First we discuss the asymptotic solution, then direct simulations are discussed. The differences between the 1d NLS and rNLS solutions are highlighted.

\subsection{Asymptotic solution.}
Using a version of the available MATLAB code~\cite{Shamp}, we solve eqs.~(\ref{eq:Wr}) for $r_j$, $j=1,\dots,4$, with initial conditions eq.~(\ref{eq:ICr}), where $u_{st}(r)$ is given by eq.~(\ref{eq:rst}) with eq.~(\ref{eq:ub}) for the constant $u_b$ in terms of constant $\rho_b>1$. A Lax-Wendroff scheme with nonlinear filtering in~\cite{Shamp} is used to evolve the Whitham system. In the limits $m \rightarrow 0$ or $m \rightarrow 1$ the terms $v_j$ and $g_j$ in eq.~(\ref{eq:Wr}) are replaced by their own limiting values; the $g_j$ limits are given at the end of section 4.
\par We take $\rho_{st}(r)$ to be nondecreasing (the analog of a simple wave here), then the corresponding root $s(x)$ of the cubic equation eq.~(\ref{eq:s})  (given by eq. (\ref{eq:s-sol})) becomes small as $x$ becomes large; mathematically $s(x)$ approaches $0$ as $x\to\infty$. Then from eqs.~(\ref{eq:rst}), (\ref{eq:RP}) we have

$$
r_1(r) = 2\sqrt{\rho_b+u_b^2}\left[-s(r/r_b) + \sqrt{2(1-s^2(r/r_b))}\right],   
$$

$$
r_4(r) = 2\sqrt{\rho_b+u_b^2}\left[-s(r/r_b) - \sqrt{2(1-s^2(r/r_b))}\right]. 
$$

\ni These functions have the following asymptotics for large $r$, i.e.~as $g\equiv g(r/r_b)\ll 1$ in eq.~(\ref{eq:s}),

$$
r_1(r) = 2\sqrt{2(\rho_b+u_b^2)}\left(1 - \frac{g}{\sqrt2} - \frac{g^2}{2} - \frac{g^3}{\sqrt2} - \frac{9g^4}{8} -  \frac{3g^5}{\sqrt2} - 4g^6 + \dots\right),
$$

$$
r_4(r) = 2\sqrt{2(\rho_b+u_b^2)}\left(-1 - \frac{g}{\sqrt2} + \frac{g^2}{2} - \frac{g^3}{\sqrt2} + \frac{9g^4}{8} -  \frac{3g^5}{\sqrt2} + 4g^6 + \dots\right).
$$

\ni Function $r_2(r)=r_{2st}(r)$ is the solution of the ODE eq.~(\ref{eq:s2st}). The initial and boundary conditions we impose for computations are given by eqs.~(\ref{eq:ICr}) and the finite maximum $r=r_{max}$ version of eq.~(\ref{eq:BCr}). We took $r_b=15$ and $r_{max}=30$ in our numerics here. It is important to note that the computation time should not exceed the time $T_{max}$ of eq.~(\ref{eq:tmax}) needed for the front edge to reach the origin since after that the character of the  solution changes. 
%Thus, the boundary conditions to impose for computations are:

%{\small$$
% \left\{ \begin{array}{l} r_1(r=0,t) = 2\sqrt2,  \\  r_1(r=r_{max},t) = 2(u_{st}(r_{max})+\sqrt{2\rho_{st}(r_{max})}),  \end{array} \right.  \qquad  \left\{ \begin{array}{l} r_2(r=0,t) = -2\sqrt2,  \\ r_2(r=r_{max},t) = r_{2st}(r_{max}),  \end{array} \right.
%$$

%\be
% \left\{ \begin{array}{l} r_3(r=0,t) = -2\sqrt2(2\sqrt{\rho_b} - 1),  \\  r_3(r=r_{max},t) = r_{2st}(r_{max}),  \end{array} \right.  \quad  \left\{ \begin{array}{l} r_4(r=0,t) = -2\sqrt2(2\sqrt{\rho_b} - 1),  \\  r_4(r=r_{max},t) = 2(u_{st}(r_{max})-\sqrt{2\rho_{st}(r_{max})}), \end{array} \right.    \label{eq:BCr}   %\eqno(BCr)
%\ee}

As for the trailing (solitonic) edge velocity, initially it is $v_{tr}(t=0)= -\sqrt{2\rho_b}$ which is the same as the constant velocity in the 1d NLS case. To our surprise numerical evidence suggests that, for the initial jump $\rho_b$ not very large, the trailing edge continues to move with almost this velocity. The small observed deflection from constancy, see Fig.~\ref{trail_edge_evolve}, could be either an order epsilon effect or a genuine weak dependence of the velocity on time following from our leading order theory. The trailing edge continues to move toward the origin, as was also observed e.g.~in~\cite{HoeferEtAl06}. 
\par After having solved the Whitham equations for $r_j$-variables, we compute the leading order rNLS solution by eqs.~(\ref{eq:2.11}) and (\ref{eq:3.6}). Hence we can compute a numerical approximation to the asymptotic solution of rNLS eq.~(\ref{eq:1.1}) using eqs. (\ref{eq:Wr})-(\ref{eq:u0r}); note that the phase is determined from eq.~(\ref{eq:Phase}). 

%%%%%%%%%%%%%%%%%%%%%%%%%%%%%%%%%%%%%%%%%%%%%%%%%
%\begin{figure} [ht]
%\centering
%\includegraphics[scale=.6]{ic_rnls.eps}
%\caption{Initial profiles of the (a) riemann invariants (\ref{eq:ICr}), and (b) density (\ref{eq:ICd}) and velocity (\ref{eq:ICu}) when $\rho_b = 2$ and $r_b$ near $15$.}
%\label{r_initial_profiles}
%\end{figure}
%%%%%%%%%%%%%%%%%%%%%%%%%%%%%%%%%%%%%%%%%%%%%%%%%

%%%%%%%%%%%%%%%%%%%%%%%%%%%%%%%%%%%%%%%%%%%%%%%%%
%\begin{figure} [ht]
%\centering
%\includegraphics[scale=.64]{compare_rho_u.eps}
%\caption{Comparison of the rNLS density and hydrodynamic velocity from direct numerics (blue) and Whitham theory (orange) at time $t = 1$ for $\epsilon = .05$\mja{; the} initial jump $r_b$ is located near $15$. The jump height parameter is $\rho_b = 2$ (left column) and $\rho_b = 3$ (right column).}
%\label{compare_rho_u}
%\end{figure}
%%%%%%%%%%%%%%%%%%%%%%%%%%%%%%%%%%%%%%%%%%%%%%%%%%
 
%the formulas given in the first section.

\par The initial profiles of the Riemann variables (\ref{eq:ICr}) are shown in Fig.~\ref{r_initial_profiles}(a). The point of discontinuity $r_b$ is located near $r = 15$, however, neither $\rho_0(r_b)$ nor $u_0(r_b)$ are explicitly computed. In our implementation the point $r_b$ is taken to be ``off-grid'' meaning there is no numerical grid point at $r_b$. Eqs.~(\ref{eq:ICd}) and (\ref{eq:ICu}) give definitions for the density and velocity, respectively; their numerical evaluations are displayed in Fig.~\ref{r_initial_profiles}(b).
\par {\it Remark.} In the case of large initial density jump $\rho_b\ge4$ (which we do not take here), for some location $r=r_v(t)<r_b$, we have $\sigma=0$, and initially $r_v=r_b$ so that

$$
\rho_b>4:  \qquad  \sigma(r,0) = \left\{ \begin{array}{cc} -1, & 0\le r<r_b \\ 1, & r>r_b \end{array}  \right.  
$$

\ni At later times, there is a vacuum point $r=r_v(t)$ determined by the condition $\lambda_1=0$ (there is only one such point in the DSW region which always lies inside $0<r\le r_b$ domain). Thus, we should take $\sigma(r,t)=-1$ to the left of this point and $\sigma(r,t)=1$ to the right of it in the radial variable $r$. 

\subsection{Comparison with the direct solution of the rNLS equation} 

%%%%%%%%%%%%%%%%%%%%%%%%%%%%%%%%%%%%%%%%%%%%%%%%%
\begin{figure} [ht]
\centering
\includegraphics[scale=.64]{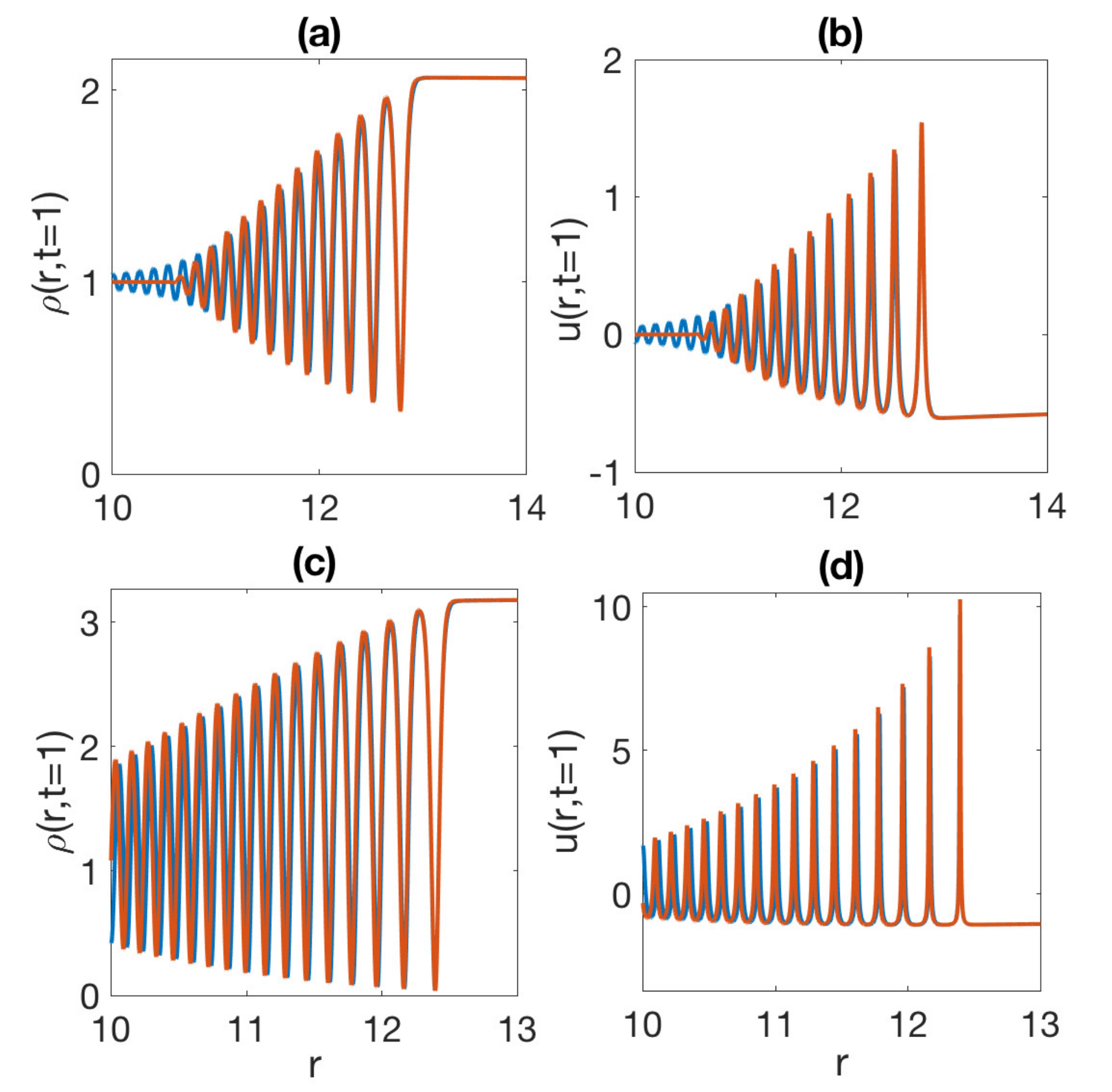}
\caption{Comparison of the rNLS density and hydrodynamic velocity from direct numerics (blue) and Whitham theory (orange) at time $t = 1$ for $\epsilon = 0.05$; the initial jump $r_b$ is located near $15$. The jump height parameter is $\rho_b = 2$ (left column) and $\rho_b = 3$ (right column).}
\label{compare_rho_u}
\end{figure}
%%%%%%%%%%%%%%%%%%%%%%%%%%%%%%%%%%%%%%%%%%%%%%%%%

The direct numerical solution of the rNLS eq.~(\ref{eq:1.1}) with step initial conditions eqs.~(\ref{eq:ICd}), (\ref{eq:ICu}) for the density $\rho=|\Psi|^2$ and hydrodynamic velocity $u=\epsilon\prt_r\text{arg}\Psi$, where functions $\rho_{st}(r)$ and $u_{st}(r)$ are given by eq.~(\ref{eq:rst}), is discussed next. The corresponding IC for the total phase $\Theta$ of the wave function $\Psi$ can then be given as

$$
\Psi = \rho^{1/2}(r,0)e^{i\Theta(r,0)}, \qquad  \Theta(r,0) = \left\{ \begin{array}{cc} 0, & 0\le r<r_b \\ \frac{\int_{r_b}^ru_{st}(z)dz}{\epsilon}, & r>r_b \end{array} \right.  \label{eq:ICwf0}   %\eqno(ICwf)
$$

\ni The BC for the wavefunction $\Psi = \rho^{1/2}e^{i\Theta}$ in the finite computation domain is

$$
\rho(0,t)\equiv 1, ~~~  \rho(r_{max},t) = \rho_{st}(r_{max}), \qquad  u(0,t) \equiv 0 , ~~~  u(r_{max},t) = u_{st}(r_{max}),
$$

\be
\Theta(0,t)\equiv -\frac{t}{\epsilon},  \quad  \Theta(r_{max},t) = -\frac{(\rho_b+u_b^2)t}{\epsilon} + \frac{\int_{r_b}^{r_{max}}u_{st}(z)dz}{\epsilon}.  \label{eq:BCmax}   %\eqno(BCmax)
\ee

%%%%%%%%%%%%%%%%%%%%%%%%%%%%%%%%%%%%%%%%%%%%%%%%%
%\begin{figure} [ht]
%\centering
%\includegraphics[scale=.64]{compare_rho_u.eps}
%\caption{Comparison of the rNLS density and hydrodynamic velocity from direct numerics (blue) and Whitham theory (orange) at time $t = 1$ for $\epsilon = 0.05$\mja{; the} initial jump $r_b$ is located near $15$. The jump height parameter is $\rho_b = 2$ (left column) and $\rho_b = 3$ (right column).}
%\label{compare_rho_u}
%\end{figure}
%%%%%%%%%%%%%%%%%%%%%%%%%%%%%%%%%%%%%%%%%%%%%%%%%

The rNLS equation (\ref{eq:1.1}) is numerically integrated by a Runge-Kutta scheme using a standard finite-difference centered discretization in space. A smoothed version of the initial functions shown in Fig.~\ref{r_initial_profiles}(b) is used; the boundary conditions are given above. In all simulations shown below $r_{max} = 30$ and $r_b$ is near 15. This value of $r_b$ is chosen close enough to the origin so that the radial term is non-trivial, yet is far enough away that it will take some time before the leading edge reaches $r = 0$. 
\par A comparison of density and velocity between the Whitham theory and direct numerics is presented in Fig.~\ref{compare_rho_u}. For the direct numerics, we compute $\rho = |\Psi|^2$ and $u = i \epsilon \left[\Psi \Psi^*_r - \Psi^* \Psi_r \right] /(2\rho)$ after integrating the rNLS eq.~(\ref{eq:1.1}). To compare with the Whitham theory results we use the asymptotic approximations $\rho_0$ and $u_0$ given in eqs.~(\ref{eq:2.11}) and (\ref{eq:3.6}), respectively. Overall the agreement between the Whitham approximation and the direct simulation is quite good. The function $\rho_0$ in eq.~(\ref{eq:2.11}) is only known up to a phase shift $\theta_*$ which is a function of space and time. Also note that $\theta_*$ is independent of the fast variable $\theta$ in eq.~(\ref{eq:2.11}). We choose this $\theta_*$ so that the lowest dips of the density near the trailing edge of the two solutions coincide with one another. The results indicate remarkably good agreement.

%To the right of the trailing edge the Whitham solution is in front of the direct solution, and vice versa to the left of this point.

%%%%%%%%%%%%%%%%%%%%%%%%%%%%%%%%%%%%%%%%%%%%%%%%%
%\comment{
\begin{figure} [ht]
\centering
\includegraphics[scale=.6]{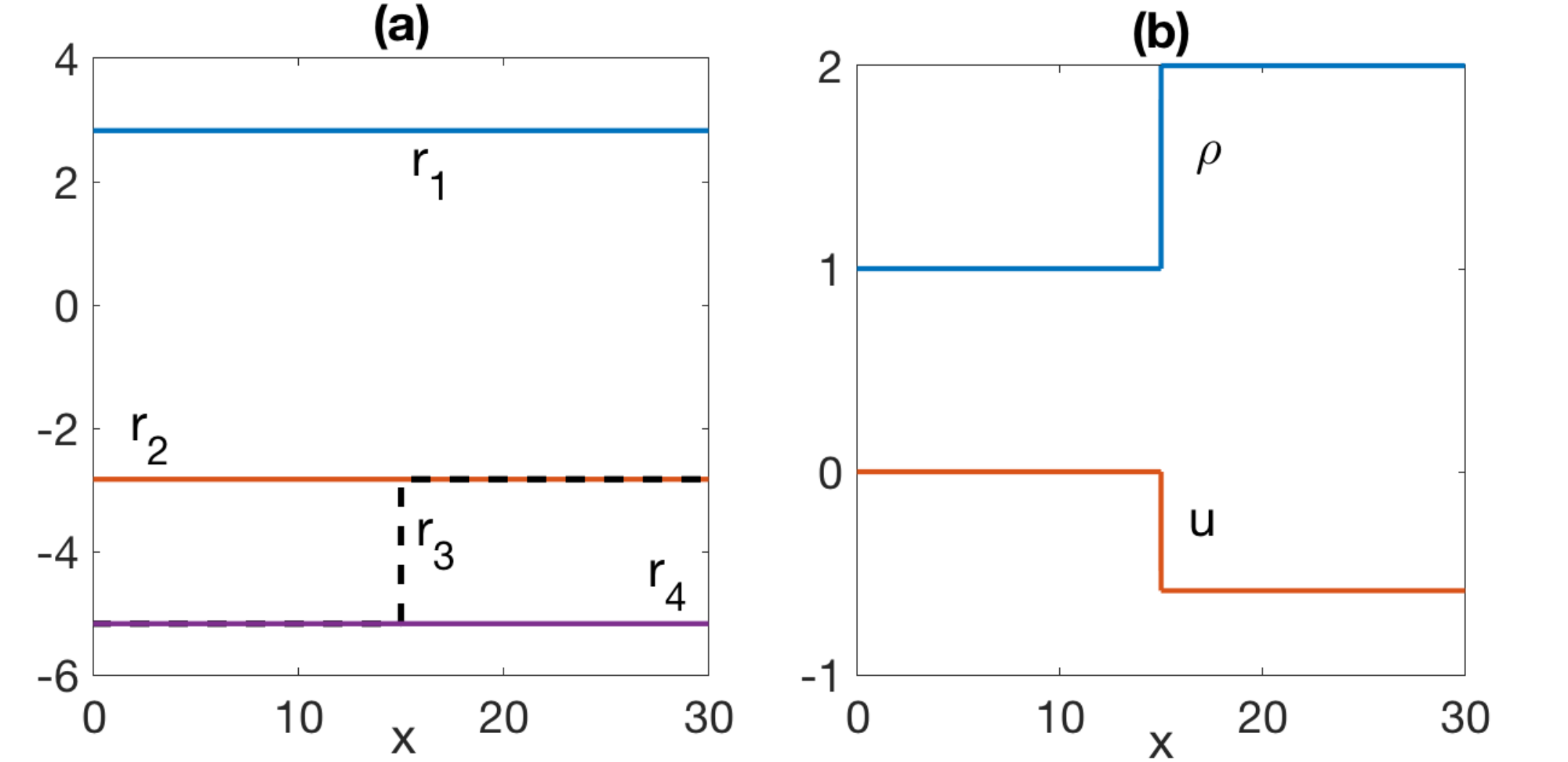}
\caption{Initial conditions for the 1d Whitham system when $\rho_b = 2$ with $r_b$ near 15.}
\label{ic_1dnls}
\end{figure}
%}
%%%%%%%%%%%%%%%%%%%%%%%%%%%%%%%%%%%%%%%%%%%%%%%%%

\subsection{Comparison with the solution of 1d NLS equation} 

Next we discuss the asymptotic solution of the corresponding 1d NLS equation,

%%%%%%%%%%%%%%%%%%%%%%%%%%%%%%%%%%%%%%%%%%%%%%%%%
\comment{
\begin{figure} [ht]
\centering
\includegraphics[scale=.6]{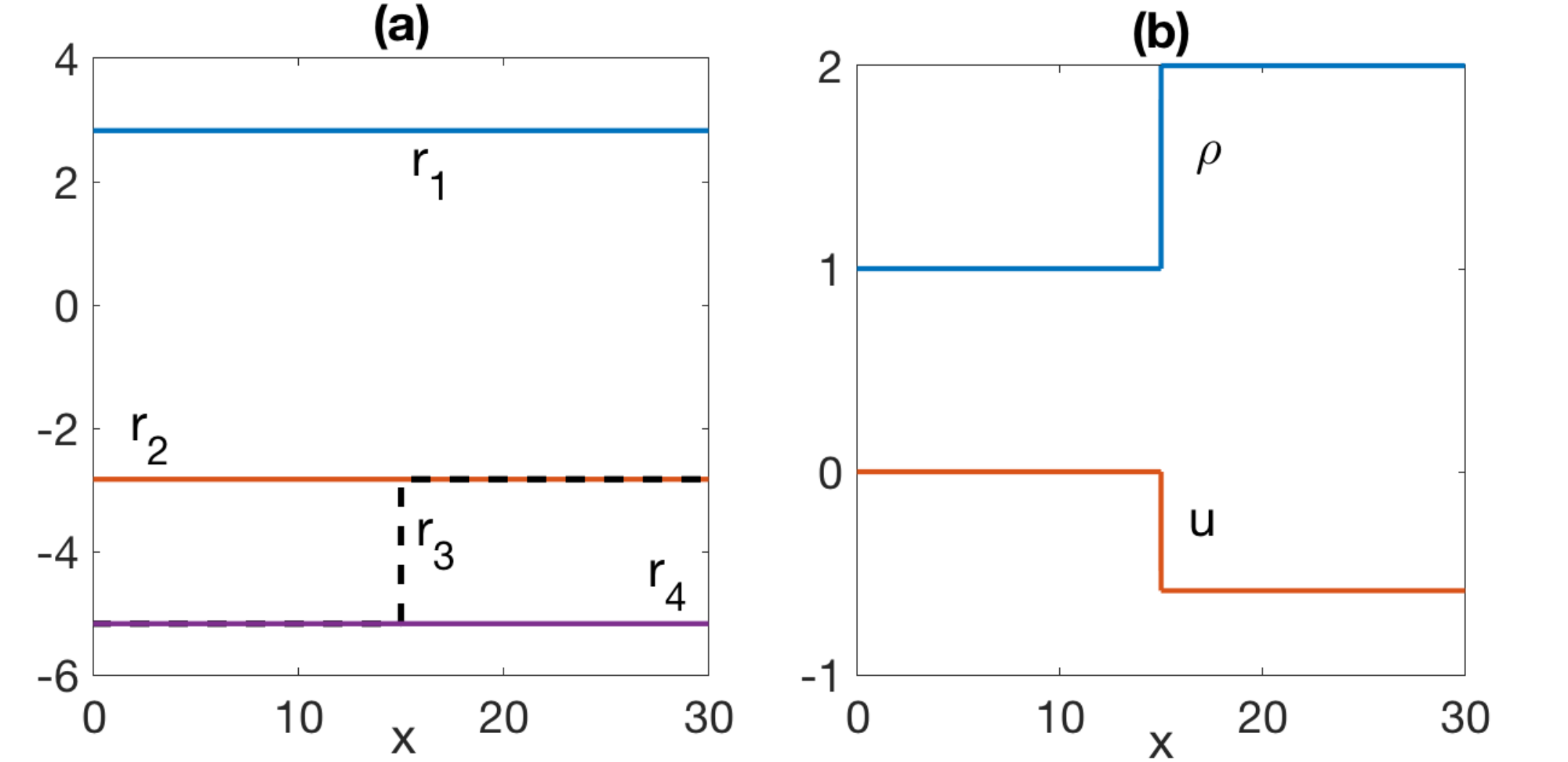}
\caption{Initial conditions for the 1d Whitham system when $\rho_b = 2$ with $r_b$ near 15.}
\label{ic_1dnls}
\end{figure}
}
%%%%%%%%%%%%%%%%%%%%%%%%%%%%%%%%%%%%%%%%%%%%%%%%%

\be
i\epsilon\prt_t\Psi + \epsilon^2\prt_{xx}\Psi - |\Psi|^2\Psi = 0,  \label{eq:1dNLS}
\ee

\ni with the step initial conditions similar to ones in~\cite{HoeferEtAl06}, i.e.~we consider the following initial conditions for the density $\rho=|\Psi|^2$ and hydrodynamic velocity $u=\epsilon\prt_x\text{arg}\Psi$,

$$
\rho(x,0) = \left\{ \begin{array}{cc} 1, & 0\le x<r_b \\ \rho_b, & x>r_b \end{array} \right.   %\eqno(ICd)
$$

$$
u(x,0) = \left\{ \begin{array}{cc} 0, & 0\le x<r_b \\ u_b<0, & x>r_b \end{array} \right.,   %\eqno(ICu)
$$

\ni where the initial jump parameters $r_b$, $\rho_b$ and $u_b$ are the same as those for the corresponding rNLS problem above. They are shown in Fig.~\ref{ic_1dnls}(b). The corresponding IC for the total phase $\Theta$ of the wave function $\Psi$ can then be given as

$$
\Psi = \rho^{1/2}e^{i\Theta},  \qquad  \Theta(x,0) = \left\{ \begin{array}{cc} 0, & 0\le x<r_b \\ \frac{u_b(x-r_b)}{\epsilon}, & x>r_b \end{array} \right.  \label{eq:ICwf1}   %\eqno(ICwf)
$$

%%%%%%%%%%%%%%%%%%%%%%%%%%%%%%%%%%%%%%%%%%%%%%%%%
\begin{figure} [ht]
\centering
\includegraphics[scale=.6]{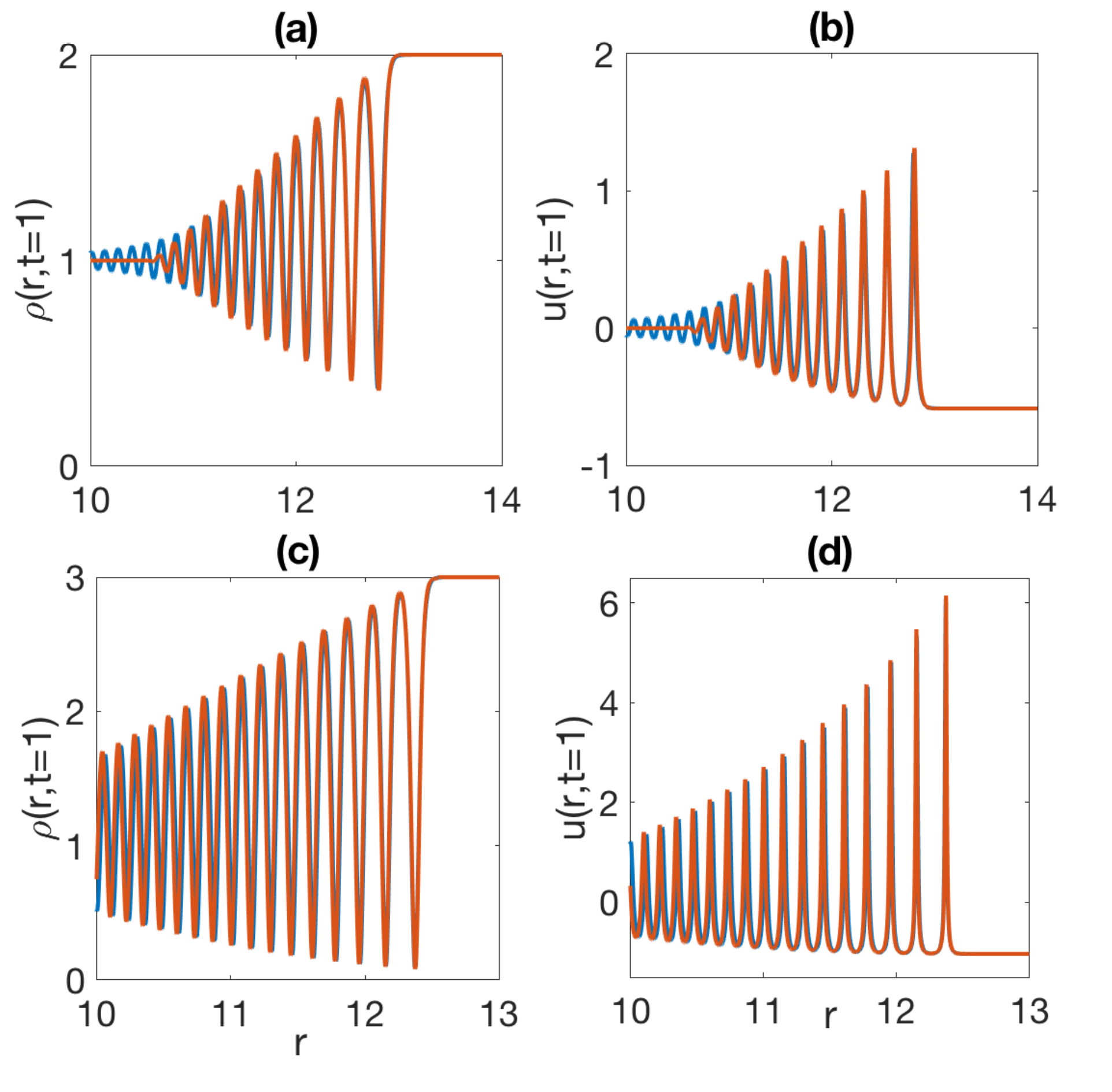}
\caption{Comparison of the 1d NLS density and hydrodynamic velocity from direct numerics (blue) and Whitham theory (orange) at time $t = 1$ for $\epsilon = 0.05$ and initial jump $r_b$ is located near $15$. The jump height parameter is $\rho_b = 2$ (left column) and $\rho_b = 3$ (right column).}
\label{compare_rho_u_1d}
\end{figure}
%%%%%%%%%%%%%%%%%%%%%%%%%%%%%%%%%%%%%%%%%%%%%%%%%

\ni The  computation domain in $x$ is the same as it is for radial variable $r$ above. However, the boundary conditions here are

$$
  \rho(0,t)\equiv 1,  ~~~ \rho(x_{max},t) = \rho_b,   \qquad   u(0,t) \equiv 0, ~~~ u(x_{max},t) = u_b,
$$
$$
\Theta(0,t)\equiv -\frac{t}{\epsilon},  \qquad  \Theta(x_{max},t) = -\frac{(\rho_b+u_b^2)t}{\epsilon} + \frac{u_b(x_{max}-r_b)}{\epsilon}.  \label{eq:BCmax1}   %\eqno(BCmax)
$$

%%%%%%%%%%%%%%%%%%%%%%%%%%%%%%%%%%%%%%%%%%%%%%%%%
\comment{\begin{figure} [ht]
\centering
\includegraphics[scale=.6]{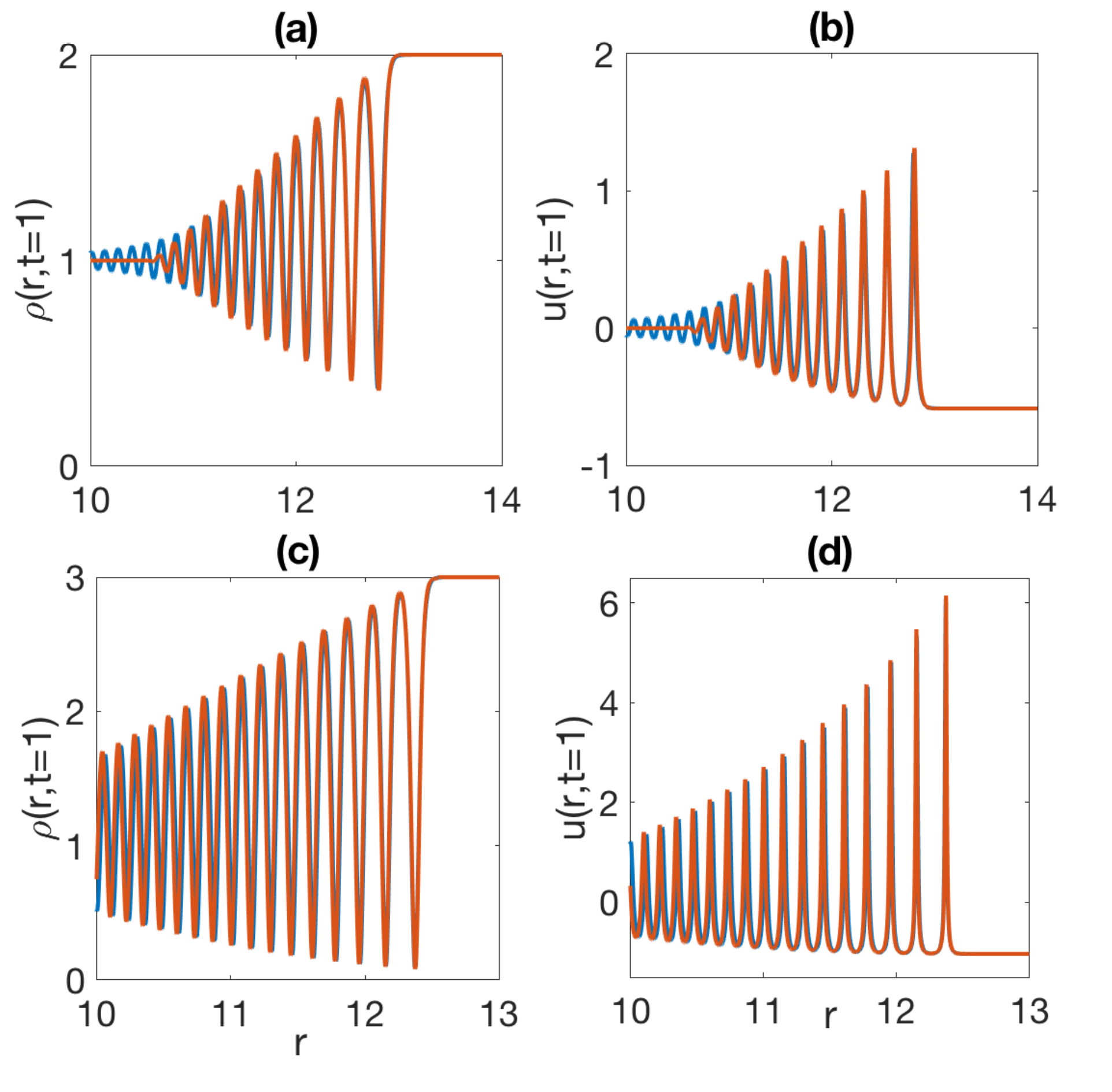}
\caption{Comparison of the 1d NLS density and hydrodynamic velocity from direct numerics (blue) and Whitham theory (orange) at time $t = 1$ for $\epsilon = 0.05$ and initial jump $r_b$ is located near $15$. The jump height parameter is $\rho_b = 2$ (left column) and $\rho_b = 3$ (right column).}
\label{compare_rho_u_1d}
\end{figure}}
%%%%%%%%%%%%%%%%%%%%%%%%%%%%%%%%%%%%%%%%%%%%%%%%%

In the 1d case the Whitham system (\ref{eq:Wr}) simplifies, with $g_j \equiv 0$ for each $j$. As such, three of the Riemann variables also simplify to constants

$$
r_1(x,t) = 2\sqrt{2},  \qquad r_2(x,t) = - 2\sqrt{2} , \qquad r_4(x,t) = -2\sqrt{2} (2 \sqrt{\rho_b}-1) ,
$$
and the other variable has a jump up located at $r_b$ with the boundary conditions 
$$ 
\begin{cases}
r_3(0,t) = -2\sqrt{2} (2 \sqrt{\rho_b} - 1) \\
r_3(x_{max},t) = -2\sqrt{2}
\end{cases} .
$$

%%%%%%%%%%%%%%%%%%%%%%%%%%%%%%%%%%%%%%%%%%%%%%%%%
%\comment{
\begin{figure} [ht]
\centering
\includegraphics[scale=.6]{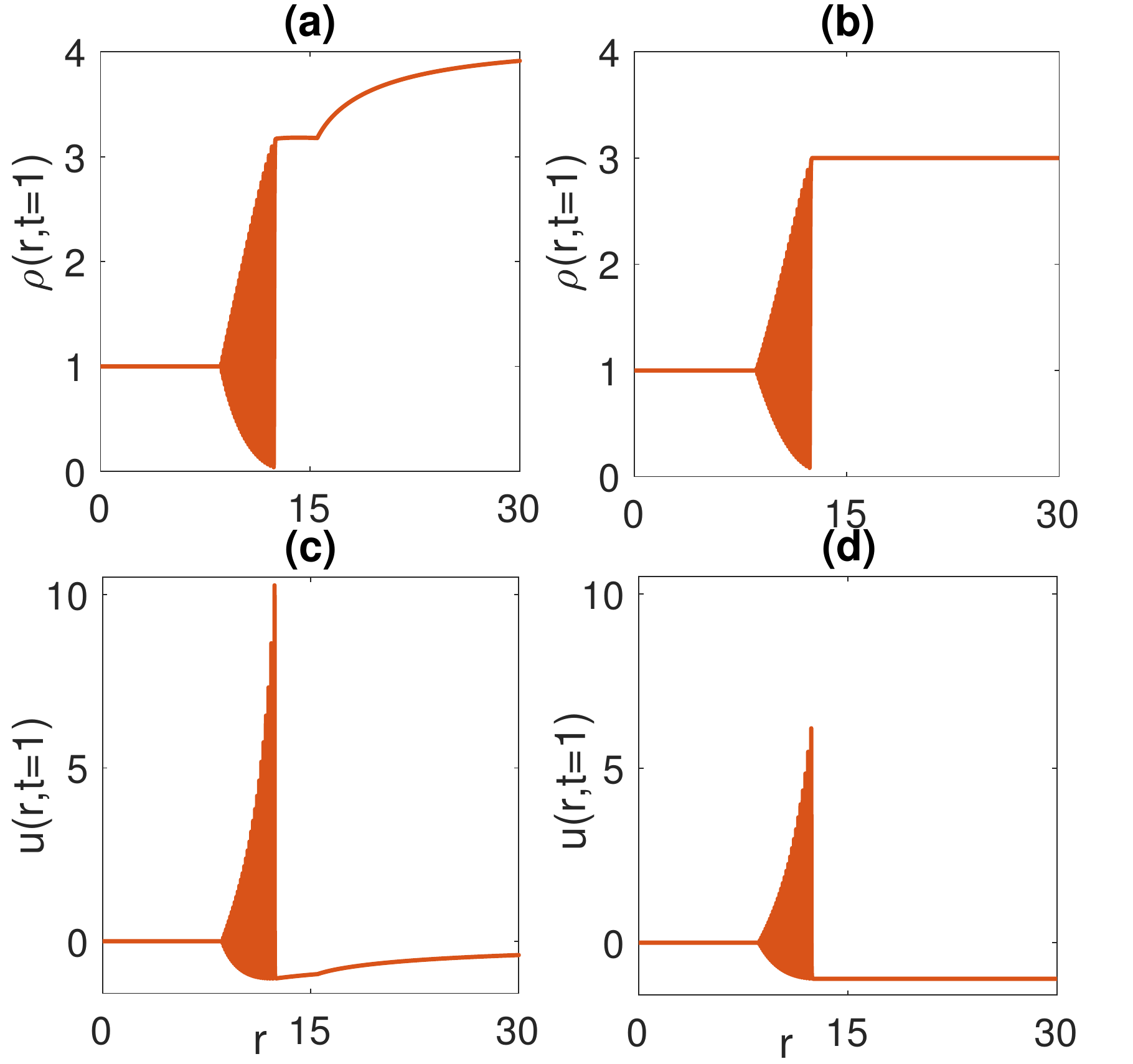}
\caption{Comparison of the rNLS (left column) and 1d NLS (right column) asymptotic density and hydrodynamic velocity at time $t = 1$ for $\epsilon = 0.05, \rho_b = 3$ and $r_b \approx 15$.}
\label{compare_rnls_1d_rhob3}
\end{figure}
%}
%%%%%%%%%%%%%%%%%%%%%%%%%%%%%%%%%%%%%%%%%%%%%%%%%

These Riemann invariants are determined by the step-up density and step-down velocity given above like in eq.~(\ref{eq:rho-u-to-r}). A typical set of initial conditions is shown in Fig.~\ref{ic_1dnls}, the reader can compare with the ICs in Fig.~\ref{r_initial_profiles} for the rNLS case.
We remark that the Riemann variable $r_3$  admits a similarity solution $r_3=r_3(x/t)$ which can be found exactly from eq.~(\ref{eq:4.1}).

%%%%%%%%%%%%%%%%%%%%%%%%%%%%%%%%%%%%%%%%%%%%%%%%%
%\begin{figure} [ht]
%\centering
%\includegraphics[scale=.6]{compare_rho_u_1d.eps}
%\caption{Comparison of the 1d NLS density and hydrodynamic velocity from direct numerics (blue) and Whitham theory (orange) at time $t = 1$ for $\epsilon = .05$ and initial jump $r_b$ is located near $15$. The jump height parameter is $\rho_b = 2$ (left column) and $\rho_b = 3$ (right column).}
%\label{compare_rho_u_1d}
%\end{figure}
%%%%%%%%%%%%%%%%%%%%%%%%%%%%%%%%%%%%%%%%%%%%%%%%%

%%%%%%%%%%%%%%%%%%%%%%%%%%%%%%%%%%%%%%%%%%%%%%%%%
\comment{
\begin{figure} [ht]
\centering
\includegraphics[scale=.6]{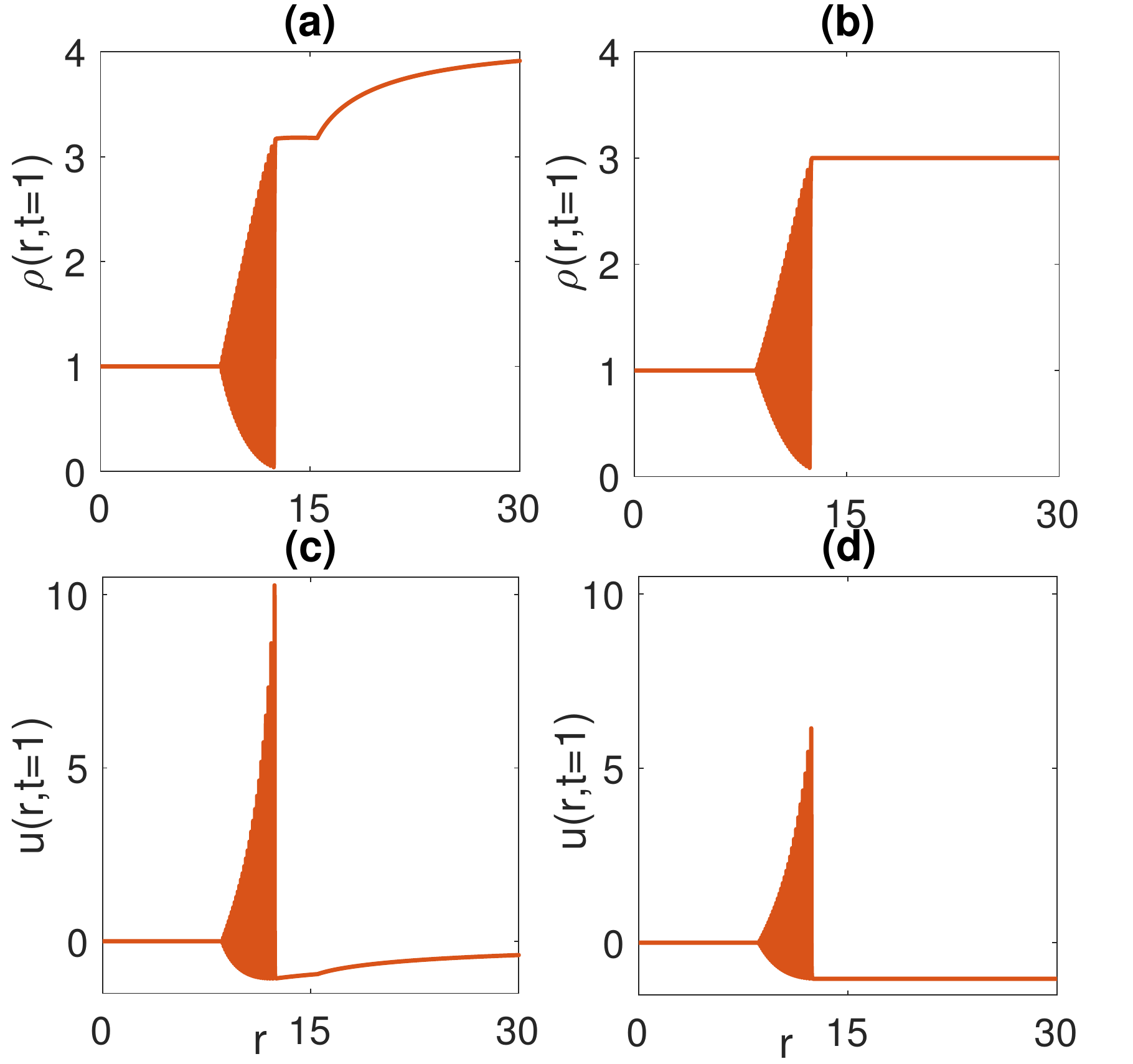}
\caption{Comparison of the rNLS (left column) and 1d NLS (right column) asymptotic density and hydrodynamic velocity at time $t = 1$ for $\epsilon = 0.05, \rho_b = 3$ and $r_b \approx 15$.}
\label{compare_rnls_1d_rhob3}
\end{figure}}
%%%%%%%%%%%%%%%%%%%%%%%%%%%%%%%%%%%%%%%%%%%%%%%%%

A comparison of the solutions found by directly solving the 1d NLS equation (\ref{eq:1dNLS}) and from its Whitham system eq.~(\ref{eq:4.1}) are shown in Fig.~\ref{compare_rho_u_1d}. The numerical schemes are the same as those used before to solve the rNLS equation. Again, the Whitham solution phase shift $\theta_*$ is chosen so that the trailing lowest dips for $\rho$ of the direct and Whitham solutions are aligned. Altogether there is very good agreement between the 1d Whitham approximation and the direct numerics. Comparing these solutions to those found in the rNLS case (see Fig.~\ref{compare_rnls_1d_rhob3}) we note that the 1d and radial solution structures do resemble each other; but there are clear numerical differences. One of the most striking differences between the two is that while density values are relatively close the hydrodynamic velocity found in rNLS is much larger. Comparing Figs.~\ref{compare_rnls_1d_rhob3}(c) and \ref{compare_rnls_1d_rhob3}(d) we note that $|u_{max}| \approx 10$ for rNLS whereas $|u_{max}| \approx 6$ for the 1d solution.

%%%%%%%%%%%%%%%%%%%%%%%%%%%%%%%%%%%%%%%%%%%%%%%%%
%\comment{
\begin{figure} [ht]
\centering
\includegraphics[scale=.6]{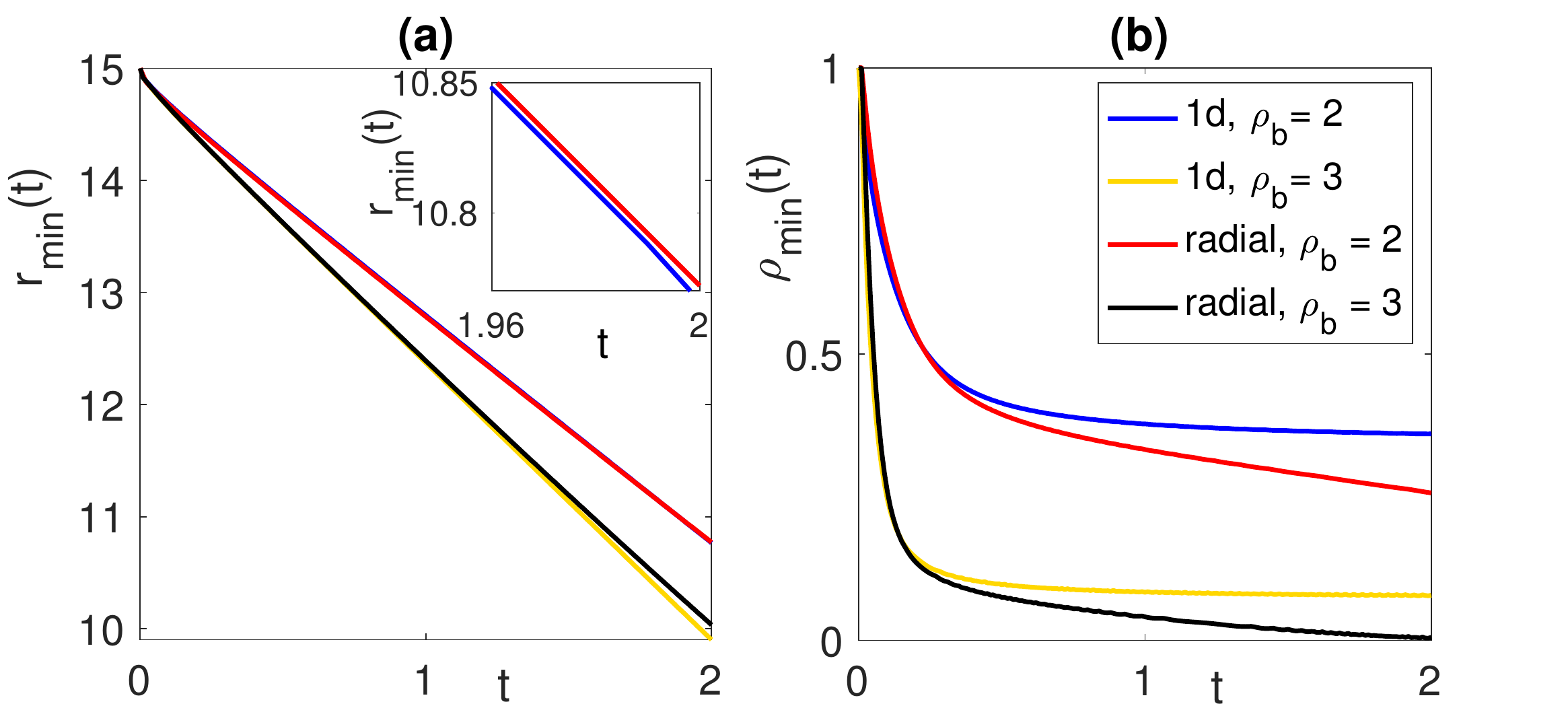}
\caption{Comparison of the radial and 1d NLS trailing edge evolution shown in Figs.~\ref{compare_rho_u} and \ref{compare_rho_u_1d}, respectively. Inset shows a slight difference between the 1d NLS and rNLS trailing edges near $t = 2$.}
\label{trail_edge_evolve}
\end{figure}
%}
%%%%%%%%%%%%%%%%%%%%%%%%%%%%%%%%%%%%%%%%%%%%%%%%%

\par One clear difference between the 1d and radial solutions is the behavior of the trailing edge, in particular its velocity and magnitude. In Fig.~\ref{trail_edge_evolve} the position $r_{\min}(t)$ of the lowest dip of $\rho$ near the trailing edge and its value $\rho_{min}(t) = \rho(r_{min}(t))  \le \rho(r,t)$ are tracked. The velocity of the rNLS trailing edge in Fig.~\ref{trail_edge_evolve}(a) is observed to be nearly constant and, to leading order, travel with the same speed as the 1d solution (see eq.~(\ref{eq:vtr1})). Next we examine the lowest dip value $\rho_{min}(t)$ in Fig.~\ref{trail_edge_evolve}(b). There we observe that the 1d solutions approach a constant at long times. This is in contrast to the rNLS trailing edges which do not appear to have any minimum (other than the obvious need to be non-negative). This appears to be one of the main differences between the 1d NLS and radial NLS modes.  

\section{Trailing edge dynamics}

%\jc{We begin this section with some motivating numerics. In Fig.~\ref{trail_edge_evolve} the position of the trailing edge $r_{\min}(t)$ and its value $\rho_{min}(t) = \rho(r_{min}(t))  \le \rho(r,t)$ are tracked. The velocity of the rNLS trailing edge in Fig.~\ref{trail_edge_evolve}(a) is observed to be constant and, to leading order, travel with the same speed as the 1d solution (see eq.~(\ref{eq:vtr1})). Next we examine the trailing edge value $\rho_{min}(t)$ in Fig.~\ref{trail_edge_evolve}(b). There we observe the 1d solutions approaches some constant at long times. This is in contrast to the rNLS trailing edges which do not appear to have any minimum (other than the obvious need to be non-negative). This appears to be one of the main differences between the 1d NLS and radial NLS modes.}

\begin{figure} [ht]
\centering
\includegraphics[scale=.6]{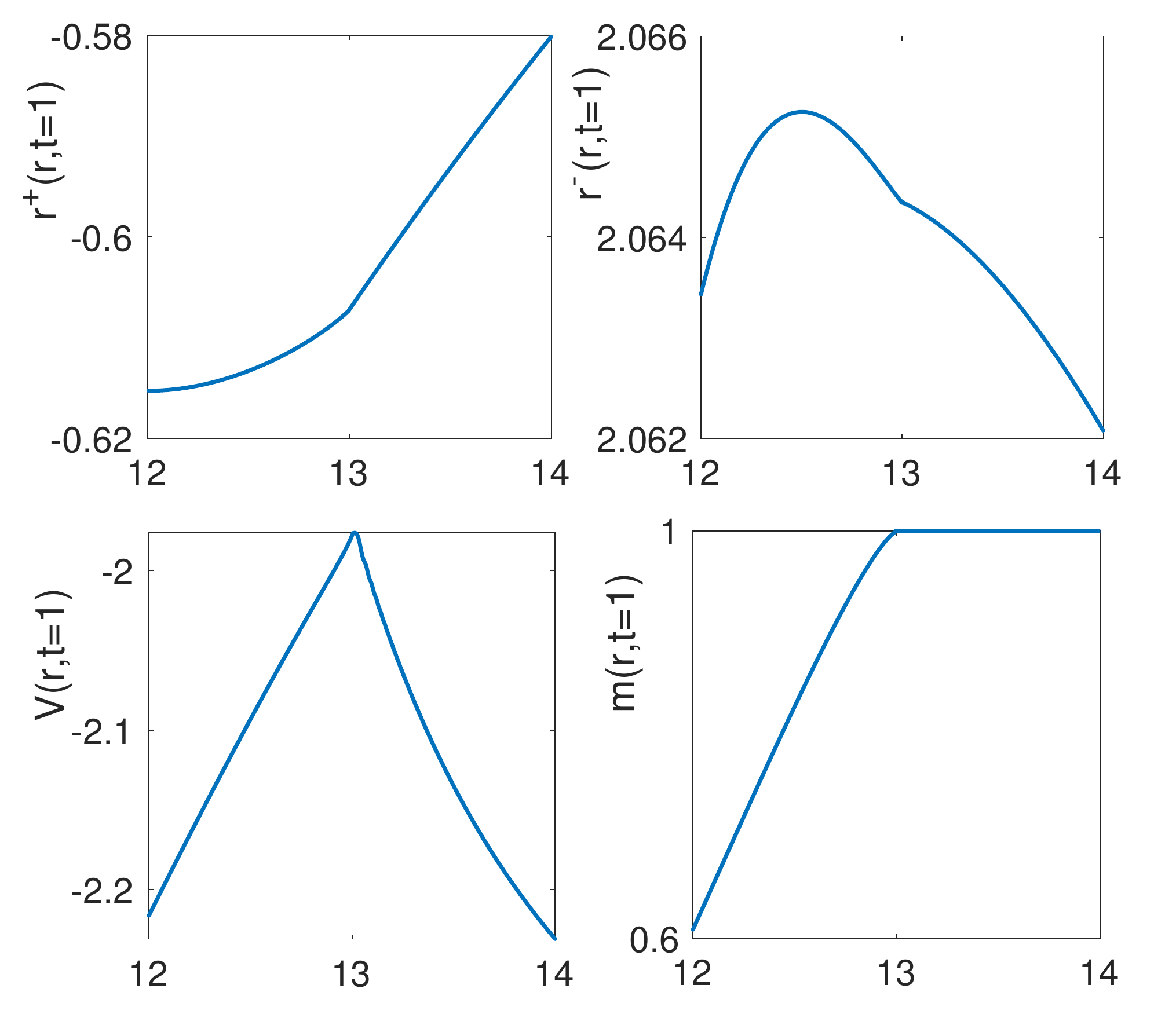}
\caption{The slow `physical' Whitham rNLS variables around the trailing edge computed at time $t=1$ when $\rho_b = 2$ and $r_b\approx15$. Top left: $r^+=(r_1+r_4)/4$; top right: $r^-=(r_1-r_4)^2/32$; bottom left: phase velocity $V=(r_1+r_2+r_3+r_4)/4$; bottom right: elliptic modulus $m=\frac{(r_1-r_2)(r_3-r_4)}{(r_1-r_3)(r_2-r_4)}$.}
\label{tr_edge_nbd}
\end{figure}

In this section we present some illuminating numerics for the spatial domain around the DSW trailing edge in Fig.~\ref{tr_edge_nbd} and give some empirical analytic considerations. 
\par We expect a rich variety of possible longer time developments for the rNLS DSW and its trailing edge structure for different initial conditions. However, we defer a thorough investigation of these dynamics to the future. It should be mentioned that there is some research in literature on the problems of the kind we face here. The complete Whitham system eq.~(\ref{eq:Wr}) has terms that depend explicitly on the radial coordinate $r$ and is nondiagonal in the $r_j$-variables. This makes it hard to analyze exactly. Similar issues were met by El, Grimshaw and Kamchatnov in~\cite{ElGrKam07} who studied undular bores in shallow water flows with variable topography and also derived a Whitham modulation equations with additional  complicated nondiagonal/nonderivative terms in the Riemann variables. They were able to understand the nonstationary `solitonic' edge dynamics by using perturbative methods near the bore front and insight from the seminal paper~\cite{GurPit73}. Other perturbative methods were also used e.g.~in~\cite{AbNiHorFr11} in studying 1d NLS dark solitons in weakly dissipative media. A significant literature exists which develops a concept of ring dark solitons~\cite{KivY94} in higher-dimensional NLS-type systems, see e.g.~\cite{FrMa99, TheoEtAl03, Fr2010} and references therein, or other ring waves~\cite{KhuZh2016}. Those structures are generally unstable but often exist for times exceeding the times of experiments and, as numerics of the cited works show, may have interesting nontrivial behavior. An approximate ring dark soliton might be a useful concept near the rNLS DSW trailing edge. We also delegate all these interesting considerations to a future research.

%%%%%%%%%%%%%%%%%%%%%%%%%%%%%%%%%%%%%%%%%%%%%%%%%
%%%%%%%%%%%%%%%%%%%%%%%%%%%%%%%%%%%%%%%%%%%%%%%%%

\par One can see in Fig.~\ref{tr_edge_nbd} that the trailing edge location is clearly indicated as the phase transition point i.e.~the location of rapid change of the Whitham slow variables. In Fig.~\ref{tr_edge_nbd}, the newly defined Whitham variables $r^+$ and $r^-$, where $r^+=(r_1+r_4)/4, r^-=(r_1-r_4)^2/32$, coincide with the hydrodynamic velocity $u$ and density $\rho$ behind the trailing edge, respectively. It is seen as a maximum point for the phase velocity $V$ and (approximately) at a point where the density from behind the DSW region reaches a zero slope in $r$.
\par Recalling the leading order solution for the density in the DSW region eq.~(\ref{eq:rho}), we note that mathematically this is the cnoidal function with the same dependence on the varying elliptic modulus $m$ as for the well-known KdV DSW~\cite{Whitham74, GurPit73} or for 1d NLS density. This implies that as $m$ reaches $1$ in the DSW region, the gradient $\prt_r\rho$ becomes close to zero with exponential accuracy. This is one of the motivations of the following approximate empirical analysis. As the numerics demonstrate, the trailing edge velocity 

\be
v_t = \frac{r_1+2r_2+r_4}{4} = u + \frac{r_2}{2} \approx -\sqrt{2\rho_b},   \label{eq:vtr1}
\ee
i.e.~is constant with good accuracy. By eq.~(\ref{eq:RP}), the system of Whitham equations eqs.~(\ref{eq:r1d})--(\ref{eq:r2d}) at and behind the trailing edge can be written in terms of variables $u$, $\rho$ and $v=u+r_2/2$ as

\be
\prt_t\rho + \left(\prt_r + \frac{1}{r}\right)(2\rho u) = 0,  \label{eq:rhot}   %\eqno(Im)
\ee

\be
\prt_tu + \prt_r(u^2 + \rho) = 0,   \label{eq:ut}
\ee

\be
\prt_tv + v\prt_rv - v\prt_ru + \prt_r(u^2 + \rho) + \frac{v^2 - vu - 2(u^2 + \rho)}{3r} = 0.   \label{eq:vt}
\ee

Assuming that the trailing edge speed is close to being constant, we can discuss analytically the dynamics there. The other crucial assumption supported by numerics is that the trailing edge location is given by the maximum of density $\rho$ between its largest dip and the slowly varying in $r$ shelf. This implies taking $\prt_r\rho(r_t, t) = 0$ where

\be
r_t(t) = r_b + v_tt   \label{eq:rtr}
\ee
is the trailing edge position as a function of time. Then eqs.~(\ref{eq:rhot})--(\ref{eq:vt}) yield a closed system of ODEs for the density $\rho_t=\rho(r_t,t)$, hydrodynamic velocity $u_t=u(r_t,t)$ and its gradient $(\prt_ru)_t=\prt_ru(r_t,t)$ at the trailing edge,

\be
\frac{d\rho_t}{dt} + 2\rho_t\left((\prt_ru)_t + \frac{u_t}{r_t}\right) = 0,  \label{eq:rhotr}   
\ee

\be
\frac{du_t}{dt} + (2u_t-v_t)(\prt_ru)_t = 0,   \label{eq:utr}
\ee

\be
(2u_t - v_t)(\prt_ru)_t + \frac{v_t^2 - v_tu_t - 2(u_t^2 + \rho_t)}{3r_t} = 0.   \label{eq:vtr}
\ee
We introduce the shifted time $\tau$ and rescaled variables as

\be
\tau = t + \frac{r_b}{v_t} < 0,   \qquad  \nu = \frac{u_t}{v_t} > 0,   \qquad  \Delta = \frac{\rho_t}{v_t^2} > 0   \label{eq:defs}
\ee
and denote by prime the derivatives w.r.t.~$\tau$. Next we substitute $(\prt_ru)_t$ from eq.~(\ref{eq:utr}),

\be
(\prt_ru)_t = -\frac{\nu'}{2\nu-1},   \label{eq:gru}
\ee
into the other equations. Then $\Delta$ can be expressed from eq.~(\ref{eq:vtr}) in terms of $\nu$,

\be
\Delta = \frac{1-\nu-2\nu^2 - 3\tau\nu'}{2} = -\frac{(\nu+1)(2\nu-1) + 3\tau\nu'}{2}.   \label{eq:del}
\ee
Finally, plugging eq.~(\ref{eq:del}) and its derivative $2\Delta' = -3\tau\nu'' - 4(\nu+1)\nu'$ into eq.~(\ref{eq:rhotr}), we obtain the second order ODE for $\nu$,

\be
3\tau^2\nu'' - \frac{6\tau^2(\nu')^2}{2\nu-1} + 2\tau(4\nu+1)\nu' + 2\nu(\nu+1)(2\nu-1)^2 = 0.   \label{eq:nuDE}
\ee
The ICs for eq.~(\ref{eq:nuDE}) are determined from the ICs $u=u_b$ and $\rho=\rho_b$ at the trailing edge,

\be
\nu_0 \equiv \nu(t=0) = \frac{u_b}{v_t} = 1 - \frac{1}{\sqrt\rho_b},   \label{eq:nu0}
\ee

\be
\tau_0\nu'_0 \equiv \tau(t=0)\nu'(t=0) = \frac{1 - \nu_0 - 2\nu_0^2 - 2\Delta(t=0)}{3} = -\left(1 - \frac{1}{\sqrt\rho_b}\right)\left(1 - \frac{2}{3\sqrt\rho_b}\right),    \label{eq:nu'0}
\ee
and $\tau_0 = -r_b/\sqrt{2\rho_b}$. Thus, for $1<\rho_b<4$, we have $0<\nu_0<1/2$ and $\nu'_0>0$. The dynamics of $u$ and $\rho$ at the trailing edge is fixed by the solution of eq.~(\ref{eq:nuDE}) with these ICs and eq.~(\ref{eq:del}). We see that initially, for $1<\rho_b<4$, the variable $\nu$ is between two of its equilibrium points, $\nu_l=0$ and $\nu_r=1/2$. It starts moving toward $\nu_r$. Reaching $\nu_r$ or its close vicinity in time not far exceeding $T_{max}$ in eq.~(\ref{eq:tmax}) means with certainty that the vacuum point $\rho = 0$ was reached since even the density at the trailing edge, which is much larger than the minimum density in the DSW, comes close to zero there as eq.~(\ref{eq:del}) implies. 

%%%%%%%%%%%%%%%%%%%%%%%%%%%%%%%%%%%%%%%%%%%%%%%%%
\begin{figure}[t!]    %[ht]
\centering
\includegraphics[width=3cm, height=3.3cm]{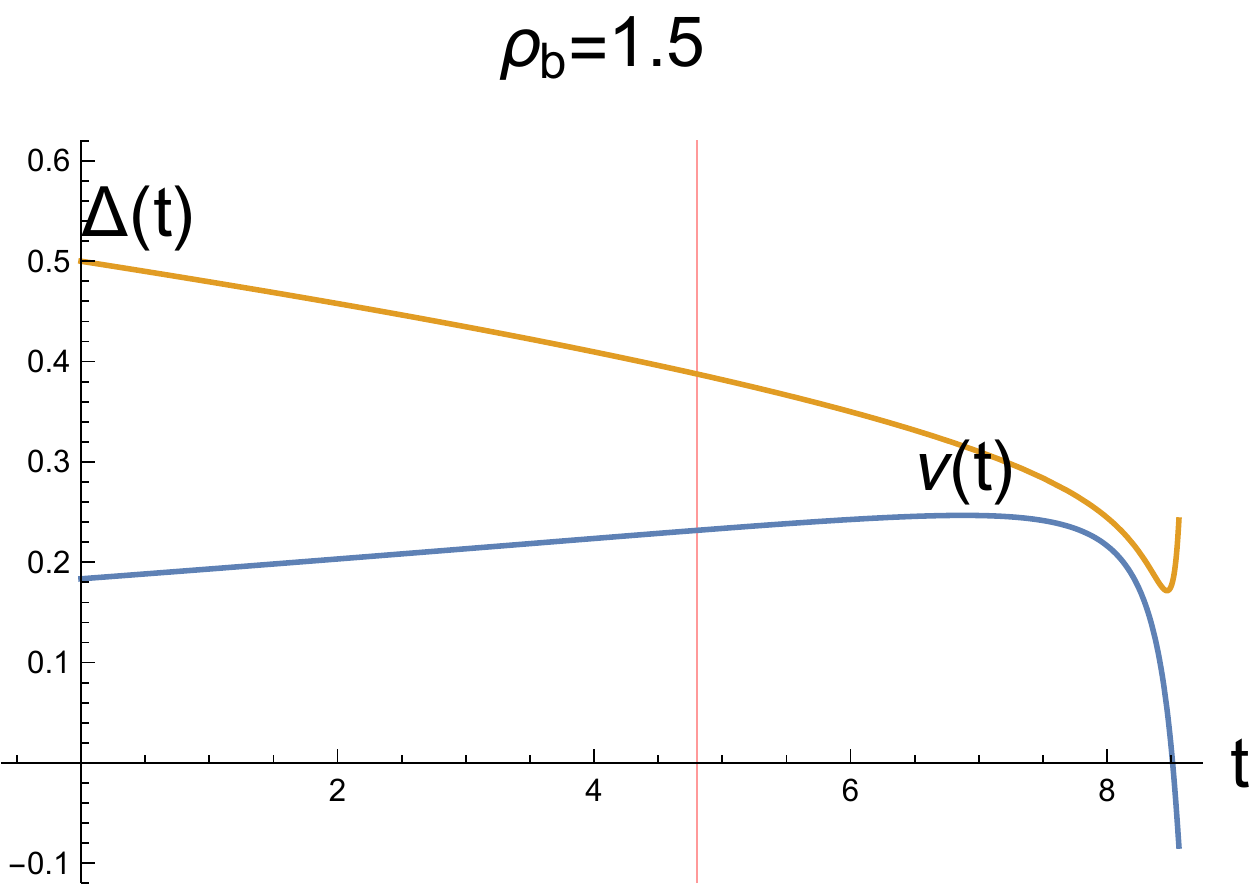} %\hglue-2em
\includegraphics[width=3cm, height=3.3cm]{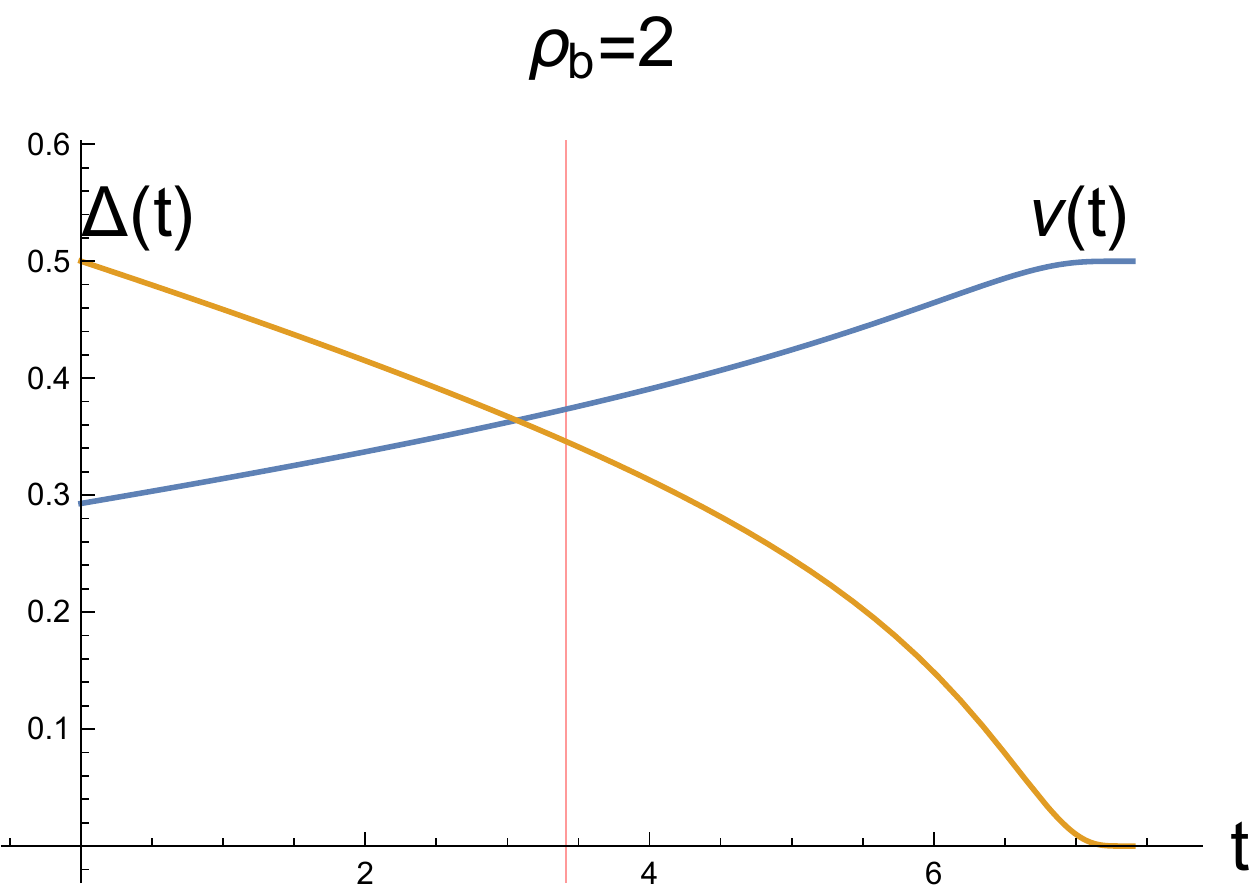} %\hglue-2em
\includegraphics[width=3cm, height=3.3cm]{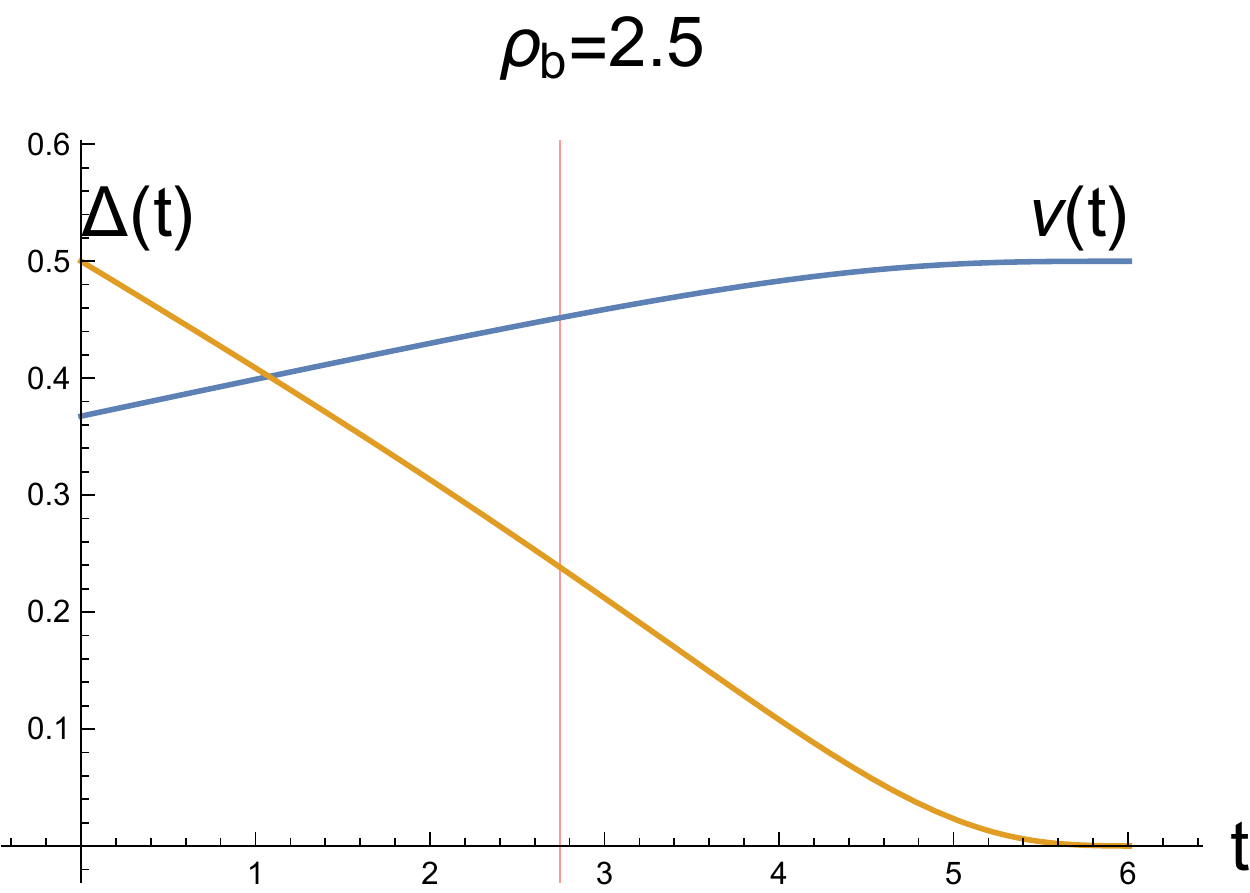} %\hglue-2em
\includegraphics[width=3cm, height=3.3cm]{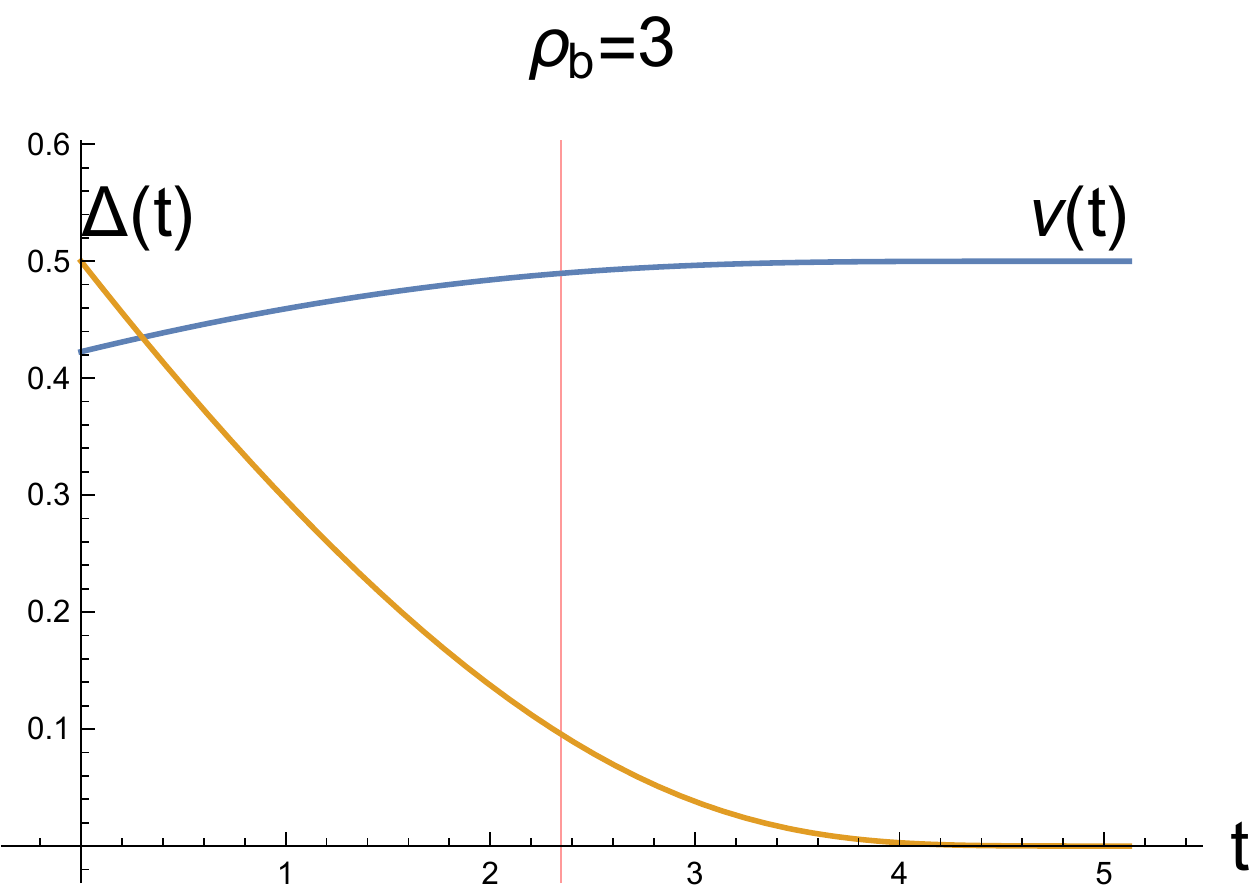} %\hglue-2em
\includegraphics[width=3cm, height=3.3cm]{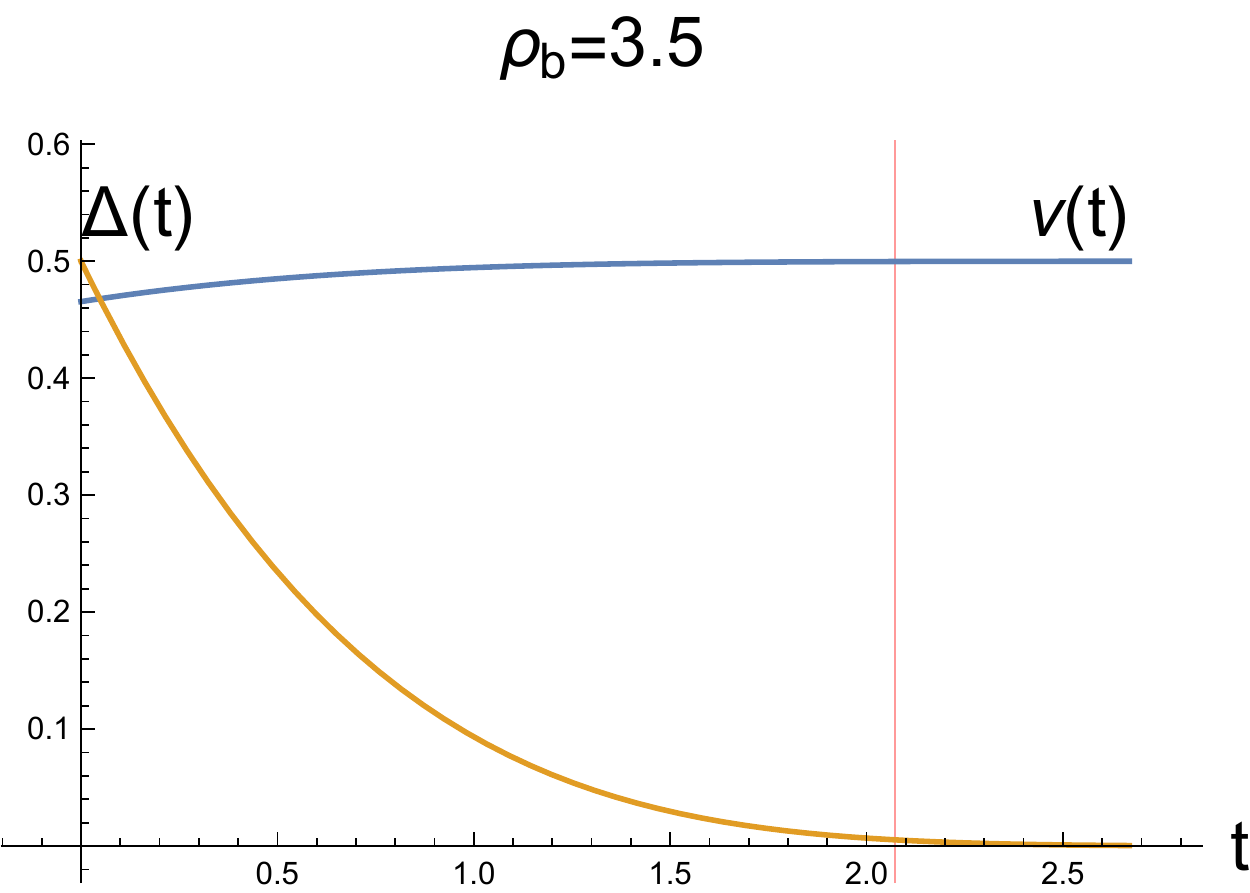}
\caption{The dynamics of $\nu(t)$ from eq.~(\ref{eq:nuDE}) (blue) and $\Delta(t)$ from eq.~(\ref{eq:del}) for $r_b=15$. From left to right: $\rho_b=1.5$; $\rho_b=2$; $\rho_b=2.5$; $\rho_b=3$; $\rho_b=3.5$. Red vertical lines are drawn at $t=T_{max}(\rho_b)$ for each $\rho_b$.}
\label{nu_del}
\end{figure}
%%%%%%%%%%%%%%%%%%%%%%%%%%%%%%%%%%%%%%%%%%%%%%%%%

We plotted in Fig.~\ref{nu_del} the solutions $\nu(t)$ of eq.~(\ref{eq:nuDE}) with IC eqs.~(\ref{eq:nu0}) and (\ref{eq:nu'0}) found by Mathematica as well as $\Delta(t)$ found from eq.~(\ref{eq:del}). In each case of different initial jump $\rho_b$, the thin red vertical line in the plot shows the corresponding time $T_{max}=T_{max}(\rho_b)$ for the leading DSW edge to reach the origin, see eq.~(\ref{eq:tmax}). Strictly speaking, after this time the theory always has to be modified. Actually one can assume that there is a time period after $T_{max}$ before perturbations from the origin reach and start influencing the slower approaching trailing edge. We see that, if the solutions were valid longer, the eventual fate for initial jumps $\rho_b\sim2$ and higher would be reaching the vacuum point $\rho=0$. In fact, however, the quantity $\Delta=\rho_t/v_t^2$ in the plots of Fig.~\ref{nu_del} changes more rapidly than the change of $\rho=\rho_t$ at the trailing edge observed in numerics of the full rNLS equation. Therefore the model of the trailing edge here can be considered as a rough approximation at best. Still the conclusions it implies are likely to be {\it qualitatively} true. The main reason to believe it is again the fact that the BEC density at the lowest dip to the left of the trailing edge $\rho_{min}(t)$ becomes at least several times smaller than the density $\rho_t(t)$ at the trailing edge. And indeed $\rho_{min}$ is already seen to be close to zero at time $t=2$ for $\rho_b=3$, see Fig.~\ref{trail_edge_evolve}. Thus, all these facts together imply a very plausible conjecture that the rNLS DSW reaches the vacuum point after some finite propagation time even for initial jumps $\rho_b\sim 2.5-3$ i.e.~substantially smaller than the cavitation threshold $\rho_b=4$ for 1d NLS~\cite{GK87, HoeferEtAl06}. This conjecture is to be verified in a future research. 

\section{Conclusion}
In this paper the Whitham theory for the radial defocusing NLS equation (rNLS) is developed. It is seen that the Whitham equations (\ref{eq:Wr}) in hydrodynamic variables differ from the 1dNLS theory by introduction of nonderivative nondiagonal terms (i.e.~the `$g_j/r$' terms). The results of Whitham theory are compared with direct numerics with very good agreement. The results show that the rNLS DSW propagation is different from the 1d NLS DSW propagation. 
\par The rNLS Whitham theory also indicates rich and novel phenomena. The detailed analytical and numerical study of these phenomena is a matter of future work. There are numerous interesting aspects to explore such as e.g.~reaching at later times a vacuum point of the BEC density for initial jumps $\rho_b$ smaller than the critical value $\rho_b=4$ at which the cavitation begins for 1d NLS DSWs. Such a possibility is expected due to the observed growth in time of the largest dip of the BEC density in the rNLS DSW. As mentioned in the previous section, perturbative methods e.g.~used~in~\cite{ElGrKam07} or in~\cite{AbNiHorFr11} might be helpful in such investigations.
\par Another direction for future research naturally emanating from the present work is the development and application of Whitham modulation theory for the full (2+1)-dimensional NLS equation and related mathematically similar models. The general nonlinear WKB approach and singular perturbation theory which we use will be an indispensable tool in this research. It has already proven to be extremely useful for (2+1)-dimensional PDEs of KP-type~\cite{ABW-KP, ABW-BO, ABR}, which are important mathematical models in many applications.

\bigskip
{\bf\large Acknowledgments} \\
This research was partially supported by the NSF under grant DMS-1712793.

\def\thesection{Appendix~\Alph{section}:}

%%%%%%%%%%%%%%%%%%%%%%%%%%%%%%%%%%%%%%%%%%%%%%%%%%%%%%%%%%%%%%%%%%%%%%%%%%%%%%%%%%%%%%%%%%%%%%%%%%%%%%%%%%%%%%%%%%%%%%%%%%

\appendix
%\section{Auxiliary formulas}
%\label{a:auxiliary}

\section{Formulas for the elliptic solution}

Consider the elliptic leading order equation (\ref{eq:2.10}) for the density $\rho_0=|\Psi_0|^2$,

\be
k^2(\rho_0')^2 = 2(\rho_0^3 - e_1\rho_0^2 + e_2\rho_0 - e_3).   \label{eq:A1}   %\eqno(A1)
\ee

\ni Its derivative in $\theta$ is

\be
k^2\rho_0'' = 3\rho_0^2 - 2e_1\rho_0 + e_2.  \label{eq:A2}   %\eqno(A2)
\ee

\ni Integrating eq.~(\ref{eq:A2}) over the period in $\theta$, one finds

\be
Q_2 \equiv \int_0^1\rho_0^2d\theta = \frac{2e_1Q-e_2}{3} = \frac{2}{3}(2\Omega Q - C_1).   \label{eq:A3}   %\eqno(A3)
\ee

\ni Dividing eq.~(\ref{eq:A1}) by $\rho_0$ and then integrating over the period in $\theta$, one obtains

$$
\int_0^1\frac{k^2(\rho_0')^2+2e_3}{\rho_0}d\theta = k^2\int_0^1\frac{4C_0^2 + (\rho_0')^2}{\rho_0}d\theta = 2(Q_2-e_1Q+e_2),
$$

\ni and, using eq.~(\ref{eq:A3}),

\be
\int_0^1\frac{k^2(\rho_0')^2+2e_3}{\rho_0}d\theta = k^2\int_0^1\frac{4C_0^2 + (\rho_0')^2}{\rho_0}d\theta = \frac{2(2e_2-e_1Q)}{3} = \frac{4(2C_1 - \Omega Q)}{3}.   \label{eq:A4}   %\eqno(A4)
\ee

\ni Another formula used in the main text may be obtained e.g.~if one takes the combination of eqs.~$($\ref{eq:A1}$) - 2\rho_0\cdot($\ref{eq:A2}$)$, divides it by $\rho_0^2$ and then integrates over the period. The result is

\be
\int_0^1\frac{k^2(\rho_0')^2-2e_3}{\rho_0^2}d\theta = k^2\int_0^1\frac{(\rho_0')^2-4C_0^2}{\rho_0^2}d\theta = 4Q - 2e_1 = 4Q - 4\Omega.   \label{eq:A5}   %\eqno(A5)
\ee

\ni We also record here another formula of the same kind (even though it is not used in the main text, it can be useful in another related context). It is the outcome of integrating the combination of eqs.~$3\cdot($\ref{eq:A1}$) - 2\rho_0\cdot($\ref{eq:A2}$)$ over the period and using again eq.~(\ref{eq:A3}):

\be
5k^2\int_0^1(\rho_0')^2d\theta = -2e_1Q_2 + 4e_2Q - 6e_3 = 4\left(e_2 - \frac{e_1^2}{3}\right)Q + \frac{2e_1e_2}{3} - 6e_3.   \label{eq:A6}   %\eqno(A6)
\ee

The variable $\alpha$ (the leading order velocity averaged over the period in $\theta$) of the main text can be expressed in terms of the third complete elliptic integral:
%The variable $\alpha$ (the leading order velocity averaged over the period in $\theta$) of the main text can be expressed in terms of incomplete elliptic integrals:

$$
\alpha = \int_0^1u_0d\theta = \frac{V}{2} - \sigma\frac{\sqrt{2\lambda_1\lambda_2\lambda_3}}{2}\int_0^1\frac{d\theta}{\rho_0} = 
$$

\be
= \frac{V}{2} - \sigma\frac{\sqrt{\lambda_2\lambda_3}}{\sqrt{2\lambda_1}}\frac{\Pi(\mu, m)}{K(m)},    \label{eq:A7}   %\eqno(A7)
\ee

%\be
%= \frac{V}{2} - \frac{\sigma\sqrt2}{2}\left[\sqrt{\frac{\lambda_1\lambda_2}{\lambda_3}} + \sqrt{\lambda_3-\lambda_1}\left(E(\chi;1-m) + \left(\frac{E(m)}{K(m)}-1\right)F(\chi;1-m)\right)\right],    \label{eq:A7}   %\eqno(A7)
%\ee

\ni where $\Pi(\mu, m)$ is the third complete elliptic integral and

$$
\mu = -\frac{\lambda_2-\lambda_1}{\lambda_1},   \qquad   m = \frac{\lambda_2-\lambda_1}{\lambda_3-\lambda_1}.
$$

%$$
%\chi = \sin^{-1}\left(\sqrt{\frac{\lambda_3-\lambda_1}{\lambda_3}}\right),
%$$

%\ni and $F(\chi;1-m)$ and $E(\chi;1-m)$ are incomplete elliptic integrals of the first and second kind, respectively.
\par The complete elliptic integrals have the following limiting expressions:

\be
m \to 0:  \qquad  K(m) = \frac{\pi}{2}\left(1 + \frac{m}{4} + \frac{9m^2}{64} + \dots\right),  \qquad E(m) = \frac{\pi}{2}\left(1 - \frac{m}{4} - \frac{3m^2}{64} + \dots\right);  \label{eq:A8}   %\eqno(A8)
\ee

\be
m \to 1:  \qquad K(m) \approx \frac{1}{2}\ln\frac{16}{1-m},  \qquad  E(m) \approx 1 + \frac{(1-m)}{4}\left(\ln\frac{16}{1-m} - 1\right).  \label{eq:A9}   %\eqno(A9)
\ee

\ni They satisfy the following differential equations in $m$:

\be
K'(m) = \frac{E-(1-m)K}{2m(1-m)},   \qquad  E'(m) = -\frac{K-E}{2m},   \label{eq:A10}   %\eqno(A10)  %(4.2)  
\ee

\ni from which it also follows e.g.~that

\be
\left(\frac{E}{K}\right)'(m) = \frac{1}{m}\left(\frac{E}{K} - \frac{1}{2} - \frac{E^2}{2(1-m)K^2}\right).   \label{eq:A11}   %\eqno(A11)  %(4.3)  
\ee

\ni Using the relation (\ref{eq:2.14}) and first formula of eq.~(\ref{eq:A10}), one finds the log-derivative of $k$ in terms of the roots,

\be
\frac{dk}{k} = \frac{E}{2(1-m)K}\frac{d(\lambda_3-\lambda_1)}{\lambda_3-\lambda_1} - \left(\frac{E}{2(1-m)K} - 1\right)\frac{d(\lambda_2-\lambda_1)}{\lambda_2-\lambda_1}.   \label{eq:A12}   %\eqno(A12)   %(4.4)  
\ee

\section{Transformation to Riemann variables}

\ni Expressing eq.~(\ref{eq:A12}) in terms of variables $R_i$, $i=1,2,3$, we find

$$
\frac{dk}{k} = \sum_{i=1}^3L_idR_i,   \qquad  L_1 = -\left(1-\frac{E}{K}\right)\frac{R_1}{R_2^2-R_1^2},  
$$

\be
 L_2 = \left(1 - \frac{E}{K}\frac{(R_3^2-R_1^2)}{(R_3^2-R_2^2)}\right)\frac{R_2}{(R_2^2-R_1^2)},  \qquad L_3 = \frac{E}{K}\frac{R_3}{R_3^2-R_2^2},   \label{eq:B1}   %\eqno(B1)  %(4.10)
\ee

\ni Then one can readily verify that the log-derivatives of $k$, $L_i$, $i=1,2,3$, satisfy the relations

\be
\sum_{i=1}^3\frac{L_i}{R_i} = 0,  \qquad \sum_{i=1}^3L_i = S = \frac{1}{R_1+R_2} - \frac{E}{K}\frac{(R_3-R_1)}{(R_1+R_2)(R_2+R_3)},  \qquad \sum_{i=1}^3R_iL_i = 1.  \label{eq:B2}   %\eqno(B2)  %(4.11)
\ee

\ni Using them, the variable $Q$ can be expressed as

\be
2Q = R_3^2-(R_3^2-R_1^2)\frac{E}{K} = (R_1+R_2)(R_2+R_3)(R_3+R_1)S - R_1R_2 - R_2R_3 - R_3R_1 \equiv P_3S-P_2,   \label{eq:B3}   %\eqno(B3)  %(4.12)
\ee

\ni and, using also eq.~(\ref{eq:A11}),

\be
2\left(dQ-\frac{Dk}{k}\right) = \sum_{i=1}^3(R_i - R_i^2L_i)dR_i = \sum_{i=1}^3(R_lL_m+R_mL_l)R_idR_i,  \quad i\neq l \neq m \neq i.   \label{eq:B4}   %\eqno(B4)  %(4.13)
\ee

\par We find some more simple relations among log-derivatives of $k$ w.r.t.~the $r_j$ and the $R_i$ variables,

\be
4\frac{dk}{k} = \sum_{j=1}^44\frac{\prt_jk}{k}dr_j \equiv \sum_{j=1}^4K_jdr_j,   \qquad  K_4 = S, \qquad K_i = 2L_i - S, \quad i=1,2,3.   \label{eq:B5}   %\eqno(B5)  %(4.15)
\ee

\ni We use also that

\be
\sum_{i=1}^3R_idR_i = \frac{1}{4}\sum_{j=1}^4r_jdr_j - VdV,   \label{eq:B6}   %\eqno(B6)  %(4.16)
\ee

\be
\sum_{i=1}^3R_lR_mdR_i = \frac{1}{8}\sum_{j=1}^4r_j^2dr_j - \frac{V}{4}\sum_{j=1}^4r_jdr_j + (V^2-\frac{\sum_{j=1}^4r_j^2}{8})dV,  \quad i\neq l \neq m \neq i.  \label{eq:B7}   %\eqno(B7)  %(4.17)
\ee

$$
\sum_{i=1}^3(R_l^2+R_m^2)R_idR_i = \frac{1}{8}\sum_{j=1}^4r_j^3dr_j - \frac{3V}{8}\sum_{j=1}^4r_j^2dr_j + 
$$

\be
+ (\frac{V^2}{2}-\frac{\sum_{j=1}^4r_j^2}{32})\sum_{j=1}^4r_jdr_j + (-2V^3+\frac{V}{2}\sum_{j=1}^4r_j^2 - \frac{1}{8}\sum_{j=1}^4r_j^3)dV,  \quad i\neq l \neq m \neq i.   \label{eq:B8}   %\eqno(B8)  %(4.18)
\ee

\be
\sum_{j=1}^4r_j^2 = 4(V^2 + \sum_{i=1}^3R_i^2)   \qquad  \sum_{j=1}^4r_j^3 = 4(V^3 + 3V\sum_{i=1}^3R_i^2 + 6R_1R_2R_3).  \label{eq:B9}   %\eqno(B9)  %(4.19)
\ee

\ni Upon substituting eq.~(\ref{eq:4.8}) into the equations (\ref{eq:E1}), (\ref{eq:Q1}) and (\ref{eq:P1}), the last take form

$$
\frac{2\left(Q\sum_{i=1}^3R_iDR_i - (DQ-Q\frac{Dk}{k})\sum_{i=1}^3R_i^2 + \frac{1}{2}\sum_{i\neq l \neq m \neq i}^3(R_l^2+R_m^2)R_iDR_i - \sum_{i\neq l}^3R_i^2R_l^2\frac{Dk}{k}\right)}{3R_1R_2R_3} +
$$

\be
 + DV + \sum_{i=1}^3R_i\prt_rR_i = 0,  \label{eq:BE1}   %\eqno(E1)
\ee

\be
D(2Q) - 2Q\frac{Dk}{k} + \prt_r(R_1R_2R_3) + \frac{V\cdot2Q + R_1R_2R_3}{r} = 0,    \label{eq:BQ1}   %\eqno(Q1)
\ee

$$
2QDV + D(R_1R_2R_3) - 2R_1R_2R_3\frac{Dk}{k} + \sum_{i\neq l \neq m \neq i}^3(R_l^2+R_m^2)R_i\prt_rR_i +
$$

\be
+ \frac{3VR_1R_2R_3 + 2\sum_{i\neq l}^3R_i^2R_l^2 - \sum_{i=1}^3R_i^2\cdot2Q}{3r} = 0.     \label{eq:BP1}   %\eqno(P1)
\ee

\ni Using formulas (\ref{eq:B1})--(\ref{eq:B4}) in eq.~(\ref{eq:BE1}), the last can be greatly simplified to give

\be
DV + \sum_{i=1}^3(R_lL_m+R_mL_l)DR_i + \sum_{i=1}^3R_i\prt_rR_i = 0,  \quad  i\neq l \neq m \neq i.   \label{eq:E*}   %\eqno(E*)
\ee

\ni Similarly, using formulas (\ref{eq:B1})--(\ref{eq:B4}) in eq.~(\ref{eq:Q1}), one can transform it to 

\be
\sum_{i=1}^3(R_lL_m+R_mL_l)R_iDR_i + \sum_{i=1}^3R_lR_m\prt_rR_i + \frac{V(P_3S-P_2) + R_1R_2R_3}{r} = 0,  \quad  i\neq l \neq m \neq i.  \label{eq:Q*}   %\eqno(Q*)
\ee

\ni At last, with the use of eqs.~(\ref{eq:B1})--(\ref{eq:B4}), we bring eq.~(\ref{eq:P1}) to the form

$$
(P_3S - P_2)DV + \sum_{i=1}^3R_lR_m(R_lL_l + R_mL_m - R_iL_i)DR_i + \sum_{i=1}^3(R_l^2+R_m^2)R_i\prt_rR_i +
$$

\be
+ \frac{3VR_1R_2R_3 + 2\sum_{i\neq l}^3R_i^2R_l^2 - \sum_{i=1}^3R_i^2(P_3S-P_2)}{3r} = 0,  \quad  i\neq l \neq m \neq i.   \label{eq:P*}   %\eqno(P*)
\ee

\par The combination $4\cdot(\ref{eq:E*}) + 4V\cdot(\ref{eq:k1})$ and using eqs.~(\ref{eq:Rr}), (\ref{eq:B5}) and (\ref{eq:B6}) yields

\be
\sum_{j=1}^4r_j(K_jDr_j + \prt_rr_j) = 0.   \label{eq:*2}   %\eqno(*2)
\ee

\ni The combination $8\cdot(\ref{eq:Q*}) + 2V\cdot(\ref{eq:*2}) + (\sum_{j=1}^4r_j^2 - 8V^2)\cdot(\ref{eq:k1})$ yields, upon using eqs.~(\ref{eq:Rr}), (\ref{eq:B5}), (\ref{eq:B7}) and the first equation of (\ref{eq:B9}),

\be
\sum_{j=1}^4r_j^2(K_jDr_j+\prt_rr_j) + \frac{8(V(P_3S-P_2) + R_1R_2R_3)}{r} = 0.   \label{eq:*3}   %\eqno(*3)
\ee

\ni Finally, the combination $8\cdot(\ref{eq:P*})+3V\cdot(\ref{eq:*3})+(\sum_{i=1}^3R_i^2-3V^2)\cdot(\ref{eq:*2}) + (4V^3-4V\sum_{i=1}^3R_i^2+24R_1R_2R_3)\cdot(\ref{eq:k1})$, after using eqs.~(\ref{eq:Rr}), (\ref{eq:B5}), (\ref{eq:B8}) and (\ref{eq:B9}), yields

$$
\sum_{j=1}^4r_j^3(K_jDr_j+\prt_rr_j) +
$$

\be
+ \frac{8[(9V^2-\sum_{i=1}^3R_i^2)V(P_3S-P_2) + 12VR_1R_2R_3 + 2\sum_{i<l}^3R_i^2R_l^2]}{3r} = 0.   \label{eq:*4}   %\eqno(*4)
\ee

\par The system of four Whitham equations (\ref{eq:k1}), (\ref{eq:E1}), (\ref{eq:Q1}) and (\ref{eq:P1}) is transformed to

\be
\sum_{j=1}^4\left(4\frac{\prt_jk}{k}Dr_j + \prt_rr_j\right) = 0,   \label{eq:kr}   %\eqno(k)
\ee

\be
\sum_{j=1}^4r_j\left(4\frac{\prt_jk}{k}Dr_j + \prt_rr_j\right) = 0,   \label{eq:*2B}   %\eqno(*2)
\ee

\be
\sum_{j=1}^4r_j^2\left(4\frac{\prt_jk}{k}Dr_j + \prt_rr_j\right) + \frac{8J_1}{r} = 0,   \label{eq:*3B}   %\eqno(*3)
\ee

\be
\sum_{j=1}^4r_j^3\left(4\frac{\prt_jk}{k}Dr_j + \prt_rr_j\right) + \frac{8(J_2+9VJ_1)}{3r} = 0,   \label{eq:*4B}   %\eqno(*4)
\ee

\ni where again $D = \prt_t + V\prt_r$, the log-derivatives $\frac{\prt_jk}{k} \equiv \frac{\prt\ln k}{\prt r_j}$ are given by eq.~(\ref{eq:4.6}), 

$$
J_1 = 2VQ + R_1R_2R_3,   \qquad  J_2 = 3VR_1R_2R_3 + 2\sum_{i<l}^3R_i^2R_l^2 - 2Q\sum_{i=1}^3R_i^2,   %\eqno(J)
$$

$$
2Q = R_3^2 - (R_3^2-R_1^2)\frac{E(m)}{K(m)},   \qquad m = \frac{R_2^2-R_1^2}{R_3^2-R_1^2}=\frac{(r_2-r_1)(r_4-r_3)}{(r_3-r_1)(r_4-r_2)},   %\eqno(Qr)
$$

\ni and eqs.~(\ref{eq:Rr}) are to be used. Then we bring the four Whitham equations to the form

\be
4\frac{\prt_jk}{k}Dr_j + \prt_rr_j + \frac{h_j(r_1,r_2,r_3,r_4)}{r} = 0,   \quad j=1,2,3,4,  \label{eq:Wh}   %\eqno(Wh)
\ee

\ni each containing $t$- and $r$-derivatives of only one variable $r_j$. The terms $h_j/r$ would be absent in the 1d NLS case. So this last transformation would diagonalize the Whitham equations for 1d NLS bringing them to the well-known form, see e.g.~\cite{HoeferEtAl06, ElKr95}. Here before the transformation the equations can be written in matrix form as

$$
\Delta_{jl}\left(4\frac{\prt_lk}{k}Dr_l + \prt_rr_l\right) + \frac{W_j}{r} = 0,
$$

\ni where $\Delta$ is the Vandermonde matrix,

$$
\Delta = \left( \begin{array}{cccc} 1 & 1 & 1 & 1 \\ r_1 & r_2 & r_3 & r_4 \\  r_1^2 & r_2^2 & r_3^2 & r_4^2 \\  r_1^3 & r_2^3 & r_3^3 & r_4^3 \end{array} \right),
$$

\ni and the vector $W$ has components

\be
W_1 = 0,  \qquad  W_2 = 0,  \qquad W_3 = 8J_1,  \qquad W_4 = \frac{8(J_2 + 9VJ_1)}{3}.  \label{eq:Wj}   %\eqno(Wj)
\ee

\ni Inverting the Vandermonde matrix, we obtain the terms $h_j$,

$$
h_1 = \frac{\Delta_{234}}{|\Delta|}((r_2+r_3+r_4)W_3 - W_4),   \qquad   h_2 = \frac{\Delta_{134}}{|\Delta|}(-(r_1+r_3+r_4)W_3 + W_4),
$$

\be
h_3 = \frac{\Delta_{124}}{|\Delta|}((r_1+r_2+r_4)W_3 - W_4),   \qquad  h_4 = \frac{\Delta_{123}}{|\Delta|}(-(r_1+r_2+r_3)W_3 + W_4),   \label{eq:hj}   %\eqno(hj)
\ee

\ni where

$$
|\Delta| = \det\Delta = \prod_{j>l}^4(r_j-r_l),  \qquad  \Delta_{jlm} = \left| \begin{array}{ccc} 1 & 1 & 1 \\ r_j & r_l & r_m \\  r_j^2 & r_l^2 & r_m^2 \end{array} \right| = (r_j-r_l)(r_l-r_m)(r_m-r_j),
$$

\ni are the determinant and the corresponding minors of the Vandermonde matrix. Thus, we get the final Whitham system eq.~(\ref{eq:Wr}) in terms of $r$-variables.
 
\section{Whitham equations: direct approach, two phases}    \label{a:dpha}

We will employ the version of Whitham theory~\cite{Whitham74} involving singular perturbations and multiple scales, see e.g.~\cite{Ab2011} and references therein. This means we will consider an $\epsilon$-expansion for $\epsilon \ll 1$ of the solution to rNLS equation of the form

\be
\Psi = U(\theta, r, t; \epsilon)e^{i\phi(r,t; \epsilon)},   \label{eq:C2.1}   %\eqno(2.1)
\ee

\ni where we impose the following conditions for two fast phases, $\theta$ and $\phi$:

\be
\prt_r\theta = \frac{k}{\epsilon},  \qquad \prt_t\theta = -\frac{\omega}{\epsilon},  \qquad   \prt_r\phi = \frac{\alpha}{\epsilon},  \qquad \prt_t\phi = -\frac{\beta}{\epsilon},   \label{eq:2.2-3}   %\eqno(2.2-2.3)
\ee

\ni and $k$, $\omega$, $\alpha$ and $\beta$ are slowly varying quantities. Then the derivatives in rNLS can be written as

$$
\prt_t\Psi = e^{i\phi}\left(-\frac{i\beta}{\epsilon} - \frac{\omega}{\epsilon}\prt_{\theta} + \tilde\prt_t\right)U,   \qquad   \prt_r\Psi = e^{i\phi}\left(\frac{i\alpha}{\epsilon} + \frac{k}{\epsilon}\prt_{\theta} + \tilde\prt_r\right)U,
$$

\ni and

$$
\prt_{rr}\Psi = e^{i\phi}\left(\frac{i\alpha}{\epsilon} + \frac{k}{\epsilon}\prt_{\theta}\right)\left(\frac{i\alpha}{\epsilon} + \frac{k}{\epsilon}\prt_{\theta} + \tilde\prt_r\right)U +
$$

$$
+ e^{i\phi}\left[ \left(\frac{i\alpha}{\epsilon} + \frac{k}{\epsilon}\prt_{\theta} + \tilde\prt_r\right)\tilde\prt_rU  + \left(\frac{i\prt_r\alpha}{\epsilon} + \frac{\prt_rk}{\epsilon}\prt_{\theta}\right)U\right].
$$

\ni We denote 
$$\omega=kV         $$  %\eqno(2.2A)$$ 
where $V$ is called the phase velocity. Also we denote $\prt_{\theta}f = f'$ here and further on. Substituting the above expressions for the derivatives, we can rewrite rNLS equation as

$$
0 = k^2U'' + i(2\alpha-V)kU' + (\beta - \alpha^2 - |U|^2)U + 
$$

\be
+ \epsilon\left(i\tilde\prt_tU + 2k\tilde\prt_rU' + \prt_rkU' + 2i\alpha\tilde\prt_rU + i\prt_r\alpha U + \frac{kU' + i\alpha U}{r}\right) + \epsilon^2\left(\tilde\prt_r^2U + \frac{\tilde\prt_rU}{r}\right).  \label{eq:2.4}   %\eqno(2.4)
\ee

\ni Now we expand $U$ in powers of small $\epsilon$,

$$
U = U_0 + \epsilon U_1 + \dots,
$$

\ni keeping only terms up to first order in $\epsilon$. At leading (zeroth) order we obtain

\be
k^2U_0'' + i(2\alpha-V)kU_0' + \left(\beta - \alpha^2 - |U_0|^2\right)U_0 = 0,   \label{eq:2.5}   %\eqno(2.5)  
\ee

\ni which is solved in terms of elliptic functions, see below. We  express $U_0$ from eq.~(\ref{eq:2.5}) in terms of amplitude and phase,
$$
U_0=A_0e^{i\Phi_0};   %\eqno(2.7A)
$$
its imaginary part gives $(A_0^2(k\Phi_0' + \alpha - V/2))'=0$ which implies

\be
\Phi_0' = \frac{C_0}{A_0^2} + \frac{V-2\alpha}{2k},  \label{eq:2.7}   %\eqno(2.7)  
\ee

\ni with a constant of integration (slow variable) $C_0$. The real part of eq.~(\ref{eq:2.5}), after substituting $\Phi_0'$ from eq.~(\ref{eq:2.7}) becomes second order ODE for the amplitude $A_0$,

\be
k^2\left(A_0'' - \frac{C_0^2}{A_0^3}\right) + \Omega A_0 - A_0^3 = 0,   \label{eq:2.8}   %\eqno(2.8)
\ee

\ni where we denoted $\Omega = \beta - V\alpha + V^2/4$. Alternatively, this can be viewed as determining $\beta$ from:
\be
\beta= V\alpha - V^2/4 - \Omega   \label{eq:2.9A} %\eqno(2.9A)
\ee
once $\alpha, \Omega,V$ are known. Multiplying eq.~(\ref{eq:2.8}) by $2A_0'$ and integrating, one obtains

$$
k^2(A_0')^2 + \frac{k^2C_0^2}{A_0^2} + \Omega A_0^2 - \frac{A_0^4}{2} = C_1,  
$$

\ni with another integration constant $C_1$. In terms of $\rho_0=A_0^2$, it takes form of eq.~(\ref{eq:2.10})

\be
2k^2(\rho_0')^2 = 4(\rho_0^3-2\Omega\rho_0^2+2C_1\rho_0-2k^2C_0^2) = 4(\rho_0-\lambda_1)(\rho_0-\lambda_2)(\rho_0-\lambda_3).  \label{eq:C2.10}  %\eqno(2.10)
\ee

\ni Its general solution can be written as eq.~(\ref{eq:2.11}).
\par From the definitions (\ref{eq:2.2-3}), it follows immediately that the slow variables satisfy the conservation laws

\be
\prt_tk + \prt_r\omega = \prt_tk + \prt_r(kV) = 0   \label{eq:Ck}   %\eqno(2.15)  
\ee
and 
\be
\prt_t\alpha + \prt_r\beta = \prt_t\alpha + \prt_r\left(V\alpha - V^2/4 - \Omega\right) = 0,   \label{eq:a}  %\eqno(2.16)  
\ee
In fact one can determine $\alpha$ from the other variables. This relation is provided by the condition that the phase $\Phi_0$ is also periodic in $\theta$ (``non-secular"). Then integration of eq.~(\ref{eq:2.7}) over the period in $\theta$ gives
\be
\alpha = \frac{V}{2} + kC_0\int_0^1\frac{d\theta}{\rho_0}   \label{eq:al}   %\eqno(al)  
\ee
\ni in terms of $\rho_0$ determined by eq.~(\ref{eq:2.11}). Then the explicit expression for $\alpha$ is given by formula (\ref{eq:A7}) of Appendix A in terms of the four key variables. 
\par At first order in $\epsilon$ we get (from now on we remove ``tilde" from $t$- and $r$-derivatives since this will not cause confusion)

$$
k^2U_1'' + i(2\alpha-V)kU_1' + \left(\beta - \alpha^2 - 2|U_0|^2\right)U_1 - U_0^2U_1^* + 
$$

\be
+ i\prt_tU_0 + 2k\prt_rU_0' + \prt_rkU_0' + 2i\alpha\prt_rU_0 + i\prt_r\alpha U_0 + \frac{kU_0' + i\alpha U_0}{r} = 0.   \label{eq:2.6}   %\eqno(2.6)  
\ee

...Eqs.~(\ref{eq:Ck}), (\ref{eq:a}) are two of the Whitham equations for rNLS. The others can be derived from secularity conditions ensuring that the solution $U_1$ of eq.~(\ref{eq:2.6}) is periodic rather than growing in its fast variable $\theta$.  
\par Consider the zero-modes $w$ of the operator $\L$ acting on $U_1$ in eq.~(\ref{eq:2.6}). They satisfy

$$
\L w \equiv k^2w'' + ik(2\alpha-V)w' + (\beta-\alpha-2|U_0|^2)w - U_0^2w^* = 0.
$$

\ni We write it as two real equations in terms of real and imaginary parts $w=w_R+iw_I$, $U_0=U_R + iU_I$,

\be
(\L w)_R \equiv k^2w_R'' - k(2\alpha-V)w_I' + (\beta-\alpha-|U_0|^2)w_R - 2U_R^2w_R - 2U_RU_Iw_I = 0,   \label{eq:wR}   %\eqno(wR)
\ee

\be
(\L w)_I \equiv k^2w_I'' + k(2\alpha-V)w_R' + (\beta-\alpha-|U_0|^2)w_I - 2U_I^2w_I - 2U_RU_Iw_R = 0.   \label{eq:wI}   %\eqno(wI)
\ee

\ni We also write the solution of eq.~(\ref{eq:2.5}) $U_0 = U_R + iU_I$ in terms of its real and imaginary parts,

\be
k^2U_R'' - k(2\alpha-V)U_I' + (\beta-\alpha-U_R^2-U_I^2)U_R = 0,    \label{eq:0R}   %\eqno(0R)
\ee

\be
k^2U_I'' + k(2\alpha-V)U_R' + (\beta-\alpha-U_R^2-U_I^2)U_I = 0,   \label{eq:0I}   %\eqno(0I)
\ee

\ni and, using eqs.~(\ref{eq:0R}) and (\ref{eq:0I}), we can readily verify that $w_R=U_I$, $w_I=-U_R$ is a solution of eqs.~(\ref{eq:wR}), (\ref{eq:wI}). Differentiating eqs.~(\ref{eq:0R}) and (\ref{eq:0I}) w.r.t.~$\theta$ gives

\be
k^2(U_R')'' - k(2\alpha-V)(U_I')' + (\beta-\alpha-|U_0|^2)U_R' - 2U_R^2U_R' - 2U_RU_IU_I' = 0,    \label{eq:0R'}   %\eqno(0R')
\ee

\be
k^2(U_I')'' + k(2\alpha-V)(U_R')' + (\beta-\alpha-|U_0|^2)U_I' - 2U_I^2U_I' - 2U_RU_IU_R' = 0.   \label{eq:0I'}   %\eqno(0I')
\ee

\ni Comparing eqs.~(\ref{eq:wR}) and (\ref{eq:wI}) with eqs.~(\ref{eq:0R'}) and (\ref{eq:0I'}) one sees that $w_R=U_R'$ and $w_I=U_I'$ is a second solution to eqs.~(\ref{eq:wR}) and (\ref{eq:wI}).
\par Now we write eq.~(\ref{eq:2.6}) in terms of its real and imaginary parts, $(\L U_1)_R + F_R = 0$ and $(\L U_1)_I + F_I = 0$, and then combine them as

$$
U_I(\L U_1)_R - U_R(\L U_1)_I + U_IF_R - U_RF_I = 0,   \qquad  U_R'(\L U_1)_R + U_I(\L U_1)_I + U_R'F_R + U_I'F_I = 0,
$$

\ni i.e.~using the two found zero-modes of $\L$. We integrate the last equations over the period in $\theta$ and integrate by parts terms with the derivatives in $(\L U_1)_R$ and $(\L U_1)_I$ making the derivatives act on the functions to the left. Finally, using eqs.~(\ref{eq:0R}), (\ref{eq:0I}) in the first equation and eqs.~(\ref{eq:0R'}), (\ref{eq:0I'}) in the second, we see that all terms depending on $U_1$ integrate to zero. Thus, we find two secularity conditions,

\be
\int_0^1(U_IF_R - U_RF_I)d\theta = 0,   \label{eq:2.17}   %\eqno(2.17)
\ee

\be
\int_0^1(U_R'F_R + U_I'F_I)d\theta = 0,   \label{eq:2.18}   %\eqno(2.18)
\ee

\ni where $F=F_R+iF_I$ is the part of eq.~(\ref{eq:2.6}) independent of $U_1$, i.e.

\be
F_R = -\prt_tU_I + 2k\prt_rU_R' + \prt_rkU_R' - 2\alpha\prt_rU_I - \prt_r\alpha U_I + \frac{kU_R' - \alpha U_I}{r},   \label{eq:2.19}   %\eqno(2.19)
\ee

\be
F_I = \prt_tU_R + 2k\prt_rU_I' + \prt_rkU_I' + 2\alpha\prt_rU_R + \prt_r\alpha U_R + \frac{kU_I' + \alpha U_R}{r}.   \label{eq:2.20}   %\eqno(2.20)
\ee

\ni Upon substituting eqs.~(\ref{eq:2.19}) and (\ref{eq:2.20}), the integrands in eqs.~(\ref{eq:2.17}) and (\ref{eq:2.18}) become

$$
-(U_IF_R - U_RF_I) = \frac{\prt_t(U_R^2+U_I^2)}{2} + 
$$

$$
+ 2k(U_R\prt_rU_I' - U_I\prt_rU_R') + \left(\prt_rk + \frac{k}{r}\right)(U_RU_I'-U_IU_R') + \left(\prt_r + \frac{1}{r}\right)\left(\alpha(U_R^2+U_I^2)\right),
$$

$$
-(U_R'F_R + U_I'F_I) = U_I'\prt_tU_R - U_R'\prt_tU_I + 
$$

$$
+ \left(\prt_r + \frac{1}{r}\right)\left(k((U_R')^2 + (U_I')^2)\right) + 2\alpha(U_I'\prt_rU_R - U_R'\prt_rU_I) + \left(\prt_r\alpha + \frac{\alpha}{r}\right)(U_RU_I'-U_IU_R').   
$$

\ni Next we transform the secularity conditions using the periodicity of $U_0$ so that all derivatives $f'(\theta)$ integrate to zero. Taking the appropriate linear combinations of eqs.~(\ref{eq:0R}) and (\ref{eq:0I}), we see that the terms appearing in the integrands of eqs.~(\ref{eq:2.17}) and (\ref{eq:2.18}) can be expressed as

\be
U_RU_I'-U_IU_R' = A_0^2\Phi_0' = C_0 + \frac{V-2\alpha}{2k}\rho_0,    \label{eq:2.21}   %\eqno(2.21)
\ee

\ni and also

$$
k^2((U_R')^2 + (U_I')^2) = k^2(A_0')^2 + k^2A_0^2(\Phi_0')^2 = 
$$

\be
= k^2\frac{(\rho_0')^2 + 4C_0^2}{4\rho_0} + (V-2\alpha)kC_0 + \frac{(V-2\alpha)^2}{4}\rho_0.   \label{eq:2.22}   %\eqno(2.22)
\ee

\ni We introduce the slow variable (the leading-order density averaged over the period)

$$
Q(t,r) = \int_0^1\rho_0(\theta; t,r)d\theta = \int_0^1|U_0|^2d\theta = \int_0^1(U_R^2+U_I^2)d\theta.
$$

\ni For the function $\rho_0$ of the form eq.~(\ref{eq:2.11}), it equals the expression in eq.~(\ref{eq:Ql}). Then we apply the identities

$$
\int_0^1(U_R\prt U_I' - U_I\prt U_R')d\theta = \int_0^1[(U_R\prt U_I - U_I\prt U_R)' - (U_R'\prt U_I - U_I'\prt U_R)]d\theta = 
$$

\be
= \int_0^1(U_I'\prt U_R - U_R'\prt U_I)d\theta = \frac{1}{2}\prt \int_0^1(U_RU_I' - U_IU_R')d\theta = \frac{1}{2}\prt \left(C_0+\frac{(V-2\alpha)Q}{2k}\right),   \label{eq:Id1}   %\eqno(Id1)
\ee

\ni where '$\prt$' stands for either $\prt_r$ or $\prt_t$. Using $A_0^2=\rho_0$, eqs.~(\ref{eq:2.21}), (\ref{eq:2.22}), (\ref{eq:Id1}) and (\ref{eq:A4}) we bring eqs.~(\ref{eq:2.17}) and (\ref{eq:2.18}) to the form 

\be
\prt_tQ + \left(\prt_r + \frac{1}{r}\right)\left(VQ + 2kC_0\right) = 0,   \label{eq:2.23}   %\eqno(2.23)
\ee

\be
\prt_t\left(C_0 + \frac{(V-2\alpha)}{2k}Q\right) + \left(\prt_r + \frac{1}{r}\right)\left(2(V-\alpha)C_0 + \frac{(V-2\alpha)}{2k}VQ + 2\frac{2C_1 - \Omega Q}{3k}\right) = 0,  \label{eq:2.24}   %\eqno(2.24)
\ee

\ni respectively. The last equations would complete the Whitham system for rNLS when considered together with eqs.~(\ref{eq:Ck}) and (\ref{eq:a}) written as

\be
\prt_tk + \prt_r(Vk) = 0,   \label{eq:Ck1}   %\eqno(k)  
\ee

\be
\prt_t\alpha + \prt_r\left(V\alpha - \frac{V^2}{4} + \Omega\right) = 0.   \label{eq:a1}  %\eqno(a)  
\ee

 Equations (\ref{eq:Ck1}), (\ref{eq:a1}), (\ref{eq:2.23}), (\ref{eq:2.24}) make another complete Whitham system; using eqs.~(\ref{eq:2.13}), (\ref{eq:2.14}), (\ref{eq:al}), they can be written in terms of $\lambda_j, j=1,2,3 \text{~and~} V$. The relations of these equations with the hydrodynamic approach of section 3 are presented below. 
 
\subsection{Relations of two-phase and hydrodynamic approaches}
 
We first express $\Psi = \sqrt\rho e^{i\Theta}$, with $\rho$ and $\Theta$ real, and %, as before, 

\be
\rho = \rho(\theta,r,t; \epsilon),  \qquad \Theta = \phi + \Phi(\theta,r,t; \epsilon),   \label{eq:C3.2}   %\eqno(3.2)
\ee

$$
\prt_r\theta = \frac{k}{\epsilon},  \qquad \prt_t\theta = -\frac{kV}{\epsilon}, \qquad \prt_r\phi = \frac{\alpha}{\epsilon}, \qquad \prt_t\phi = -\frac{\beta}{\epsilon},
$$

\ni The hydrodynamic velocity $u$ is introduced as

\be
u = \epsilon\prt_r\Theta = \alpha + \epsilon\prt_r\Phi.     \label{eq:C3.3}    %\eqno(3.3)
\ee

\ni Then the imaginary part of eq.~(\ref{eq:1.1}) is

\be
\epsilon\prt_t\rho + \left(\epsilon\prt_r + \frac{\epsilon}{r}\right)(2\rho u) = 0,  \label{eq:C3.4}   %\eqno(3.4)
\ee

\ni while its real part can be written as

\be
\epsilon\prt_t\Theta + u^2 + \rho = \frac{\epsilon^2\prt_{rr}\rho}{2\rho} - \frac{\epsilon^2(\prt_r\rho)^2}{4\rho^2} + \frac{\epsilon^2\prt_r\rho}{2\rho r}.  \label{eq:C3.5}  %\eqno(3.5)
\ee

\ni Unlike the imaginary part, it contains the total phase $\Theta$, besides the hydrodynamic density $\rho$ and velocity $u$. Using multiscale form of the Whitham approach, we present the derivatives as sums of fast and slow ones, 
$$\epsilon\prt_rf = (k\prt_{\theta} + \epsilon\tilde\prt_r)f,    \qquad   \epsilon\prt_tf = (-kV\prt_{\theta} + \epsilon\tilde\prt_t)f$$
for $f=\rho$ or $f=u$, 
while 
$$\epsilon\prt_r\Theta = \alpha + (k\prt_{\theta} + \epsilon\tilde\prt_r)\Phi,   \qquad  \epsilon\prt_t\Theta = -\beta + (-kV\prt_{\theta} + \epsilon\tilde\prt_t)\Phi.$$
Let also $f' \equiv \prt_\theta f$ as before. 
\par Then we expand  in $\epsilon$, $\rho = \rho_0 + \epsilon\rho_1 + \dots$, $u = u_0 + \epsilon u_1 + \dots$ and $\Phi = \Phi_0 + \epsilon\Phi_1 + \dots$. At the leading order eq.~(\ref{eq:C3.4}) yields $-kV\rho_0' + 2k(\rho_0u_0)' = 0$, so that its first integral is

\be
u_0 = \frac{V}{2} + \frac{kC_0}{\rho_0} = \frac{V}{2} - \sigma\frac{\sqrt{2\lambda_1\lambda_2\lambda_3}}{2\rho_0},   \qquad \sigma \equiv -\text{sign}(kC_0),  \label{eq:C3.6}   %\eqno(3.6)  
\ee

\ni where the integration constant (slow variable) $C_0$ is introduced to match the same quantity in section 2. We now turn to the real part of rNLS, eq.~(\ref{eq:C3.5}). Its leading order yields 

\be
-\beta - kV\Phi_0' + u_0^2 + \rho_0 = k^2\left(\frac{\rho_0''}{2\rho_0} - \frac{(\rho_0')^2}{4\rho_0^2}\right).  \label{eq:C3.10}   %\eqno(3.10)  
\ee

\ni Upon using eqs.~(\ref{eq:C3.6}), (\ref{eq:2.7}) and $\rho_0=A_0^2$, it is equivalent to eq.~(\ref{eq:2.8}) and therefore also integrates to eq.~(\ref{eq:2.10}). From eq.~(\ref{eq:C3.3}), it follows also that

\be
\Phi_0' = \frac{u_0-\alpha}{k} = \frac{C_0}{\rho_0} + \frac{V-2\alpha}{2k},   \label{eq:3.7}  %\eqno(3.7)  
\ee

\ni which is eq.~(\ref{eq:2.7}).
\par The next order of eq.~(\ref{eq:C3.5}) is (using $u_1 = k\Phi_1' + \prt_r\Phi_0$ following from eq.~(\ref{eq:C3.3}))

$$
\prt_t\Phi_0 + V\prt_r\Phi_0 + (2u_0-V)u_1 + \rho_1 = k^2\left(\frac{\rho_1''}{2\rho_0} - \frac{\rho_0'\rho_1'}{2\rho_0^2} - \frac{\rho_0''\rho_1}{2\rho_0^2} + \frac{(\rho_0')^2\rho_1}{2\rho_0^3}\right) + 
$$

\be
+ k\left(\frac{\prt_r\rho_0'}{\rho_0} - \frac{\rho_0'\prt_r\rho_0}{2\rho_0^2} + \frac{\rho_0'}{2\rho_0r}\right) + \prt_rk\frac{\rho_0'}{2\rho_0}.   \label{eq:C3.11}   %\eqno(3.11)  
\ee

\ni If we subtract eqs.~(\ref{eq:C3.11}) taken at two values of $\theta$ different by a period, we observe that all terms except the difference of terms $(\prt_t+V\prt_r)\Phi_0$ cancel out, from where we obtain the periodicity of $(\prt_t+V\prt_r)\Phi_0$. This implies the periodicity of $\Phi_0$ itself. Knowing this, integration of eq.~(\ref{eq:2.7}) over the period yields $\alpha$ as the averaged leading-order hydrodynamic velocity,

$$
\alpha = \int_0^1u_0d\theta = \frac{V}{2} + kC_0\int_0^1\frac{d\theta}{\rho_0},     %\eqno(3.12)  
$$

\ni which is eq.~(\ref{eq:al}).
\par We recall that the leading order of eq.~(\ref{eq:3.13}) is a total derivative in $\theta$ which integrates to
$$
\rho_0 + u_0^2 - Vu_0 + H_0 = k^2\left(\frac{\rho_0''}{2\rho_0} - \frac{(\rho_0')^2}{4\rho_0^2}\right),
$$
\ni where $H_0$ is the integration constant. By consistency with the leading order of eq.~(\ref{eq:C3.5}), eq.~(\ref{eq:C3.10}), one finds that 
$$H_0 = V\alpha-\beta= V^2/4 - \Omega.$$
The next order of eq.~(\ref{eq:3.13}) reads
$$
-kVu_1' + k(2u_0u_1+\rho_1)'  - k\left[\frac{\prt_{rr}\rho}{2\rho} - \frac{(\prt_r\rho)^2}{4\rho^2}\right]_1' - k\left(\frac{k\rho_0'}{2\rho_0r}\right)' +
$$
\be
+ \prt_tu_0 + \prt_r(u_0^2+\rho_0) - \prt_r\left(\frac{k^2\rho_0''}{2\rho_0} - \frac{k^2(\rho_0')^2}{4\rho_0^2}\right) = 0,   \label{eq:Cu1}
\ee

\ni where $[...]_1$ means terms order one in $\epsilon$ of the expression in the brackets. Integrating over the period and using the leading order equation gives 

$$
0 = \prt_t\int_0^1u_0d\theta + \prt_r(V\int_0^1u_0d\theta - H_0) =  \prt_t\int_0^1u_0d\theta + \prt_r\left(V\int_0^1u_0d\theta - \frac{V^2}{4} + \Omega\right),
$$

\ni which is exactly eq.~(\ref{eq:a}) with the identification eq.~(\ref{eq:al}).   
\par The $\theta$-derivative of eq.~(\ref{eq:C3.5}) reads

\be
\epsilon\prt_t\Phi' + (u^2+\rho)' = \epsilon^2\left(\frac{\prt_{rr}\rho}{2\rho} - \frac{(\prt_r\rho)^2}{4\rho^2} + \frac{\prt_r\rho}{2\rho r}\right)'.  \label{eq:3.14}    %\eqno(3.14)  
\ee

\ni With it, eq.~(\ref{eq:2.24}) also can be derived in the current approach. We now combine eq.~(\ref{eq:3.14}) with eq.~(\ref{eq:C3.4}) and, after some rearrangement of terms and cancellations, get

\be
\epsilon\prt_t(\rho\Phi') + \epsilon\left(\prt_r + \frac{1}{r}\right)\left(2u\rho\Phi' + \frac{\rho'\epsilon\prt_r\rho}{2\rho} - \frac{\epsilon\prt_r\rho'}{2}\right) + \rho\rho' = 0.  \label{eq:3.25}   %\eqno(3.25)  
\ee

\ni The leading order of eq.~(\ref{eq:3.25}) is another total $\theta$-derivative which integrates to

\be
\frac{k^2}{2}\left(\rho_0'' - \frac{(\rho_0')^2}{\rho_0}\right) = \frac{\rho_0^2}{2} + \rho_0k\Phi_0'(2u_0-V) + G_1,  \label{eq:3.26}    %\eqno(3.26)  
\ee

\ni where the integration ``constant" $G_1$ is fixed by comparison with eq.~(3.16) as $2G_1 = 4kC_0\alpha + H_1 = 2kC_0(2\alpha-V) - 2C_1$ (also eq.~(\ref{eq:2.7}) is used). At the next-to-leading order one finds from eq.~(\ref{eq:3.25}) (using the same notation as in eq.~(\ref{eq:Cu1})) %(3.23))

$$
-kV[\rho\Phi']_1' + k\left[2u\rho\Phi' + \frac{\rho'\epsilon\prt_r\rho}{2\rho} - \frac{\epsilon\prt_r\rho'}{2}\right]_1' + \left[\frac{\rho^2}{2}\right]_1' +
$$

\be
+ \prt_t(\rho_0\Phi_0') + \left(\prt_r + \frac{1}{r}\right)\left(2u_0\rho_0\Phi_0' + \frac{k(\rho_0')^2}{2\rho_0} - \frac{k\rho_0'}{2}\right) = 0.  \label{eq:3.27}   %\eqno(3.27)  
\ee

\ni Again, after integration of eq.~(\ref{eq:3.27}) in $\theta$ over the period, its first line gives zero while the second line integrates to a secularity condition,

\be
\prt_t\int_0^1\rho_0\Phi_0'd\theta + \left(\prt_r + \frac{1}{r}\right)\int_0^1\left(2u_0\rho_0\Phi_0' + \frac{k(\rho_0')^2}{2\rho_0}\right)d\theta = 0,   \label{eq:3.28}   %\eqno(3.28)  
\ee

\ni the last terms involving $\rho_0'$ also going away. After using eqs.~(\ref{eq:3.7}) and (\ref{eq:A4}), eq.~(\ref{eq:3.28}) becomes identical to eq.~(\ref{eq:2.24}).

\par Further consistency relations are found when we combine all the Whitham equations obtained so far, eqs.~(\ref{eq:Ck1}), (\ref{eq:a1}), (\ref{eq:(Q)}), (\ref{eq:P}), (\ref{eq:E}) and (\ref{eq:2.24}), and recall that $\alpha$ is determined by eq.~(\ref{eq:al}). Let 

\be
\alpha = \frac{V}{2} + \frac{\gamma}{2},  \qquad  \gamma = 2kC_0\int_0^1\frac{d\theta}{\rho_0},  \label{eq:defg}   %\eqno(defg)
\ee

\ni and rewrite all the Whitham equations we derived. From now on we will use the elementary symmetric functions $e_1$, $e_2$ and $e_3$ of the roots of the cubic in eq.~(\ref{eq:2.10}) which were introduced in eq.~(\ref{eq:2.13}), instead of the variables $\Omega$, $C_0$ and $C_1$. Then we have six Whitham PDEs,

$$
\prt_tk + \prt_r(Vk) = 0,   \eqno(\ref{eq:Ck1})   
$$

\be
\prt_t(V+\gamma) + \prt_r\left(V\gamma + \frac{V^2}{2} + e_1\right) = 0,   \label{eq:g}   %\eqno(g)   
\ee

\be
\prt_tQ + \left(\prt_r + \frac{1}{r}\right)\left(VQ - \sigma\sqrt{2e_3}\right) = 0,   \label{eq:CQ}  %\eqno(Q)   
\ee

\be
\prt_t(VQ - \sigma\sqrt{2e_3}) + \left(\prt_r + \frac{1}{r}\right)(V^2Q - 2V\sigma\sqrt{2e_3} + e_2) + \frac{e_2 - 2e_1Q}{3r} = 0,  \label{eq:CP}   %\eqno(P)   
\ee

$$
\prt_t\left((V^2+\frac{2e_1}{3})Q - 2V\sigma\sqrt{2e_3} + \frac{2e_2}{3}\right) + 
$$

\be
+ \left(\prt_r + \frac{1}{r}\right)\left(V(V^2+\frac{2e_1}{3})Q - (3V^2+2e_1)\sigma\sqrt{2e_3} + \frac{8Ve_2}{3}\right) = 0,   \label{eq:CE}   %\eqno(E)   
\ee

$$
-\prt_t\left(\frac{\sigma\sqrt{2e_3}}{2k} + \frac{Q}{2k}\gamma\right) + 
$$

\be
+ \left(\prt_r + \frac{1}{r}\right)\left(-(V-\gamma)\frac{\sigma\sqrt{2e_3}}{2k} - \frac{Q}{2k}V\gamma + \frac{2e_2 - e_1Q}{3k}\right) = 0.  \label{eq:F}   %\eqno(F)   
\ee

\ni This system of six PDEs may look overdetermined but it must be consistent. Therefore let us see which further nontrivial relations among the dependent variables it implies. 

\subsection{Transformation of the ``overdetermined" system}   \label{a:transf}

Upon using the operator $D = \prt_t + V\prt_r$ instead of partial time derivative $\prt_t$, the derived six PDEs take form, respectively,

\be
\frac{Dk}{k} + \prt_rV = 0,   \label{eq:Dk}   %\eqno(k)   
\ee

\be
D(V+\gamma) + \gamma\prt_rV + \prt_re_1 = 0,   \label{eq:Dg}  %\eqno(g)   
\ee

\be
DQ + Q\prt_rV - \prt_r(\sigma\sqrt{2e_3}) + \frac{VQ - \sigma\sqrt{2e_3}}{r} = 0,   \label{eq:DQ}   %\eqno(Q)   
\ee

$$
D(VQ - \sigma\sqrt{2e_3}) + (VQ - 2\sigma\sqrt{2e_3})\prt_rV - V\prt_r(\sigma\sqrt{2e_3}) + \prt_re_2 + 
$$

\be
+ \frac{(3V^2-2e_1)Q - 6V\sigma\sqrt{2e_3} + 4e_2}{3r} = 0,  \label{eq:DP}  %\eqno(P)   
\ee

$$
D\left((V^2+\frac{2e_1}{3})Q - 2V\sigma\sqrt{2e_3} + \frac{2e_2}{3}\right) + \left((V^2+\frac{2e_1}{3})Q - 4V\sigma\sqrt{2e_3} + \frac{8e_2}{3}\right)\prt_rV  -
$$

\be
- 2\sigma\sqrt{2e_3}\prt_re_1 - (V^2+2e_1)\prt_r(\sigma\sqrt{2e_3}) + 2V\prt_re_2 + \frac{V(V^2+\frac{2e_1}{3})Q - (3V^2+2e_1)\sigma\sqrt{2e_3} + \frac{8Ve_2}{3}}{r} = 0.   \label{eq:DE}  %\eqno(E)   
\ee

$$
-D\left(\frac{\sigma\sqrt{2e_3}}{2k} + \frac{Q}{2k}\gamma\right) - \left(\frac{\sigma\sqrt{2e_3}}{2k} + \frac{Q}{2k}\gamma\right)\prt_rV +
$$

\be
+ \prt_r\left(\frac{\sigma\sqrt{2e_3}}{2k}\gamma + \frac{2e_2 - e_1Q}{3k}\right) + \frac{1}{r}\left(-(V-\gamma)\frac{\sigma\sqrt{2e_3}}{2k} - \frac{Q}{2k}V\gamma + \frac{2e_2 - e_1Q}{3k}\right) = 0.  \label{eq:DF}   %\eqno(F)   
\ee

\ni Taking combination $(\ref{eq:DQ}) - Q\cdot(\ref{eq:Dk})$, we get

\be
DQ - Q\frac{Dk}{k} - \prt_r(\sigma\sqrt{2e_3}) + \frac{VQ - \sigma\sqrt{2e_3}}{r} = 0.   \label{eq:DQ1}   %\eqno(Q1)   
\ee
This and some other equations here and at the end of section 3 contain relatively simple $r$-derivative terms polynomial when expressed in Riemann $r_j$-variables which makes them useful intermediate equations for diagonalization of the Whitham system with respect to the derivatives of $r_j$-variables (completed in appendix B).
\par Taking combination $(\ref{eq:DP}) - V\cdot(\ref{eq:DQ1}) - (VQ-2\sigma\sqrt{2e_3})\cdot(\ref{eq:Dk})$, we get, after some cancellations,

\be
QDV - D(\sigma\sqrt{2e_3}) + 2\sigma\sqrt{2e_3}\frac{Dk}{k} + \prt_re_2 + \frac{-3V\sigma\sqrt{2e_3} + 4e_2 - 2e_1Q}{3r} = 0.   \label{eq:DP1}   %\eqno(P1)
\ee

\ni Taking combination \\
$(\ref{eq:DE}) - 2V\cdot(\ref{eq:DP1}) - (V^2+2e_1)\cdot(\ref{eq:DQ1}) - \left((V^2+\frac{2e_1}{3})Q - 4V\sigma\sqrt{2e_3} + \frac{8e_2}{3}\right)\cdot(\ref{eq:Dk})$, we get, after a number of cancellations,

\be
-2\sigma\sqrt{2e_3}DV + \frac{2Q}{3}De_1 - \frac{4e_1}{3}(DQ - Q\frac{Dk}{k}) + \frac{2}{3}(De_2 - 4e_2\frac{Dk}{k}) - 2\sigma\sqrt{2e_3}\prt_re_1 = 0.   \label{eq:DE1}   %\eqno(E1)
\ee

\ni Taking combination $(\ref{eq:Dg}) - \gamma\cdot(\ref{eq:Dk})$, we get

\be
D(V+\gamma) - \gamma\frac{Dk}{k} + \prt_re_1 = 0.   \label{eq:Dg1}   %\eqno(g1)   
\ee

\par Comparing eq.~(\ref{eq:DE1}) with eq.~(\ref{eq:Dg1}) yields a nontrivial relation between $D$-derivatives of $\gamma$ and other quantities,

\be
kD\left(\frac{\gamma}{k}\right) = D\gamma - \gamma\frac{Dk}{k} = -\frac{QDe_1 - 2e_1(DQ - Q\frac{Dk}{k}) + De_2 - 4e_2\frac{Dk}{k}}{3\sigma\sqrt{2e_3}}.   \label{eq:gE}   %\eqno(gE)
\ee

\ni Finally, the combination $2\cdot(\ref{eq:DF}) + (Q/k)\cdot(\ref{eq:Dg1}) - (1/k)\cdot(\ref{eq:DP1}) + (\gamma/k)\cdot(\ref{eq:DQ1})$, after a number of cancellations and also using eq.~(\ref{eq:Dk}) yields 

\be
-\sigma\sqrt{2e_3}\prt_r\left(\frac{\gamma}{k}\right) = \frac{Q}{3k}\prt_re_1 - \frac{2e_1}{3}\prt_r\left(\frac{Q}{k}\right) - \frac{\prt_re_2}{k} + \frac{4}{3}\prt_r\left(\frac{e_2}{k}\right). \label{eq:DF1}   %\eqno(F1)
\ee

\ni The last equation is readily seen to be the same relation for the $\prt_r$-derivatives as eq.~(\ref{eq:gE}) is for the $D$-derivatives. Together they imply

\be
kd\left(\frac{\gamma}{k}\right) = d\gamma - \gamma\frac{dk}{k} = -\frac{Qde_1 - 2e_1(dQ - Q\frac{dk}{k}) + de_2 - 4e_2\frac{dk}{k}}{3\sigma\sqrt{2e_3}},   \label{eq:dg}   %\eqno(dg)
\ee

\ni as a general relation for the differentials, which allows one to effectively eliminate variable $\gamma$. (E.g.~this identity allows one to avoid using an explicit expression such as eq.~(\ref{eq:A7}) and simplify manipulations with the 1dNLS or rNLS Whitham systems.) This identity illuminates why different Whitham systems in `physical' variables can be obtained here with different approaches to their derivation.

\section{Cylindrical KdV and its Whitham system}    \label{a:cKdV}

The cylindrical KdV (cKdV) equation

\be
\prt_tu + u\prt_xu + \epsilon^2\prt_{xxx}u + \frac{u}{2(t+t_0)} = 0,    \label{eq:cKdV}
\ee
where $\epsilon\ll 1$ is small dispersion parameter and $t_0$ is a constant, leads to the following Whitham modulation system~\cite{ADM} which we rewrite here in a compact form:

\be
\prt_tr_i + v_i\prt_xr_i + \frac{2r_i - v_i}{2(t+t_0)} = 0,   \qquad  i=1,2,3,   \label{eq:WcKdV}
\ee
where $r_i$, $r_1\le r_2 \le r_3$, are the Riemann variables and $v_i$ are the KdV velocities~\cite{Whitham74} as functions of $r_i$,

\be
v_i = V + \frac{1}{3\prt_ik/k} = \frac{r_1+r_2+r_3}{3} + \frac{1}{3\prt_ik/k},   \label{eq:vKdV}
\ee
where $\prt_ik/k \equiv {\prt \ln k}/{\prt r_i}$ are logarithmic derivatives of $k$ with respect to the Riemann variables $r_1,r_2,r_3$, given by 
\be
\frac{\prt_1k}{k} = -\frac{(1 - E/K)}{2(r_2-r_1)},   \quad  \frac{\prt_2k}{k} = \frac{1}{2}\left( \frac{1-E/K}{r_2-r_1} - \frac{E/K}{r_3-r_2} \right),  \quad  \frac{\prt_3k}{k} = \frac{E/K}{2(r_3-r_2)},   \label{eq:djk}
\ee
and $K=K(m)$, $E=E(m)$ are the first and second complete elliptic integrals, respectively, with $m=(r_2-r_1)/(r_3-r_1)$. We find analytically the cKdV DSW edge dynamics which was not done in~\cite{ADM}. These results agree with numerics of~\cite{ADM} for the initial conditions $r_1(x,0) = 0$, $r_3(x,0) = 1$ and $r_2(x,0)=0$ for $x<0$, $r_2(x,0)=1$ for $x>0$, taken there.
\par At the leading (solitonic) edge, where $m=1$, we have

\be
r_2=r_3,   \qquad  v_1=r_1,  \qquad   v_2=v_3=\frac{r_1+2r_3}{3},   \label{eq:m1}
\ee
while at the trailing (linear) edge, where $m=0$, we have

\be
r_2=r_1,   \qquad  v_3=r_3,  \qquad   v_2=v_1=2r_1-r_3.   \label{eq:m0}
\ee

\ni Substituting eq.~(\ref{eq:m1}) into the Whitham equations (\ref{eq:WcKdV}) leads to the two different equations,

\be
\prt_tr_1 + r_1\prt_xr_1 + \frac{r_1}{2(t+t_0)} = 0,   \qquad  \prt_tr_3 + \frac{r_1+2r_3}{3}\prt_xr_3 + \frac{4r_3-r_1}{6(t+t_0)} = 0.   \label{eq:Wse}   
\ee
Their solutions at the leading edge for the initial conditions $r_1(x,0) = 0$ and $r_3(x,0) = 1$, taken in~\cite{ADM}, are found by first noting that first equation for $r_1$ is independent of $r_3$ and is such that $r_1$ remains zero being identically zero initially. Then one solves the second equation for $r_3$, which, ahead of the leading edge where $\prt_xr_3=0$, becomes a simple ODE,

$$
\frac{dr_3}{dt} = \prt_tr_3 = -\frac{2r_3}{3(t+t_0)}.
$$

%$$
%\frac{dr_3(x_+(t), t)}{dt} = \prt_tr_3(x_+(t), t) + v_3(x_+(t), t)\prt_xr_3(x_+(t), t) = -\frac{2r_3(x_+(t), t)}{3(t+t_0)}.
%$$
\ni Thus, one obtains

\be
r_1(x_+(t), t) = 0,   \qquad   r_3(x_+(t), t) = \frac{t_0^{2/3}}{(t+t_0)^{2/3}}.   \label{eq:solse}
\ee

\ni The solitonic edge $x_+(t)$ moves with velocity $v_+(t) = v_2(x_+(t), t)=v_3(x_+(t), t)$ which implies (since $x_+(0)=0$)

\be
x_+(t) =  2t_0^{2/3}[(t+t_0)^{1/3} - t_0^{1/3}],   \qquad  v_+(t) \equiv \frac{dx_+}{dt} = \frac{2}{3}\frac{t_0^{2/3}}{(t+t_0)^{2/3}}.   \label{eq:se}
\ee

\ni Similarly, substituting eq.~(\ref{eq:m0}) into the Whitham equations (\ref{eq:WcKdV}) leads to the two different equations,

\be
\prt_tr_1 + (2r_1-r_3)\prt_xr_1 + \frac{r_3}{2(t+t_0)} = 0,   \qquad  \prt_tr_3 + r_3\prt_xr_3 + \frac{r_3}{2(t+t_0)} = 0,   \label{eq:Wle}
\ee
whose solutions are determined considering that behind the trailing edge $\prt_xr_1=\prt_xr_3=0$. One solves the second ODE containing only $r_3$, then substitutes it and solves the first ODE for $r_1$. Thus, also at the trailing edge, for the initial conditions $r_1(x,0) = 0$ and $r_3(x,0) = 1$, the solutions are

\be
r_1(x_-(t), t) = \frac{t_0^{1/2}}{(t+t_0)^{1/2}} - 1,   \qquad   r_3(x_-(t), t) = \frac{t_0^{1/2}}{(t+t_0)^{1/2}}.   \label{eq:solle}
\ee

\ni The linear edge $x_-(t)$ moves with velocity $v_-(t) = v_2(x_-(t), t)$ which implies (since $x_-(0)=0$)

\be
x_-(t) =  2(t+t_0)^{1/2}[t_0^{1/2} - (t+t_0)^{1/2}],   \qquad  v_-(t) \equiv \frac{dx_-}{dt} = \frac{t_0^{1/2}}{(t+t_0)^{1/2}} - 2.    \label{eq:le}
\ee

\ni Thus, both leading and trailing edges of the cKdV DSW move with time-dependent speeds.

\end{document}